\documentclass[12pt]{article}
\pdfoutput=1 

\usepackage[utf8]{inputenc}

\usepackage{amsmath,amsfonts,amssymb,epsfig}
\usepackage{slashed}
\usepackage{cite}
\usepackage{multirow}
\usepackage{multicol}
\usepackage{cite}

\usepackage{xcolor}
\usepackage[
   bookmarks=true,
   linktocpage=true,
   colorlinks=true,
   allbordercolors=white, allcolors=blue
]{hyperref}

\usepackage{calc}
\usepackage{rotating}
\usepackage[english]{babel}
\usepackage{pifont}

\usepackage{graphicx}
\usepackage{subfig}
\usepackage{float}
\usepackage{amsmath}
\usepackage{amssymb}
\usepackage{amsthm}
\usepackage{latexsym}
\usepackage{dcolumn}
\usepackage{geometry}
\usepackage{mciteplus}
\usepackage{fancyhdr}

\usepackage[normalem]{ulem}
\usepackage[utf8]{inputenc}
\usepackage{caption}
\usepackage{sectsty}

\allsectionsfont{\boldmath}

\def\gtwid{\mathrel{\raise.3ex\hbox{$>$\kern-.75em\lower1ex\hbox{$\sim
$}}}}
\def\vio{\mathrel{\hbox{$E$\kern-.60em\hbox{$/
$}}}}
\newcommand{\newc}{\newcommand*}

\long\def\begincomment#1\endcomment{%
        \begingroup\sf\baselineskip12pt#1\endgroup}

\newc{\etal}{\textrm{et al.}} 
\newc{\eg}{\textrm{e.g.}} 
\newc{\ie}{\textrm{i.e.}}
\newc{\etc}{\textrm{etc}}
\newc\vs{\textrm{vs.}}
\newc{\cl}{\rm {C.L.}}

\newc{\ev}{\ensuremath{\,\mathrm{eV}}}
\newc{\kev}{\ensuremath{\,\mathrm{keV}}}
\newc{\mev}{\ensuremath{\,\mathrm{MeV}}}
\newc{\gev}{\ensuremath{\,\mathrm{GeV}}}
\newc{\tev}{\ensuremath{\,\mathrm{TeV}}}
\newc{\MeV}{\mev} 
\newc{\TeV}{\tev}
\newc{\invpb}{\ensuremath{/\text{pb}}}
\newc{\invfb}{\ensuremath{/\text{fb}}}
\newc\nb{\ensuremath{\,\mathrm{nb}}} \newc\pb{\ensuremath{\,\mathrm{pb}}} \newc\fb{\ensuremath{\,\mathrm{fb}}}
\newc\pc{\ensuremath{\,\mathrm{pc}}}
\newc\kpc{\ensuremath{\,\mathrm{kpc}}}
\newc\mpc{\ensuremath{\,\mathrm{Mpc}}}
\newc\ps{\ensuremath{\,\mathrm{ps}}} 
\newc\cmeter{\ensuremath{\,\mathrm{cm}}} 
\newc\meter{\ensuremath{\,\mathrm{m}}} 
\newc\kmeter{\ensuremath{\,\mathrm{km}}}
\newc\second{\ensuremath{\,\mathrm{s}}}
\newc\msecond{\ensuremath{\,\mathrm{ms}}}
\newc\nsecond{\ensuremath{\,\mathrm{ns}}}
\newc\psecond{\ensuremath{\,\mathrm{ps}}}

\newc{\chisqmin}{\ensuremath{\chi^2_{\mathrm{min}}}}
\newc{\Delchisq}{\ensuremath{\Delta\chi^2}}
\newc{\chisq}{\ensuremath{\chi^2}}
\newc{\like}{\ensuremath{\mathcal{L}}}

\newc\lsim{\ensuremath{\mathrel{\rlap{\lower4pt\hbox{\hskip1pt$\sim$}}\raise1pt\hbox{$<$}}}}
\newc\gsim{\ensuremath{\mathrel{\rlap{\lower4pt\hbox{\hskip1pt$\sim$}}\raise1pt\hbox{$>$}}}}
\newc{\VEV}[1]{\ensuremath{\langle #1 \rangle}}
\newc{\dl}{\ensuremath{\stackrel{\leftarrow}{D}}}
\newc{\dr}{\ensuremath{\stackrel{\rightarrow}{D}}}

\newc{\bcenter}{\begin{center}}   
\newc{\ecenter}{\end{center}}
\newc{\bfl}{\begin{flushleft}}    
\newc{\efl}{\end{flushleft}}
\newc{\bfr}{\begin{flushright}}   
\newc{\efr}{\end{flushright}}

\newc{\bi}{\begin{itemize}}
\newc{\ei}{\end{itemize}}
\newc{\bed}{\begin{description}}
\newc{\eed}{\end{description}}
\newc{\ben}{\begin{enumerate}}
\newc{\een}{\end{enumerate}}

\newc{\be}{\begin{equation}}
\newc{\ee}{\end{equation}}
\newc{\bea}{\begin{eqnarray}}
\newc{\eea}{\end{eqnarray}}
\newc{\bfle}{\begin{flalign}}
\newc{\efle}{\end{flalign}}
\newc{\ra}{\rightarrow}

\newc{\alphas}{\ensuremath{\alpha_s}}
\newc{\alphatwo}{\ensuremath{\alpha_2}}
\newc{\alphaone}{\ensuremath{\alpha_1}}
\newc{\alphai}[1]{\ensuremath{\alpha_{#1}}}
\newc{\alphaem}{\ensuremath{\alpha_{\mathrm{em}}}}
\newc{\alphaeff}{\ensuremath{\alpha_{\mathrm{eff}}}}
\newc{\sineff}{\ensuremath{\sin \theta_{\mathrm{eff}}}}
\newc{\sinsqeff}{\ensuremath{\sin^2 \theta_{\mathrm{eff}}}}
\newc{\dalphahad}{\ensuremath{\Delta \alpha_{\mathrm{had}}}}
\newc{\yt}{\ensuremath{h_t}} \newc{\yb}{\ensuremath{h_b}} \newc{\ytau}{\ensuremath{h_{\tau}}}
\newc\mz{\ensuremath{M_Z}} 
\newc\mw{\ensuremath{m_W}}
\newc\mZ{\mz}        \newc\mW{\mw}
\newc\mhsm{\ensuremath{ m_{H_{\mathrm{SM}}}}}
\newc{\mtop}{\ensuremath{ m_t}}               \newc{\mtpole}{\ensuremath{ M_t}}
\newc{\mbottom}{\ensuremath{ m_b}} 
\newc{\mtau}{\ensuremath{ m_{\tau}}}
\newc{\mt}{\mtpole}
\newc{\mb}{\mbottom} 
\newc{\rtwogg}{\ensuremath{R_{h_2}(\gamma\gamma)}}
\newc{\rtwozz}{\ensuremath{R_{h_2}(ZZ)}}
\newc{\ronegg}{\ensuremath{R_{h_1}(\gamma\gamma)}}
\newc{\ronezz}{\ensuremath{R_{h_1}(ZZ)}}
\newc{\rsiggg}{\ensuremath{R_{h_\textrm{sig}}(\gamma\gamma)}}
\newc{\rsigzz}{\ensuremath{R_{h_\textrm{sig}}(ZZ)}}
\newc{\llbar}{\ensuremath{\ell\bar{\ell}}}
\newc{\tauptaum}{\ensuremath{ \tau^+\tau^-}}
\newc{\qqbar}{\ensuremath{ q\bar{q}}} \newc{\ppbar}{\ensuremath{ p\bar{p}}}
\newc{\bbbar}{\ensuremath{ b\bar{b}}} \newc{\ttbar}{\ensuremath{ t\bar{t}}}
\newc{\ffbar}{\ensuremath{ f\bar{f}}} \newc{\tautaubar}{\ensuremath{ \tau\bar{\tau}}}

\newc{\mchi}{\ensuremath{m_\neutone}}
\newc{\squark}{\ensuremath{\tilde{q}}}
\newc{\slepton}{\ensuremath{\tilde{l}}}
\newc{\gluino}{\ensuremath{\tilde{g}}} 
\newc{\mgluino}{\ensuremath{{m_{\gluino}}}}
\newc{\wino}{\ensuremath{\tilde{W}}} 
\newc{\mwino}{\ensuremath{{m_{\wino}}}}
\newc{\tone}{\ensuremath{{\tilde{t}_1}}}
\newc{\Hone}{\ensuremath{{\tilde{H}_{1}}}}
\newc{\Htwo}{\ensuremath{{\tilde{H}_{2}}}}
\newc{\Hhtwo}{\ensuremath{{H_{2}}}}
\newc{\qli}{\ensuremath{{\tilde{Q}_{i}}}}
\newc{\uri}{\ensuremath{{\tilde{u}_{i}}}}
\newc{\dri}{\ensuremath{{\tilde{d}_{i}}}}
\newc{\lli}{\ensuremath{{\tilde{L}_{i}}}}
\newc{\eri}{\ensuremath{{\tilde{e}_{i}}}}

\newc{\sthw}{\ensuremath{ \sin\theta_W}}              \newc{\cthw}{\ensuremath{\cos\theta_W}}
\newc{\tanthw}{\ensuremath{ \tan\theta_W}}              \newc{\cotthw}{\ensuremath{\cot\theta_W}}
\newc{\ssqthw}{\ensuremath{\sin^2 \theta_W}}
\newc{\msbar}{\ensuremath{\overline{MS}}} \newc{\drbar}{\ensuremath{\overline{DR}}}
\newc{\mtmtsmmsbar}{\ensuremath{ m_t(m_t)^{\msbar}_{{\mathrm{SM}}}}}
\newc{\mtmtsmdrbar}{\ensuremath{ m_t(m_t)^{\drbar}_{{\mathrm{SM}}}}}
\newc{\mtmtmssmdrbar}{\ensuremath{ m_t(m_t)^{\drbar}_{{\mathrm{SUSY}}}}}
\newc{\mbmbmsbar}{\ensuremath{ m_b(m_b)^{\msbar} }}
\newc{\mbmbsmmsbar}{\ensuremath{ m_b(m_b)^{\msbar}_{{\mathrm{SM}}}}}
\newc{\mbmzsmmsbar}{\ensuremath{ m_b(\mz)^{\msbar}_{{\mathrm{SM}}}}}
\newc{\mbmzsmdrbar}{\ensuremath{ m_b(\mz)^{\drbar}_{{\mathrm{SM}}}}}
\newc{\mbmzmssmdrbar}{\ensuremath{ m_b(\mz)^{\drbar}_{{\mathrm{SUSY}}}}}
\newc{\mtaumzsmmsbar}{\ensuremath{ m_{\tau}(\mz)^{\msbar}_{{\mathrm{SM}}}}}
\newc{\mtaumzsmdrbar}{\ensuremath{ m_{\tau}(\mz)^{\drbar}_{{\mathrm{SM}}}}}
\newc{\mtaumzmssmdrbar}{\ensuremath{ m_{\tau}(\mz)^{\drbar}_{{\mathrm{SUSY}}}}}
\newc{\alphasmzms}{\ensuremath{\alpha_s(M_Z)^{\overline{MS}}}}
\newc{\alphaimzms}[1]{\ensuremath{\alpha_{#1}(M_Z)^{\overline{MS}}}}
\newc{\alphaemmz}{\ensuremath{\alpha_{\mathrm{em}}(M_Z)^{\overline{MS}}}}

\newc{\mzero}{\ensuremath{{m_0}}}
\newc{\mhalf}{\ensuremath{ m_{1/2}}}
\newc{\tanb}{\ensuremath{\tan\beta}}
\newc{\azero}{\ensuremath{ A_0}}
\newc{\signmu}{\ensuremath{\rm{sgn}\,\mu}}
\newc{\atau}{\ensuremath{{A_{\tau}}}}
\newc{\mueff}{\ensuremath{\mu_{\rm{eff}}}}
\newc{\lam}{\ensuremath{{\lambda}}}
\newc{\kap}{\ensuremath{{\kappa}}}
\newc{\alam}{\ensuremath{{A_{\lambda}}}}
\newc{\akap}{\ensuremath{{A_{\kappa}}}}
\newc{\hs}{\ensuremath{ H_s}}      
\newc{\mhs}{\ensuremath{ m_{H_s}}} 
\newc{\mgut}{\ensuremath{ M_{\rm GUT}}}
\newc{\mvl}{\ensuremath{ M_{\rm VL}}}
\newc{\gut}{\ensuremath{{\rm GUT}}}
\newc{\mplanck}{\ensuremath{ M_{\rm P}}}      \newc{\mpl}{\ensuremath{ M_{\rm Pl}}}
\newc{\msusy}{\ensuremath{ M_{\rm SUSY}}}      \newc{\ms}{\ensuremath{ M_{\rm S}}}
 \newc{\hu}{\ensuremath{ H_u}}       \newc{\hd}{\ensuremath{ H_d}}
 \newc{\mhu}{\ensuremath{ m_{H_u}}}       \newc{\mhd}{\ensuremath{ m_{H_d}}}
 \newc{\mhuew}{\ensuremath{ m^{\ast}_{H_u}}}       \newc{\mhdew}{\ensuremath{ m^{\ast}_{H_d}}}
 \newc{\mhuewsq}{\ensuremath{ m^{\ast\, 2}_{H_u}}}       \newc{\mhdewsq}{\ensuremath{ m^{\ast\, 2}_{H_d}}}
 \newc{\mhl}{\ensuremath{m_\hl}} 
 \newc{\mhone}{\ensuremath{m_{h_1}}} 
 \newc{\mhtwo}{\ensuremath{m_{h_2}}} 
 \newc{\mhi}{\ensuremath{m_{\tilde{h}}}} 
 \newc{\mul}{\ensuremath{m_{\tilde{u}_L}}} 
 \newc{\mtone}{\ensuremath{m_{\tilde{t}_1}}} 
 \newc{\ma}{\ensuremath{m_A}} 
 \newc{\mH}{\ensuremath{m_H}} 
 \newc{\maone}{\ensuremath{m_{a_1}}} 
 \newc{\matwo}{\ensuremath{m_{a_2}}}
 \newc{\hone}{\ensuremath{h_1}}
 \newc{\htwo}{\ensuremath{h_2}}
 \newc{\aone}{\ensuremath{a_1}}
 \newc{\atwo}{\ensuremath{a_2}}
 \newc{\mqthree}{\ensuremath{m_{\tilde{Q}_3}^2}}
 \newc{\muthree}{\ensuremath{m_{\tilde{u}_3}^2}}
 \newc{\mqli}{\ensuremath{m_{\tilde{Q}_{i}}}}
 \newc{\muri}{\ensuremath{m_{\tilde{u}_{i}}}}
 \newc{\mdri}{\ensuremath{m_{\tilde{d}_{i}}}}
 \newc{\mlli}{\ensuremath{m_{\tilde{L}_{i}}}}
 \newc{\meri}{\ensuremath{m_{\tilde{e}_{i}}}}
 \newc{\ts}{\ensuremath{T_{SUSY}}}

\newc{\sigsip}{\ensuremath{\sigma^{\rm SI}_{p}}}	\newc{\sigsin}{\ensuremath{\sigma^{\rm SI}_{n}}}
\newc{\sigsdp}{\ensuremath{\sigma^{\rm SD}_{p}}}	\newc{\sigsdn}{\ensuremath{\sigma^{\rm SD}_{n}}}
\newc{\sigsi}{\ensuremath{\sigma^{\rm SI}}}	\newc{\sigsd}{\ensuremath{\sigma^{\rm SD}}}
\newc{\abund}{\ensuremath{ \Omega h^2}}
\newc{\omegadm}{\ensuremath{ \Omega_{{\rm DM}}}}     \newc{\abunddm}{\ensuremath{ \Omega_{{\rm DM}} h^2}} 
\newc{\omegam}{\ensuremath{ \Omega_{{\rm m}}}}       \newc{\abundm}{\ensuremath{ \Omega_{{\rm m}} h^2}}
\newc{\omegab}{\ensuremath{ \Omega_{{\rm b}}}}	\newc{\abundb}{\ensuremath{ \Omega_{{\rm b}} h^2}}
\newc{\omegatot}{\ensuremath{ \Omega_{{\rm TOT}}}}
\newc{\omegacdm}{\ensuremath{ \Omega_{{\rm CDM}}}}   \newc{\abundcdm}{\ensuremath{ \Omega_{{\rm CDM}} h^2}}
\newc{\omegalambda}{\ensuremath{ \Omega_{\Lambda}}} \newc{\abundlambda}{\ensuremath{ \Omega_{\Lambda} h^2}}
\newc{\omegarad}{\ensuremath{ \Omega_{{\rm rad}}}}  \newc{\abundrad}{\ensuremath{ \Omega_{{\rm rad}} h^2}}
\newc{\rhocrit}{\ensuremath{ \rho_{\rm crit}}}
\newc{\rhochi}{\ensuremath{ \rho_{\chi}}}
\newc{\abunchi}{\ensuremath{\Omega_\chi h^2}}
\newc{\abundlsp}{\ensuremath{\Omega_{\rm LSP}h^2}}

\newc{\amu}{\ensuremath{ a_{\mu}}}        \newc{\amususy}{\ensuremath{ a_{\mu}^{\mathrm{SUSY}}}}
\newc{\amuexpt}{\ensuremath{ a_{\mu}^{\mathrm{expt}}}}        \newc{\amusm}{\ensuremath{ a_{\mu}^{\mathrm{SM}}}}
\newc\deltaamu{\ensuremath{\Delta a_{\mu}}} \newc{\deltaamususy}{\ensuremath{\delta a_{\mu}^{\mathrm{SUSY}}}}
\newc\gmtwo{\ensuremath{ (g-2)_{\mu}}} 
\newc{\deltagmtwomususy}{\ensuremath{\delta\left(g-2\right)_{\mu}^{\mathrm{SUSY}}}}
\newc{\deltagmtwomu}{\ensuremath{\delta\left(g-2\right)_{\mu}}}
\newc{\deltagmtwoe}{\ensuremath{\delta\left(g-2\right)_{e}}}
\newc\BR{\ensuremath{\rm BR}}
\newc\bsgamma{\ensuremath{ b\rightarrow s \gamma }}
\newc\bxsgamma{\ensuremath{\overline{B}\rightarrow X_{s}\gamma}}
\newc\brbsgamma{\ensuremath{\BR\left(\bsgamma\right)}}
\newc\brbxsgamma{\ensuremath{\BR\left(\bxsgamma\right)}}
\newc\bsmumu{\ensuremath{B_s\to\mu^+\mu^-}}
\newc\brbsmumu{\ensuremath{\BR\left(B_s\to\mu^+\mu^-\right)}}
\newc\bdmmumu{\ensuremath{\overline{B}_d\to\mu^+\mu^-}}
\newc\bbbarmix{\ensuremath{\overline{B}_s\mbox{-}B_s}}      
\newc\delmbs{\ensuremath{\Delta M_{B_s}}}
\newc{\butaunu}{\ensuremath{B_u \rightarrow \tau \nu}}
\newc{\brbutaunu}{\ensuremath{\BR\left(B_u \rightarrow \tau \nu\right)}}

\newcommand*{\reftable}[1]{Table~\ref{#1}}         
       
\newcommand*{\reffig}[1]{Fig.~\ref{#1}}
 
        \newcommand*{\refeq}[1]{Eq.~(\ref{#1})}
     \newcommand*{\refsec}[1]{Sec.~\ref{#1}}

\newcommand*{\neutone}{\ensuremath{\tilde{\chi}^0_1}}

\let\oldcite\cite
\renewcommand*{\cite}{~\oldcite}

\newcommand*{\hl}{\ensuremath{h}}

\restylefloat{figure}

\newcommand{\email}[1]{\href{mailto:#1}{#1}}
 
\begin{document}

\title{\LARGE {\bf Stabilizing dark matter with quantum scale symmetry}}

\author{\\ Abhishek Chikkaballi,\footnote{\email{abhishek.chikkaballiramalingegowda@ncbj.gov.pl}}\,\,
Kamila Kowalska,\footnote{\email{kamila.kowalska@ncbj.gov.pl}}\\
Rafael R.~Lino dos Santos,\footnote{\email{rafael.santos@ncbj.gov.pl}}\,
and Enrico Maria Sessolo\footnote{\email{enrico.sessolo@ncbj.gov.pl}}\\[2ex]
\small {\em National Centre for Nuclear Research}\\
\small {\em Pasteura 7, 02-093 Warsaw, Poland  }\\
}
%
\date{}
\maketitle
\thispagestyle{fancy}
\begin{abstract}
In the context of gauge-Yukawa theories with trans-Planckian asymptotic safety, quantum scale symmetry 
can prevent the appearance in the Lagrangian of couplings that would otherwise be allowed by the gauge symmetry. Such couplings correspond to irrelevant 
Gaussian fixed points of the renormalization group flow. Their absence in the theory implies that different sectors of the gauge-Yukawa theory are secluded from one another, in similar fashion to the effects of a global or a discrete symmetry. As an example, we impose the trans-Planckian scale symmetry on a model of Grand Unification based on the gauge group SU(6), showing that it leads to the emergence of several fermionic WIMP dark matter candidates whose coupling strengths are entirely predicted by the UV completion.
\end{abstract}
\newpage 

\tableofcontents

\setcounter{footnote}{0}

\section{Introduction\label{sec:intro}}

Asymptotic safety~(AS) is the property of a 
quantum field theory to develop
fixed points of the renormalization group~(RG) flow of the action\cite{inbookWS}. 
Following the development of functional renormalization group~(FRG) techniques\cite{WETTERICH199390,Morris:1993qb}, 
it was shown in several papers that AS may 
arise in quantum gravity and provide the key ingredient for the non-perturbative renormalizability of the theory\cite{Reuter:1996cp,Lauscher:2001ya,Reuter:2001ag,Lauscher:2002sq,Litim:2003vp,Codello:2006in,Machado:2007ea,Codello:2008vh,Benedetti:2009rx,Dietz:2012ic,Falls:2013bv,Falls:2014tra}. Subsequent developments\cite{Percacci:2002ie,Percacci:2003jz,Zanusso:2009bs,Daum:2009dn,Daum:2010bc,Folkerts:2011jz,Dona:2013qba,Meibohm:2015twa,Oda:2015sma,Eichhorn:2016esv,Christiansen:2017cxa,Eichhorn:2017eht,Pawlowski:2020qer,Eichhorn:2022gku} further revealed that, given a particle theory in four space-time dimensions 
coupled to the gravitational action, the full system of gravity and matter may 
feature ultraviolet~(UV) fixed points in the energy regime where gravitational interactions become strong, thus planting the seeds for the exploration of AS in models of particle physics.

In a theory with RG fixed points, the coefficients of the operators of the quantum effective action may be divided in two broad classes on the basis of their scaling behavior in the vicinity of the fixed point. 
\textit{Irrelevant} couplings are infrared~(IR)-attractive, they are 
properly scale-invariant at the quantum level so that in one interpretation\cite{Wetterich:1987fm,Shaposhnikov:2008xb,Wetterich:2019qzx,Wetterich:2020cxq} 
they may be thought of as the fundamental couplings of the microscopic theory. \textit{Relevant} couplings, on the other hand, are IR-repulsive and thus drive 
deviations from the scaling solution near the fixed point. In the RG flow relevant parameters can identify -- at the point where they roughly become larger than~1 --
a characteristic scale of the low-energy theory. Typical examples include
the dimensionless Newton constant in asymptotically safe quantum gravity, the dimensionless Higgs potential mass-squared parameter in the Standard Model~(SM), and the strong gauge coupling in quantum
chromodynamics~(QCD), which generate, respectively, the Planck scale, the electroweak symmetry-breaking~(EWSB) scale, 
and $\Lambda_{\textrm{QCD}}$. Since relevant couplings are free parameters of the macroscopic theory, 
they essentially play the role of \textit{effective} quantities that cannot be predicted from first principles and thus must be determined experimentally. 

Vice versa, the \emph{a priori} free couplings of the matter Lagrangian that correspond to irrelevant directions become predictable in AS. The predictions of the scale-invariant 
theory can then be tested by phenomenological means. Early successes in this approach include a gravity-driven solution to the triviality problem in U(1) gauge 
theories\cite{Harst:2011zx,Christiansen:2017gtg,Eichhorn:2017lry}; a ballpark prediction for the value of the Higgs mass (more precisely, of the quartic coupling of the Higgs potential) obtained a few 
years ahead of its discovery\cite{Shaposhnikov:2009pv} (see also Refs.\cite{Eichhorn:2017als,Kwapisz:2019wrl,Eichhorn:2021tsx}); and the retroactive ``postdiction'' of the top-quark mass
value\cite{Eichhorn:2017ylw}. 

When it comes to gauge-Yukawa theories of phenomenological interest, 
a gravity-driven prediction of the 
top/bottom mass ratio of the SM was extracted in Ref.\cite{Eichhorn:2018whv}. 
Possible imprints of UV fixed points on the flavor structure of the SM and, in particular, 
the Cabibbo-Kobayashi-Maskawa matrix, were sought in Ref.\cite{Alkofer:2020vtb}, and an equivalent analysis for the Pontecorvo-Maki-Nakagawa-Sakata matrix elements 
can be found in Ref.\cite{Kowalska:2022ypk}. The impact of asymptotically safe quantum-gravity calculations on the renormalization group equations~(RGEs) of the Majorana mass term was 
investigated in detail in Refs.\cite{DeBrito:2019rrh,Hamada:2020vnf,Domenech:2020yjf}.
Predictions were also extracted for 
several models in relation to neutrino masses\cite{Grabowski:2018fjj,deBrito:2025ges}, flavor anomalies\cite{Kowalska:2020gie,Chikkaballi:2022urc},
the muon anomalous magnetic moment\cite{Kowalska:2020zve}, baryon number\cite{Boos:2022jvc,Boos:2022pyq}, and the asymptotically safe SM\cite{Pastor-Gutierrez:2022nki,Pastor-Gutierrez:2024sbt}.

In a couple of recent papers\cite{Kowalska:2022ypk,Chikkaballi:2023cce} (see also Ref.\cite{Eichhorn:2022vgp})
some of us showed that, in theories that include some 
Yukawa couplings constrained by low-scale observations to extremely small numbers 
(for example, the $\mathcal{O}(10^{-13})$ Yukawa coupling of a purely Dirac neutrino receiving its mass via the Higgs mechanism or, in more exotic cases, the tiny Yukawa coupling of a sterile-neutrino dark matter~(DM) candidate whose relic abundance is determined 
via the freeze-in mechanism\cite{McDonald:2001vt,Choi:2005vq,Kusenko:2006rh,Petraki:2007gq,Hall:2009bx}), 
such minuscule values may arise dynamically  
if the RG flow of the Yukawa coupling develops a Gaussian irrelevant fixed point. 
In that case, in fact, the coupling in question receives
a ``natural'' exponential suppression along the RG flow. A straightforward, yet unexplored consequence of the same idea is that the presence of an irrelevant Gaussian fixed point in the gauge-Yukawa sector can 
forbid the appearance of Lagrangian interactions that, while allowed by the gauge and global symmetries of the theory, may be in strong tension with observations.
In other words, in certain situations of phenomenological interest the fundamental scale symmetry of the UV completion may act ``transversely'' to the gauge and other symmetries to switch off some of the unwanted couplings. 

The presence of Gaussian fixed points 
can induce the emergence of DM in theories that have \textit{a priori} no dark sector. 
We consider here a theory of Grand Unification~(GUT) 
as a framework to accommodate the desired particle content beyond the SM~(BSM). It has long been known that most non-supersymmetric GUTs do not usually feature a fermionic DM candidate. 
The group (and its breaking chain) $\textrm{SO(10)}\to\textrm{SU(5)}\times\textrm{U(1)}\to\textrm{SM}$
does potentially contain fermion DM\cite{Ferrari:2018rey}, because the particles of the dark and visible sectors are separated by their U(1) charges. 
But this is typically not the case in GUTs based on $\textrm{SU($N$)}\to\textrm{SU(5)}\times\textrm{SU($N-5$)}\to\textrm{SM}$\cite{Rizzo:2022lpm}. In those cases, DM is either a pseudo-Nambu-Goldstone boson\cite{Cacciapaglia:2019ixa,Cai:2020njb,Otsuka:2022zdy,Chiang:2023omu}, or one is forced to introduce extra symmetries\cite{Ma:2020hyy}. In this paper, we show that quantum scale symmetry, in the form of irrelevant Gaussian fixed points of the trans-Planckian RGEs, can act as \textit{the} symmetry driving the emergence of DM in SU(6) and, by extension, higher rank SU($N$) GUTs. Moreover, in this scenario the coupling strengths of the DM become entirely predicted by the asymptotically safe UV completion. Once the correct value of the relic abundance is factored in, the DM mass scale becomes a prediction too, thus providing a path for the complete testability of the theory.

It is worth pointing out that, while AS may be a viable alternative to Grand Unification, in the sense that one does not really need a GUT to guarantee that all of the SM couplings remain finite up to arbitrary energies, 
a GUT provides nevertheless some advantages with respect to GUT-less scenarios with AS. On the one hand, it obviously creates an elegant framework for incorporating the BSM particle content. But, on a more subtle level, it also provides a theory that, unlike Refs.\cite{Kowalska:2022ypk,Chikkaballi:2023cce}, 
features a completely relevant, asymptotically free gauge sector. As we shall see below, this quality allows one to derive phenomenological predictions that depend to great extent solely on the scaling behavior of the matter couplings and not on the fixed-point value of the coefficients of the gravitational action.   
 
The paper is organized as follows. In \refsec{sec:FRG} we recall basic notions of trans-Planckian AS. In \refsec{sec:mech} we describe the way quantum scale symmetry may act transversely to the gauge symmetry 
to prevent some of the Yukawa couplings from appearing in the Lagrangian at any scale.  In Sec.~\ref{sec:su6} we discuss a particular realization of this mechanism in the framework of an SU(6) GUT. After defining the particle content of the model, we perform the UV fixed-point analysis of its gauge-Yukawa sector, derive predictions for the low-scale values of the gauge and Yukawa couplings, and comment on the qualitative UV features of the scalar potential. In \refsec{sec:DM} we discuss the DM properties of the model. We finally summarize our findings in \refsec{sec:sum}. Appendices contain, respectively, discussion of the scalar sector of the SU(6) model, trans-Planckian RGEs for the SU(6) gauge and Yukawa couplings, some analytical formulae for the DM relic abundance, and the most generic form of the low-scale Yukawa Lagrangian.

\section{General notions of asymptotic safety\label{sec:FRG}}

In AS, quantum gravity effects become important at approximately the Planck scale, where it is expected that graviton fluctuations start to contribute to the RG flow of the matter couplings. In a (renormalizable) matter theory with gauge and Yukawa interactions, the corresponding RGEs receive trans-Planckian contributions that at the leading order look like
\bea
\frac{d g_i}{d t}&=&\beta_i^{(\textrm{matter})}-f_g\, g_i \label{eq:betag} \\
\frac{d y_j}{d t}&=&\beta_j^{(\textrm{matter})}-f_y\, y_j \label{eq:betay}\,,
\eea
where $t=\ln\mu$ denotes the renormalization scale, $g_i$ and $y_j$ (with $i,j=1,2,3\dots$) are the set of gauge and Yukawa couplings, respectively,
and $\beta_{i,j}^{(\textrm{matter})}$ stand for the matter beta functions without gravity. 

The trans-Planckian objects $f_g$ and $f_y$, which parameterize leading gravitational corrections to the running of the gauge and Yukawa couplings, are expected to be universal
(in the sense that they multiply linearly all matter couplings of the same kind), as gravity does not distinguish between the internal degrees of freedom in the matter sector. In the framework of the FRG, they are determined by the fixed points of the operators of the gravitational action, and can be computed from first principles.  
While the derivation of $f_g$ and $f_y$ is subject to extremely large uncertainties, including the choice of truncation in the gravity/matter action, the selected renormalization scheme, and the gauge-fixing parameters\cite{Reuter:2001ag,Lauscher:2002sq,Percacci:2002ie,Percacci:2003jz,Codello:2007bd,Benedetti:2009rx,Narain:2009qa,Dona:2013qba,Falls:2017lst,Falls:2018ylp,deBrito:2022vbr},
it was nonetheless established that $f_g$ is likely not negative, irrespective of the chosen RG scheme\cite{Folkerts:2011jz,Christiansen:2017cxa}. While $f_g=0$ corresponds to respecting a particular classical symmetry of the gravitational action\cite{Folkerts:2011jz}, $f_g \neq 0$ may arise not just in mass-dependent schemes like the FRG, but in dimensional regularization too\cite{Toms:2008dq,Toms:2009vd}. Incidentally, as $f_g\geq 0$ supports asymptotic freedom in the non-abelian gauge sector of the SM, these results lend credit to the consistency of AS.

Unlike $f_g$, the gravitational contribution to the Yukawa coupling, $f_y$, is subject to somewhat greater uncertainty. It was investigated in a set of simplified models\cite{Rodigast:2009zj,Zanusso:2009bs,Oda:2015sma,Eichhorn:2016esv}, 
but no general results and definite conclusions regarding its size and sign are available (see, however, Ref.\cite{Pastor-Gutierrez:2022nki} for the most recent determination of $f_y$ in the SM).

Once the flow of the gravitational action develops a fixed point dynamically, 
a trans-Planckian fixed point in the matter sector may emerge as well.  
Such a fixed point, which we henceforth indicate with an asterisk, $\{g_i^\ast,y_j^\ast\}$, corresponds to a zero of the beta functions of the system of Eqs.~(\ref{eq:betag}) and (\ref{eq:betay}), given by the condition: $\beta_{i(j)}^{(\textrm{matter})}(g_i^\ast,y_j^\ast)-f_{g(y)}\,g_i^{\ast}(y_j^{\ast})=0$. One defines the stability matrix, $M_{ij}$, by linearizing the RG flow of the couplings $\{\alpha_k\}\equiv\{g_i,y_j\}$ around the fixed point,
\be\label{stab}
M_{ij}=\partial\beta_i/\partial\alpha_j|_{\{\alpha^{\ast}_k\}}\,.
\ee
Critical exponents $\theta_i$ are then defined as opposite of eigenvalues of the stability matrix and  they characterize the power-law evolution of the matter couplings in the vicinity of the fixed point. $\theta_i>0$ corresponds to a relevant and UV-attractive eigendirection.  All RG trajectories along this direction asymptotically reach the fixed point, so any deviation of a relevant coupling from the fixed point introduces a free parameter in the theory. It also means that the high-scale value of the coupling can always be adjusted to match its eventual measurement at the experimentally accessible energies.  

If $\theta_i<0$, the corresponding eigendirection is irrelevant and IR-attractive. In this case only one trajectory exists that the coupling's flow can follow towards the IR, thus either potentially providing a specific prediction for its value at the scale of phenomenological interest (if the fixed point is nonzero), or preventing the coupling from appearing in the theory (if the fixed point is zero). Finally, $\theta_i=0$ corresponds to a \textit{marginal} eigendirection. The RG running is logarithmically slow along this direction and an analysis beyond the linear order is required in order to determine whether a fixed point is UV-attractive or IR-attractive.

In this study, we are going to work with $\beta_{i,j}^{(\textrm{matter})}$ at the one-loop level. When it comes to the predictions of AS for the matter theory, it was shown in Ref.\cite{Kotlarski:2023mmr} that uncertainties stemming from neglecting higher-order contributions are extremely small, especially when considering couplings that remain in the perturbative regime along the entire RG flow.  
Note also that, at the first order in perturbation theory, the parameters of the scalar potential do not affect the gauge-Yukawa system, as they only enter \refeq{eq:betag} from the third-loop level up, and \refeq{eq:betay} from the second loop. Therefore, the gauge-Yukawa sector can be treated independently.

As a matter of fact, the quartic couplings of the scalar potential should also be subject to gravitational contributions to the beta function. 
In the trans-Planckian regime, the scalar beta functions are modified similarly to Eqs.~(\ref{eq:betag}) and (\ref{eq:betay}), with the gravitational corrections parameterized by $f_{\lam}$. One should keep in mind, however, that  gravitational corrections to the running couplings of the scalar potential do not need to be multiplicative, as some truncations of the gravitational action can generate contributions which are additive\cite{Eichhorn:2020kca}.
The impact of UV boundary conditions 
on the scalar potential of some realistic BSM theories was investigated, 
with respect to 
the relic abundance of DM in Refs.\cite{Reichert:2019car,Eichhorn:2020kca,deBrito:2023ydd}; for axion models in Ref.\cite{deBrito:2021akp}; in the context of dark energy and gravitational waves in Refs.\cite{Eichhorn:2022ngh,Eichhorn:2023gat}; and in GUTs in Refs.\cite{Eichhorn:2019dhg,Held:2022hnw}.

We conclude this section by emphasizing that 
assessing to what extent the FRG and other mass-dependent schemes can be considered the most appropriate tools for investigating/enforcing AS in quantum gravity is not trivial and it remains a very exciting topic of debate in the literature\cite{Donoghue:2019clr,Bonanno:2020bil,Branchina:2024xzh,Branchina:2024lai,Bonanno:2025xdg,Held:2025vkd}. It must be pointed out, however, that significant progress has recently been made on the Lorentzian RG approach to asymptotically safe gravity\cite{Fehre:2021eob,Pawlowski:2023gym,DAngelo:2023wje,Pastor-Gutierrez:2024sbt}. 

\section{Vanishing couplings from quantum scale symmetry\label{sec:mech}}

Let us consider a UV model that includes 
several copies of Weyl-fermion and scalar multiplets belonging to the same representation of a unified gauge group $\mathcal{G}$: $F_i=F_1, F_2, ... \simeq N\in  \mathcal{G}$ and  
$S_j=S_1, S_2, ... \simeq M\in  \mathcal{G}$. The model also includes at least one more fermion multiplet $G$
belonging to the representation $\bar{N}\times \bar{M}$, and/or a scalar multiplet $H$ included in $\bar{N}\times \bar{N}$.
All Yukawa couplings of the form 
$G F_i S_j$ (or $F_i F_j H$) become allowed by the symmetry. 

Let us focus on the $G F_i S_j$ case. One can schematically write down the Yukawa interaction part of the Lagrangian,
\be\label{eq:gen_lag}
\mathcal{L}\supset \sum_{i,j=1,2, ...} y_{ij} G F_i S_j
+\textrm{other Yukawa terms}+\textrm{H.c.}\,,
\ee
where ``other Yukawa terms'' include all unspecified matter content that features a Yukawa coupling with one 
or the other $F_i$, $S_j$ fields. 

It is easy to show that differences in the fixed-point scaling behavior of the $y_{ij}$ Yukawa couplings can emerge even though the Yukawa interactions in \refeq{eq:gen_lag} 
feature identical gauge quantum numbers. Let us first write down the generic trans-Planckian RGE for Yukawa couplings~$y_{ij}$ ($i,j=1,2,...$) at one loop,
\be\label{eq:betasys}
\frac{d y_{ij}}{dt}=\frac{1}{16\pi^2}\left[\left(\sum_{k,l=1,2,...} a_{kl}^{(ij)} y_{kl}^2+\sum_{y_r\in \textrm{other}} b_{r}^{(ij)} y_{r}^2+c^{(ij)} g^2_{\mathcal{G}}\right) y_{ij} +\sum_{m,p,q\neq (ij)} d_{mpq}^{(ij)}\, y_m y_p y_q\right]-f_y y_{ij}\,,
\ee
where $f_y$ parameterizes the leading-order graviton contribution to the beta function, calculated at the fixed point of the gravitational action,
$a_{kl}^{(ij)}, b_{r}^{(ij)}, c^{(ij)}$, and $d_{mql}^{(ij)}$ are real coefficients specific to the matter beta function of coupling $y_{ij}$,
the sum in $y_{r}$ spans all the Yukawa couplings that are not indicated explicitly in \refeq{eq:gen_lag}, $g_{\mathcal{G}}$~is the gauge coupling of the theory, and the last addend inside square-brackets
parameterizes the terms of the beta function that are not multiplicative in $y_{ij}$.  As we shall see, system~(\ref{eq:betasys}) in general admits several real fixed points. However, 
to make our case we can focus for the moment on the most minimal non-trivial solution, which is characterized by one single 
interactive Yukawa coupling, $y_{\hat{k}\hat{l}}^{\ast}\neq 0$, while all others are set at the Gaussian fixed point: $y_{m}^{\ast}=0$, with $m\neq (\hat{k}\hat{l})$. It is straightforward to see that such a solution always exists,
\be\label{eq:sol}
y_{\hat{k}\hat{l}}^{\ast}=\sqrt{\frac{16\pi^2 f_y- c^{(\hat{k}\hat{l})}g_{\mathcal{G}}^{\ast 2} }{  a_{\hat{k}\hat{l}}^{(\hat{k}\hat{l})}}}\,,
\ee
and that it can be real. We can further ground the discussion by assuming that the sole interactive Yukawa coupling of \refeq{eq:sol} is,
\textit{e.g.}, $y_{22}^{\ast}\neq 0$, and consider the case with $g_{\mathcal{G}}^{\ast}=0$, which is typical of asymptotically free gauge theories. 
The critical exponents of two Gaussian Yukawa couplings,
$y_{11}^{\ast}=y_{12}^{\ast}=0$, may thus be approximated as
\bea
\theta_{11}\approx - \frac{\partial \beta_{11}}{\partial y_{11}} &=& - \frac{a_{22}^{(11)}}{16\pi^2} y_{22}^{\ast 2} + f_y=\left(1-\frac{a_{22}^{(11)}}{a_{22}^{(22)}} \right) f_y\label{eq:cri11}\\
\theta_{12} \approx - \frac{\partial \beta_{12}}{\partial y_{12}} & = &- \frac{a_{22}^{(12)}}{16\pi^2} y_{22}^{\ast 2} + f_y=\left(1-\frac{a_{22}^{(12)}}{a_{22}^{(22)}} \right) f_y\label{eq:cri12}
\eea
(the off-diagonal terms of the stability matrix can be neglected as they do not contribute to $\theta_{11}$ and $\theta_{12}$ at the fixed point $y_{11}^{\ast}=y_{12}^{\ast}=0$).
It follows straightforwardly that, depending on the relative size of $a_{22}^{(11)}$, $a_{22}^{(12)}$, and $a_{22}^{(22)}$, Yukawa couplings 
$y_{11}$ and $y_{12}$ may emerge from the Gaussian fixed point with different scaling behavior. If, for example, $f_y>0$ and $a_{22}^{(11)}<a_{22}^{(22)}$, then 
$\theta_{11}>0$  and $y_{11}$ emerges from the fixed point along a relevant direction, effectively being a free parameter of the theory. 
At the same time, there is no guarantee that $a_{22}^{(12)}<a_{22}^{(22)}$ in \refeq{eq:cri12}, 
since we expect typically that $a_{22}^{(12)}> a_{22}^{(11)}$.

\begin{figure}[t]
\centering
\includegraphics[width=0.30\textwidth]{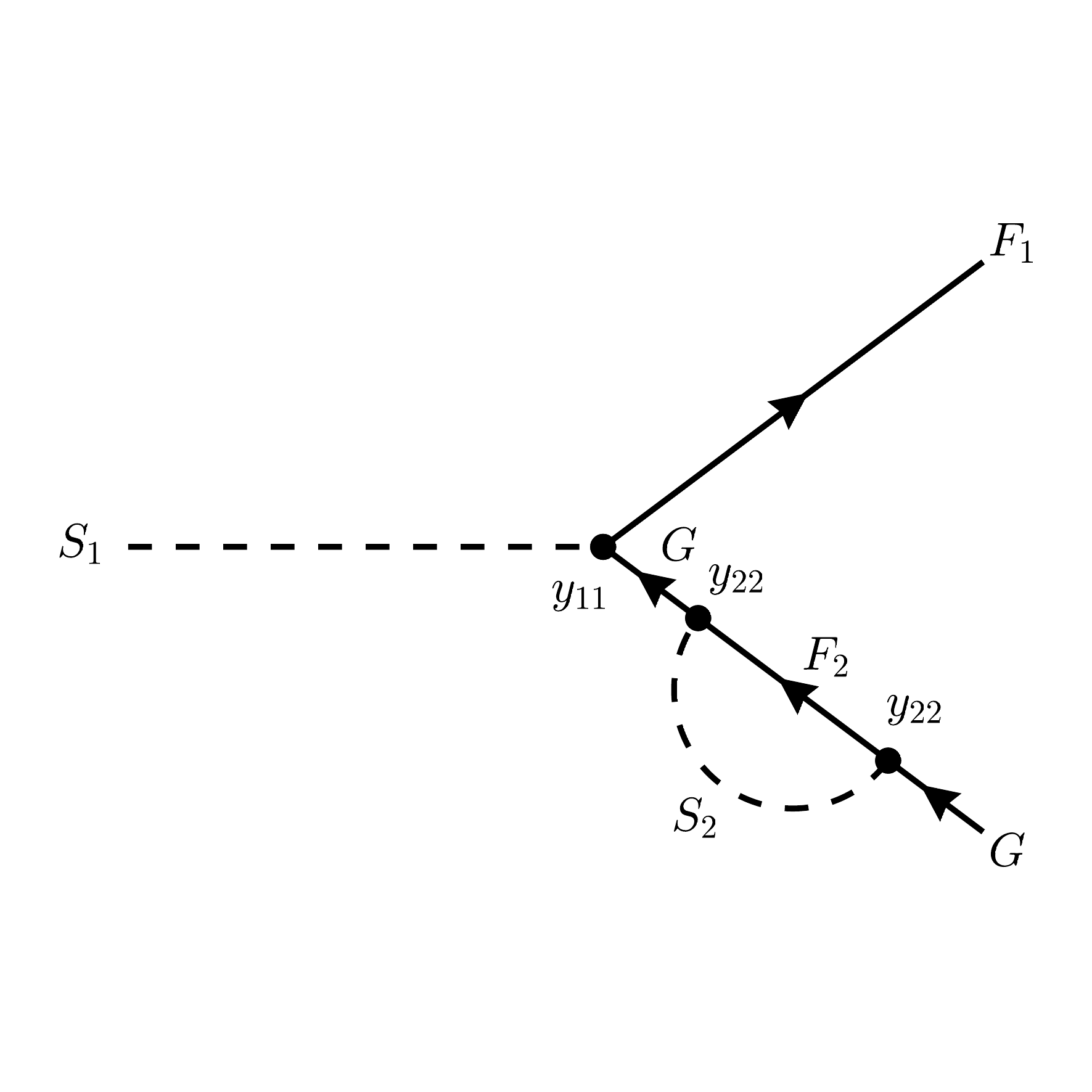}
\hspace{0.02\textwidth}
\includegraphics[width=0.30\textwidth]{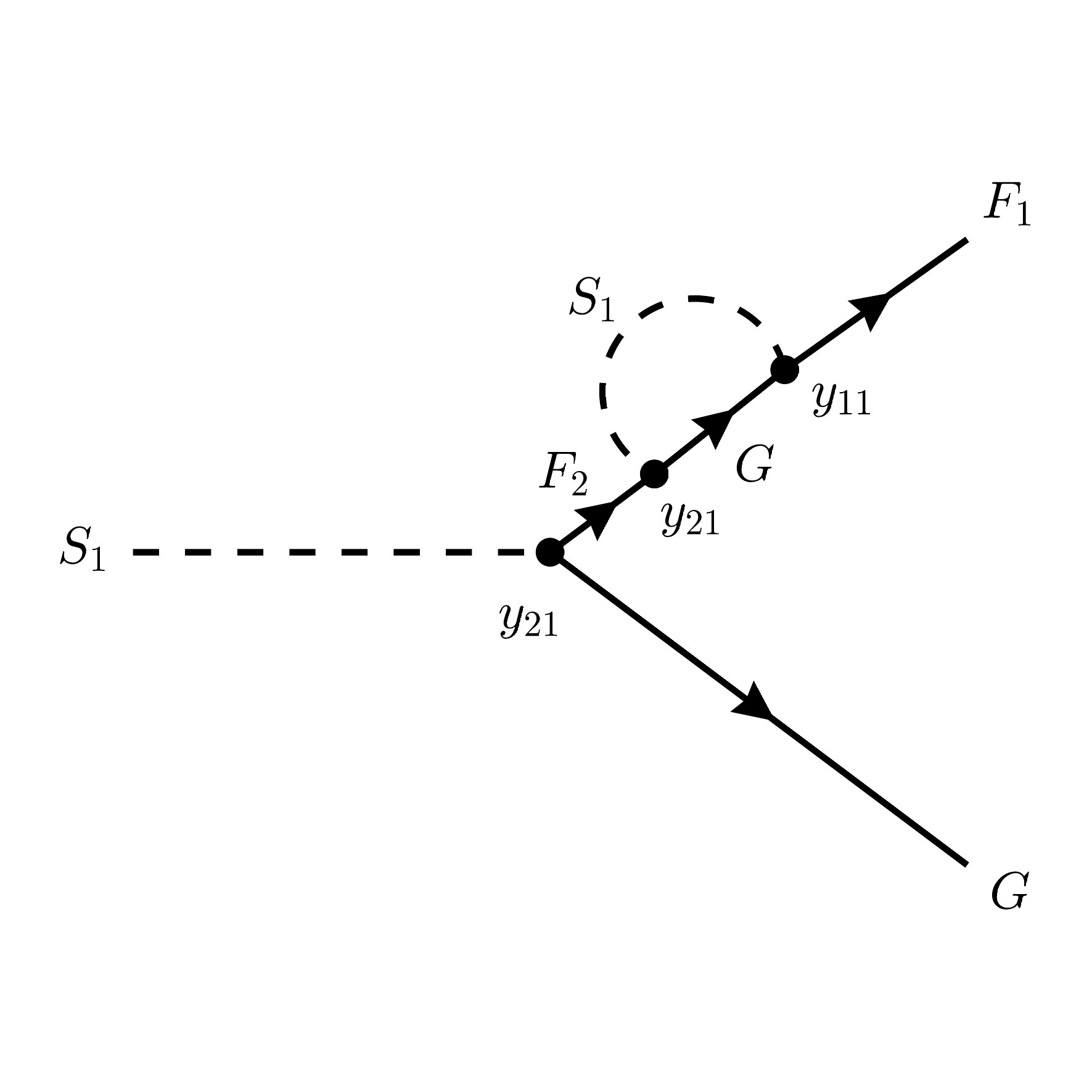}
\hspace{0.02\textwidth}
\includegraphics[width=0.30\textwidth]{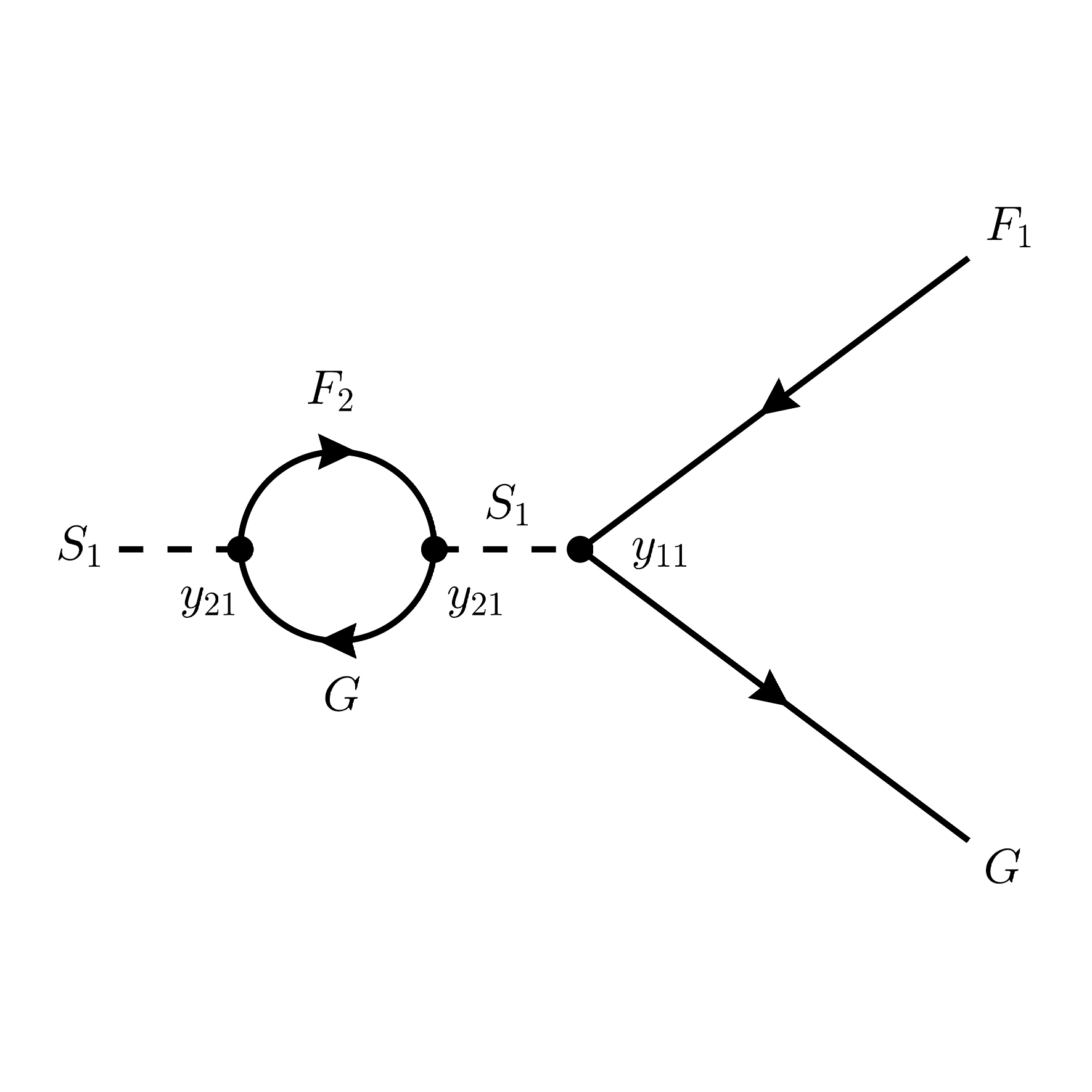}
\\
\subfloat[]{%
\includegraphics[width=0.30\textwidth]{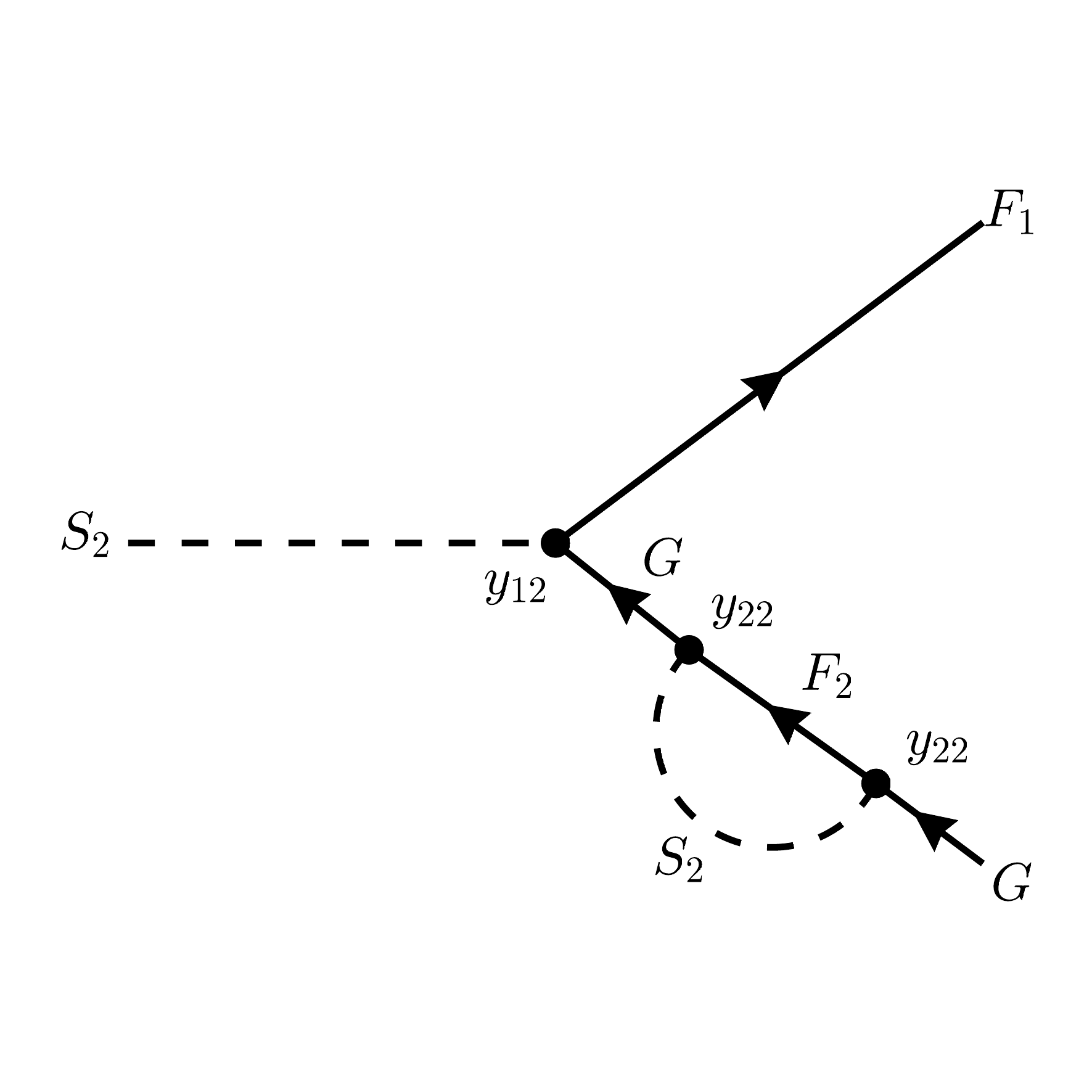}
}%
\hspace{0.02\textwidth}
\subfloat[]{%
\includegraphics[width=0.30\textwidth]{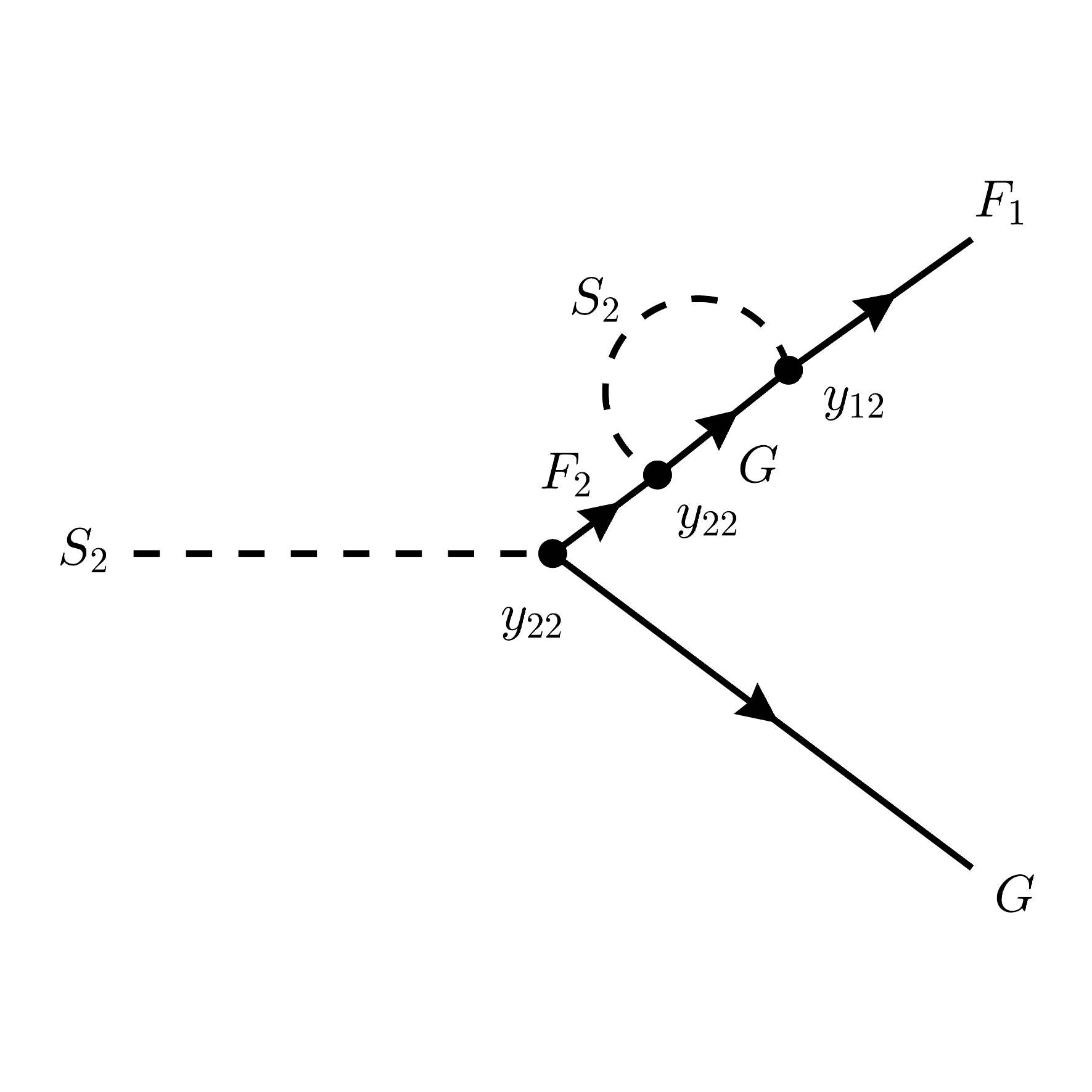}
}%
\hspace{0.02\textwidth}
\subfloat[]{%
\includegraphics[width=0.30\textwidth]{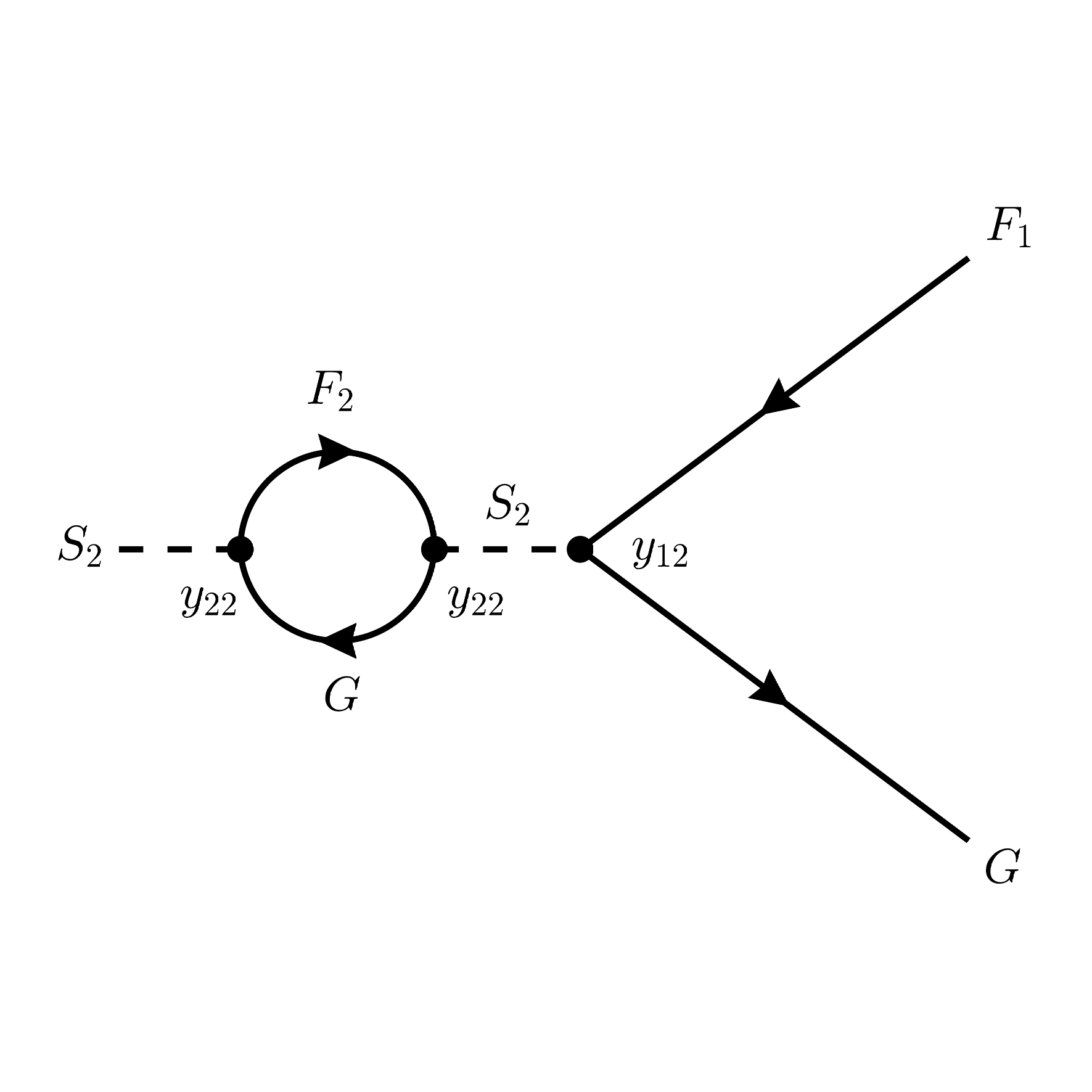}
}%
\caption{The three diagrams of the upper row contribute to the beta function of Yukawa coupling $y_{11}$. The three diagrams of the lower row contribute to the beta function of Yukawa coupling $y_{12}$. }  
\label{fig:diag}
\end{figure}

In order to clarify why this is the case, 
we present in the upper row of \reffig{fig:diag} some of the diagrams contributing to the one-loop beta function of Yukawa coupling~$y_{11}$. The lower row shows the corresponding 
diagrams for $y_{12}$.  At the Gaussian fixed point $y_{11}^{\ast}=y_{12}^{\ast}=y_{21}^{\ast}=0$,
Yukawa coupling $y_{22}^{\ast}\neq 0$ contributes equally to the two diagrams in \reffig{fig:diag}(a), which give $a_{22}^{(11)}$ and $a_{22}^{(12)}$. However, while diagrams~(b) and (c) 
of the lower row contribute to 
$a_{22}^{(12)}$, increasing its value, the corresponding diagrams of the upper row do not contribute to
$a_{22}^{(11)}$.  One can prove that, in the absence of other interactive couplings
at the fixed point, it will be $ a_{22}^{(11)}< a_{22}^{(12)} =a_{22}^{(22)}$, which leads to $\theta_{12}=0$. Yukawa coupling $y_{12}$ will in this case be marginal. However, by considering solutions with 
more interactive couplings at the fixed point it is possible to push $\theta_{12}$ to negative values, while $\theta_{11}$ remains positive.  As a result, $y_{12}$ will follow an irrelevant direction and 
will be forbidden to appear in the Lagrangian at any scale. Even if the two Yukawa couplings are identically allowed by the gauge symmetry, quantum scale symmetry is forbidding one of them. 

One interesting consequence of the described mechanism is that sectors of the theory that are in principle allowed to mix by the gauge symmetry, remain in fact secluded. 
For example, suppose scalar fields $S_1$, $S_2$ acquire vacuum expectation values~(vevs), $v_1$ and $v_2$ respectively, 
thus breaking the gauge symmetry. A schematic expression for the mass matrix of fermions 
$f_{L,1},f_{L,2}\in G$, $f_{R,1}\in F_1$, $f_{R,2}\in F_2$ will read 
\be
M_f=\frac{1}{\sqrt{2}}\left(\begin{array}{cc}
y_{11} v_1 & y_{12} v_2  \\
y_{21} v_1 &  y_{22} v_2
\end{array}\right)\,.
\ee
If quantum scale invariance protects $y_{12}=y_{21}=0$ from appearing at any scale, 
the fermion mass matrix becomes diagonal and mixing is forbidden. Incidentally, according to the discussion above, since $y_{11}$ is relevant, it is a free parameter of the theory. 
$y_{22}\neq 0$, on the other hand, is a fundamental prediction of the UV completion.

\section{SU(6) GUT\label{sec:su6}}

\subsection{General introduction\label{sec:mech_gi}}

The mechanism described in \refsec{sec:mech} applies to any UV model featuring more particles in copies of the same representation. This is typical in GUTs, where the requirement of anomaly cancellation often leads to the presence of a number of fermion multiplets indistinguishable from one another under the gauge symmetry. Moreover, as a means to seclude different sectors of the Yukawa theory, quantum scale symmetry
finds a particularly useful realization in non-supersymmetric SU($N$) GUTs, as they typically are not endowed with a dark sector\cite{Rizzo:2022lpm}. Once the requirement of anomaly cancellation is imposed, all SU($N$) groups can be given an equivalent fermionic content, for example: $N-4$ copies of the antifundamental and one antisymmetric representation per generation. This implies that the secluding strategy described in \refsec{sec:mech}
can be applied in principle to the Yukawa sector of SU($N$) theories of any rank. On the other hand, 
the low-scale phenomenology of the model will be strongly characterized by the mass hierarchies in the scalar sector and thus has to be analyzed on a case-by-case basis. For illustrative purposes, in this section we focus on one particular example, SU(6), which is the most minimal SU($N$) group that can accommodate a DM candidate.

\subsection{The model\label{sec:model}}

Let us consider a simple modification of the SU(6) DM scenario introduced originally
in Ref.\cite{Ma:2020hyy}. The fermionic content of the model is determined by anomaly cancellation: for each fermion generation it includes three Weyl multiplets, 
$\mathbf{\bar{6}_1}^{(F)}$, $\mathbf{\bar{6}_2}^{(F)}$, and $\mathbf{15}^{(F)}$. To make sure that all fermions acquire masses after the gauge symmetries are spontaneously broken,  
at least three scalar multiplets need to be introduced. In the following we are going to consider 
four scalar multiplets, $\mathbf{6_1}^{(S)}$, $\mathbf{6_2}^{(S)}$, $\mathbf{15}^{(S)}$, and $\mathbf{21}^{(S)}$, 
which is 
a simpler scalar content than in Ref.\cite{Ma:2020hyy} where the two scalar $\mathbf{6}$'s 
were replaced by two scalar $\mathbf{84}$'s.\footnote{It is certainly advantageous to be able to 
obtain the same DM as in Ref.\cite{Ma:2020hyy} 
with a smaller number of fields, however, the most 
important reason to replace the multiplets has to do with the nature of our fixed points. 
Not all systems of gauge-Yukawa RGEs admit solutions that are real. In particular, the system obtained from Lagrangian terms involving the $\mathbf{84}$ scalars admits fixed points with complex Yukawa couplings. While the implications for DM of CP~violation in the Yukawa sector 
would be an interesting subject of investigation \textit{per se}, its thorough analysis exceeds the purpose of this paper. Real fixed-point solutions with the properties we are seeking emerge from the RGE system with the scalar $\mathbf{6}$ multiplets. As we shall see in \refsec{sec:DM}, 
we also differ from Ref.\cite{Ma:2020hyy} in our choice of DM annihilation mechanism.}
One also needs the adjoint
$\mathbf{35}^{(S)}$, which breaks $\textrm{SU(6)}\to\textrm{SU(5)}_{\textrm{SM}}\times\textrm{U(1)}_X$.

The model in Ref.\cite{Ma:2020hyy} admits the existence of a SM-singlet, Majorana fermion weakly interacting massive particle~(WIMP) belonging to a combination of the multiplets $\mathbf{\bar{6}}^{(F)}$. 
The WIMP in Ref.\cite{Ma:2020hyy} is not stable, 
but its lifetime is longer than the age of the universe. Such a long lifetime is a consequence of decay channels mediated by loop-induced couplings with the SM 
that are naturally suppressed by powers of the GUT-scale mass. On the other hand, the long lifetime is also facilitated by, first, the Author's choice 
of an extra $\mathbb{Z}_2$ symmetry 
that forbids a certain number of gauge-allowed Yukawa couplings; and second, by the Author adjusting
the size of some of the remaining $\mathbb{Z}_2$-allowed Yukawa coupling to very small values. 
As we shall see below, in the AS-based framework that we described in \refsec{sec:mech} the same outcome may be 
obtained without the need of introducing a $\mathbb{Z}_2$ symmetry, or of adjusting any of the gauge-Yukawa parameters, which instead are either forbidden by quantum scale symmetry or emerge as naturally suppressed from the UV~completion. However, as in any GUT construction, loop-induced effective couplings are present, and for a large enough GUT mass they guarantee that the WIMP DM particle is metastable.    

Let us now move to define the gauge-Yukawa sector of the SU(6) GUT. For each fermion generation, the Yukawa part of the Lagrangian reads,
\bea\label{eq:yuk}
\mathcal{L} & \supset & 
y_{11} \mathbf{15}^{(F)} \mathbf{\bar{6}_1}^{(F)} \mathbf{\bar{6}_1}^{(S)}+ y_{12} \mathbf{15}^{(F)} \mathbf{\bar{6}_1}^{(F)} \mathbf{\bar{6}_2}^{(S)}+ y_{21} \mathbf{15}^{(F)} \mathbf{\bar{6}_2}^{(F)} \mathbf{\bar{6}_1}^{(S)}+y_{22} \mathbf{15}^{(F)} \mathbf{\bar{6}_2}^{(F)} \mathbf{\bar{6}_2}^{(S)}\nonumber\\
 & &+\,\tilde{y}_{11}\, \mathbf{\bar{6}_1}^{(F)} \mathbf{\bar{6}_1}^{(F)} \mathbf{15}^{(S)} +\tilde{y}_{12}\, \mathbf{\bar{6}_1}^{(F)} \mathbf{\bar{6}_2}^{(F)} \mathbf{15}^{(S)} +\tilde{y}_{22}\, \mathbf{\bar{6}_2}^{(F)} \mathbf{\bar{6}_2}^{(F)} \mathbf{15}^{(S)}\nonumber \\ 
 & &+\,\hat{y}_{11}\, \mathbf{\bar{6}_1}^{(F)} \mathbf{\bar{6}_1}^{(F)} \mathbf{21}^{(S)} + \hat{y}_{12}\, \mathbf{\bar{6}_1}^{(F)} \mathbf{\bar{6}_2}^{(F)} \mathbf{21}^{(S)}+ \hat{y}_{22}\, \mathbf{\bar{6}_2}^{(F)}  \mathbf{\bar{6}_2}^{(F)} \mathbf{21}^{(S)}\nonumber\\
 & &+\,y_u\, \mathbf{15}^{(F)} \mathbf{15}^{(F)} \mathbf{15}^{(S)}+\textrm{H.c.}
\eea
We assume for simplicity that the three SM fermion generations do not mix.

The scalar potential of the model is presented in Appendix~\ref{app:scap}. In this work,
we do not investigate the scalar sector of the theory from the point of view of AS. Suffices to say that
masses and trilinear couplings 
are canonically relevant, therefore we will treat them as free parameters that can assume any needed value, positive or negative.
In the context of fundamental scale invariance these couplings, rescaled by their mass dimension, have the role of effective parameters that break
the quantum scale symmetry and dynamically 
generate a physical scale when they reach values greater than~1 in their flow to the IR.

We adopt the following SU(6) GUT breaking chain:
\begin{itemize}
\item $v_6=\langle \mathbf{35}^{(S)}\rangle\approx 10^{16}\gev$ breaks $\textrm{SU(6)}\to\textrm{SU(5)}_{\textrm{SM}}\times\textrm{U(1)}_X$
\item $v_5=\langle \mathbf{24}_0^{(S)}\rangle\approx 10^{16}\gev$ breaks $\textrm{SU(5)}_{\textrm{SM}}\to \textrm{SU(3)}_\textrm{c}\times \textrm{SU(2)}_L\times\textrm{U(1)}_Y$
\item $\mathbf{\bar{6}_1}^{(S)}\supset\mathbf{\bar{5}}^{(S)}_{-1}\supset\left(\mathbf{1},\mathbf{\bar{2}},-\frac{1}{2};-1\right)+\left(\mathbf{\bar{3}},\mathbf{1},\frac{1}{3};-1\right)$, where we have indicated
in parentheses the quantum numbers in the extended SM group, $\textrm{SU(3)}_\textrm{c}\times \textrm{SU(2)}_L\times\textrm{U(1)}_Y\times \textrm{U(1)}_X$. 
Doublet-triplet splitting follows from the tuning of some scalar potential couplings (see Appendix~\ref{app:gsb}) and it guarantees that the SU(2) doublet field remains light
\item $\mathbf{\bar{6}_2}^{(S)}\supset\mathbf{1}^{(S)}_{5}=\left(\mathbf{1},\mathbf{1},0;5\right)$ is a SM-singlet scalar field that remains light
\item $\mathbf{15}^{(S)}\supset\mathbf{5}^{(S)}_{-4}\supset\left(\mathbf{1},\mathbf{2},\frac{1}{2};-4\right)+\left(\mathbf{3},\mathbf{1},-\frac{1}{3};-4\right)$. Doublet-triplet splitting guarantees that the SU(2) doublet field remains light
\item $\mathbf{21}^{(S)}\supset\mathbf{1}^{(S)}_{-10}=\left(\mathbf{1},\mathbf{1},0; -10\right)$ is a SM-singlet scalar that remains light.
\end{itemize}
At low energy one is thus left with a two-Higgs doublet model~(2HDM) 
extended by the addition of two complex scalar SM-singlets. 

We rename the light scalar fields 
in the following way:
\bea\label{eq:lightf}
H_d=\left(\mathbf{1},\mathbf{\bar{2}},-\frac{1}{2};-1\right)\,, & \quad & H_u=\left(\mathbf{1},\mathbf{2},\frac{1}{2};-4\right)\,,\nonumber\\
s_6=\left(\mathbf{1},\mathbf{1},0;5\right)\,, & \quad & s_{21}=\left(\mathbf{1},\mathbf{1},0;-10\right)\,,
\eea
where we shall henceforth indicate weak isospin doublets with capital letters, and weak singlets in lowercase. 
Since no scalar field is neutral under the U(1)$_X$ group, the heaviest of the acquired vevs will approximately determine the
mass of the $Z'$ gauge boson, modulo the size of gauge coupling $g_X$ and U(1)$_X$ charge of the scalars. The vevs of the light scalar fields, $v_d$, $v_u$, $v_{s_6}$, $v_{s_{21}}$, give mass to the light fermions in the model. As is customary in the 2HDM, we define $\tanb\equiv v_u/v_d$. 

We can define the low-scale, left-chiral Weyl fermion multiplets of the extended SM group
$\textrm{SU(3)}_\textrm{c}\times \textrm{SU(2)}_L\times\textrm{U(1)}_Y\times \textrm{U(1)}_X$. For each generation,
\be\label{eq:fer1}
Q:\left(\mathbf{3},\mathbf{2},\frac{1}{6};2\right),\quad u:\left(\mathbf{\bar{3}},\mathbf{1},-\frac{2}{3};2 \right),\quad  d_1, d_2:\left(\mathbf{\bar{3}},\mathbf{1},\frac{1}{3};-1 \right),\quad 
d':\left(\mathbf{3},\mathbf{1},-\frac{1}{3};-4 \right),
\ee
\be\label{eq:fer2}
L_1, L_2:\left(\mathbf{1},\mathbf{2},-\frac{1}{2};-1\right),\quad L':\left(\mathbf{1},\mathbf{\bar{2}},\frac{1}{2};-4\right),\quad
e:\left(\mathbf{1},\mathbf{1},1;2 \right),\quad \nu_1, \nu_2:\left(\mathbf{1},\mathbf{1},0;5\right),
\ee
where, again, we have indicated weak isospin doublets in capital letters, and isospin singlets in lowercase. For completeness, we indicate the original SU(6) representations from which the fermions of \refeq{eq:fer1} and \refeq{eq:fer2} emerge upon the GUT-symmetry breaking,
\bea
&\mathbf{\bar{6}_1}^{(F)}\supset\mathbf{\bar{5}}^{(\textrm{SM})}_{-1}\supset d_1,\,L_1& \qquad \mathbf{\bar{6}_1}^{(F)}\supset\mathbf{1}^{(F)}_{5}\supset \nu_1\nonumber\\
&\mathbf{\bar{6}_2}^{(F)}\supset\mathbf{\bar{5}}^{(F)}_{-1}\supset d_2,\,L_2& \qquad \mathbf{\bar{6}_2}^{(F)}\supset\mathbf{1}^{(F)}_{5}\supset \nu_2\nonumber\\
&\mathbf{15}^{(F)}\supset\mathbf{10}^{(\textrm{SM})}_{2}\supset Q,\, u,\, e&\qquad \mathbf{15}^{(F)}\supset\mathbf{5}^{(F)}_{-4}\supset d',\, L'\,.
\eea 

We do not concern ourselves in this work with the 
details of gauge-coupling unification. It is well known\cite{Ma:2020hyy} that unification can be achieved at the scale $M_{\textrm{GUT}}\approx 4\times 10^{16}\gev$, if one adds to the low-energy model some extra fields: a color-octet fermion and a weak isospin-triplet fermion, 
which above the GUT scale should belong to an adjoint $\mathbf{35}$.  
We neglect in
\refeq{eq:yuk}
possible Yukawa couplings involving the adjoint needed exclusively to guarantee unification. In the context of AS, 
this is equivalent to assuming that those Yukawa couplings are zero in the trans-Planckian regime, and they are thus either protected from appearing at the low scale (irrelevant) or fine tuned to negligible values (relevant). 

\subsection{Trans-Planckian fixed points\label{sec:z2}}

\begin{table}[t]
\centering
\begin{tabular}{|c c c c||c|c|c|c|c c | c|}
\hline
$y_u^{\ast}$ & $y_{22}^{\ast}$ & $\hat{y}_{11}^\ast$ & $\hat{y}_{22}^{\ast}$ & $y_{11}^{\ast}$   & $\tilde{y}_{11}^{\ast}$ & $\tilde{y}_{12}^{\ast}$ & $\tilde{y}_{22}^{\ast}$ & 
$\hat{y}_{12}^{\ast}$ & $y_{12}^{\ast}$ & $y_{21}^{\ast}$\\
\hline
$0.25$ & $0.35$ & $0.38$ & $0.32$ & $0.0$  &  $0.0$ & $0.0$ & $0.0$ & $0.0$ & $0.0$ & $0.0$  \\
\hline
$\theta_u$ & $\theta_{22}$ & $\hat{\theta}_{11}$ & $\hat{\theta}_{22}$ & $\theta_{11}$ & $\tilde{\theta}_{11}$ & $\tilde{\theta}_{12}$& $\tilde{\theta}_{22}$  & 
$\hat{\theta}_{12}$ & $\theta_{12}$ &   $\theta_{21}$   \\
\hline
$-5.1$ & $-1.1$ & $-4.7$ & $-3.7$ & $0.63$ & $-0.27$ & $-0.27$ & $-0.27$ & $-3.7$ & $0$ &  $0$  \\
\hline
\end{tabular}
\caption{Upper line: Trans-Planckian fixed points of the SU(6) Yukawa couplings for $f_y=0.016$. Lower line: The corresponding critical exponents times $16\pi^2$.
Critical exponents grouped together in the same box identify non-diagonal eigendirections in theory space. The diagonal directions are presented with the corresponding exponent in individual boxes.
\label{tab:fixpoint}}
\end{table}

The full set of trans-Planckian RGEs of the gauge and Yukawa couplings of the 
SU(6) model is presented in Appendix~\ref{app:rges}. 
They are derived with the public tool~\texttt{PyR@TE 3}\cite{Poole:2019kcm,Sartore:2020gou}. As was explained in \refsec{sec:FRG}, 
the current status of quantum gravity calculations, using both the FRG and other schemes, strongly favors $f_g\geq 0$. As a consequence of this choice,  
$g_6^{\ast}=0$ must be a relevant Gaussian fixed point. The RGE system admits several fixed points. Some of them are presented in Appendix~\ref{app:new_fp}.  Since we expect, by construction, the UV completion to be maximally predictive, 
we present here the fixed point with the maximal number of irrelevant couplings in the Yukawa sector. It features eight irrelevant eigendirections, two marginal, and one relevant.  
The fixed-point values are reported, together with their critical exponents, in \reftable{tab:fixpoint}.

\begin{figure}[t]
	\centering%
 	    \subfloat[]{%
		\includegraphics[width=0.45\textwidth]{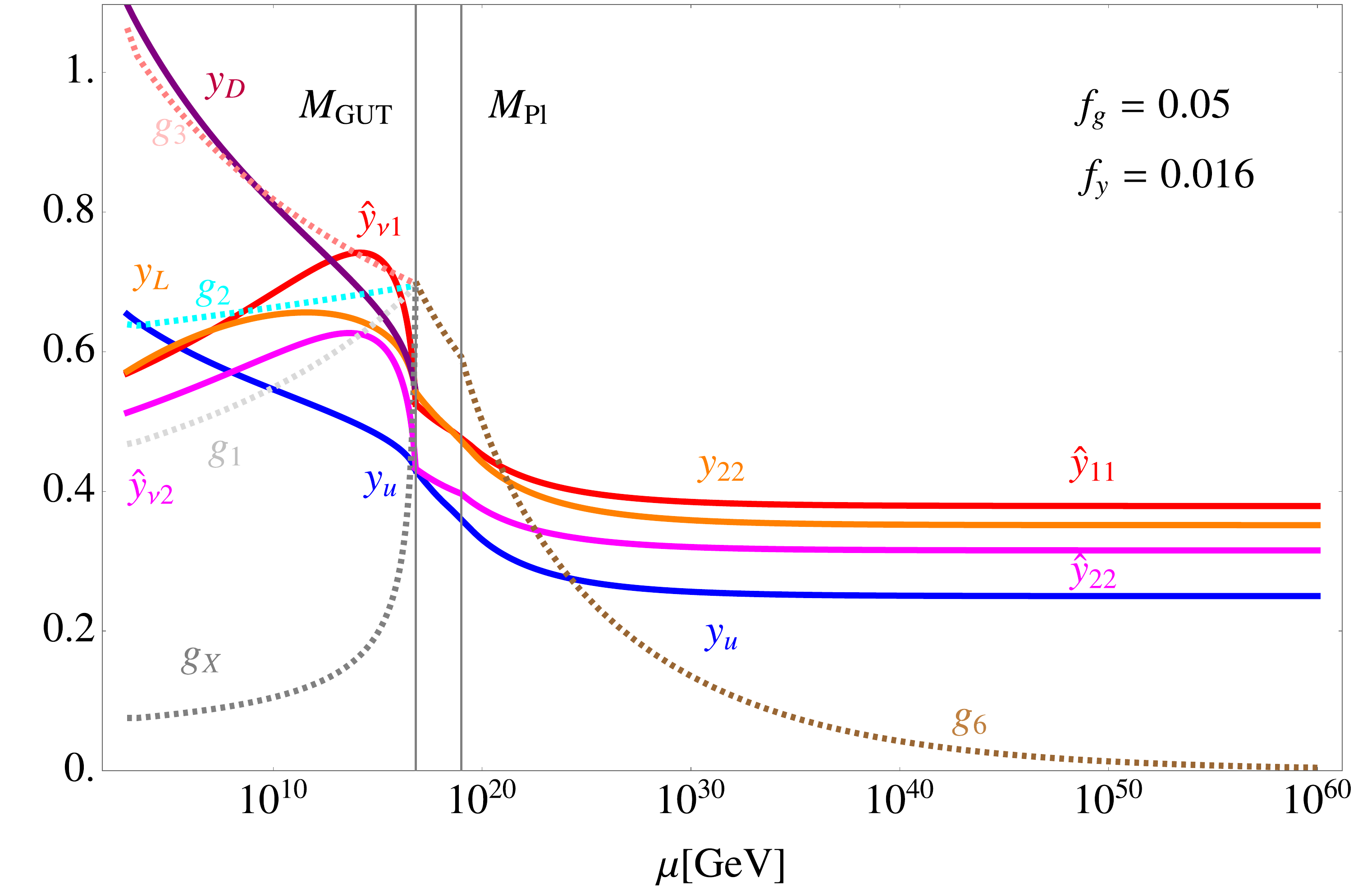}}
  		\hspace{0.8cm}
    	\subfloat[]{%
  		\includegraphics[width=0.45\textwidth]{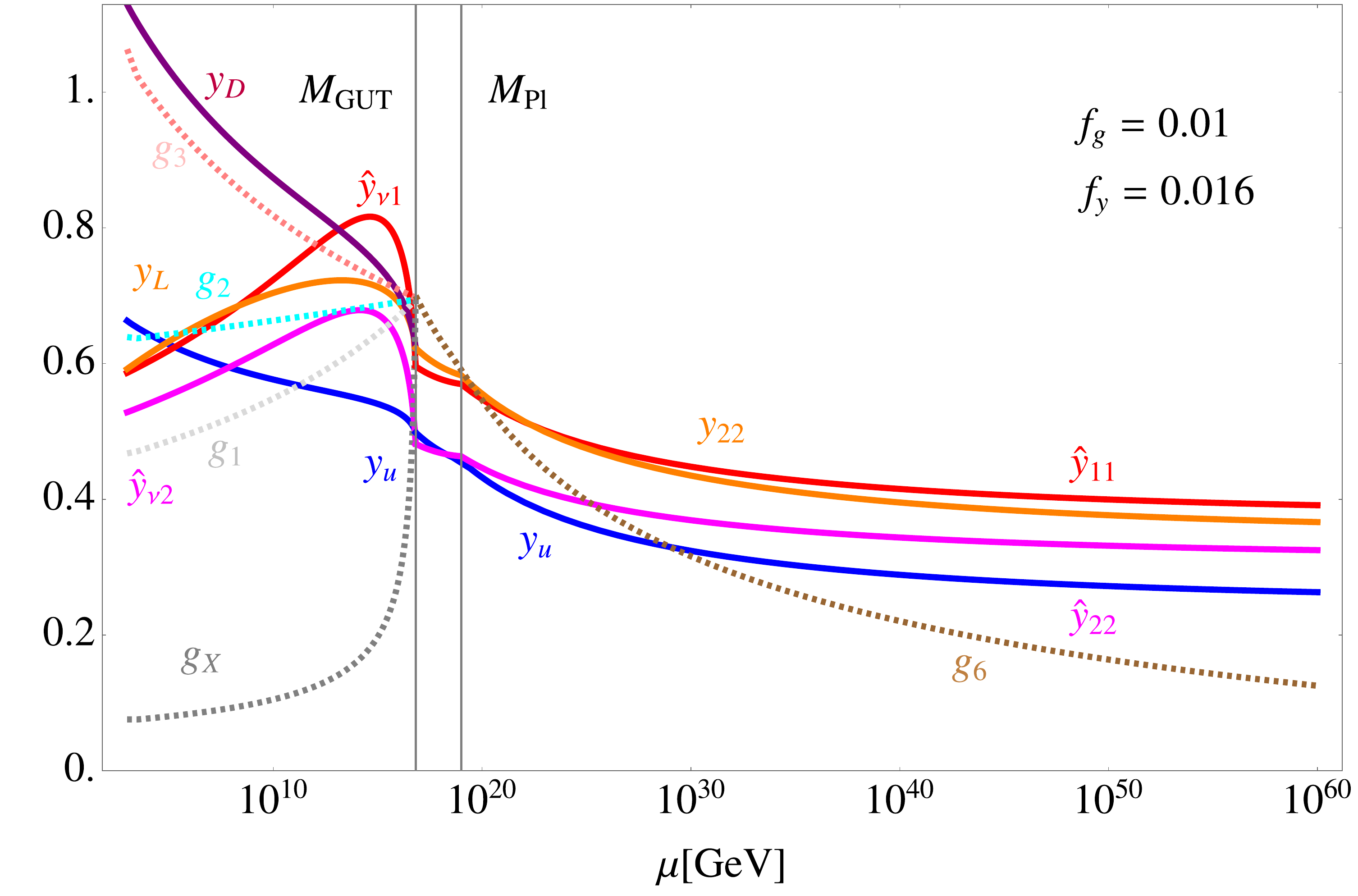}}\\
    	\subfloat[]{%
  		\includegraphics[width=0.45\textwidth]{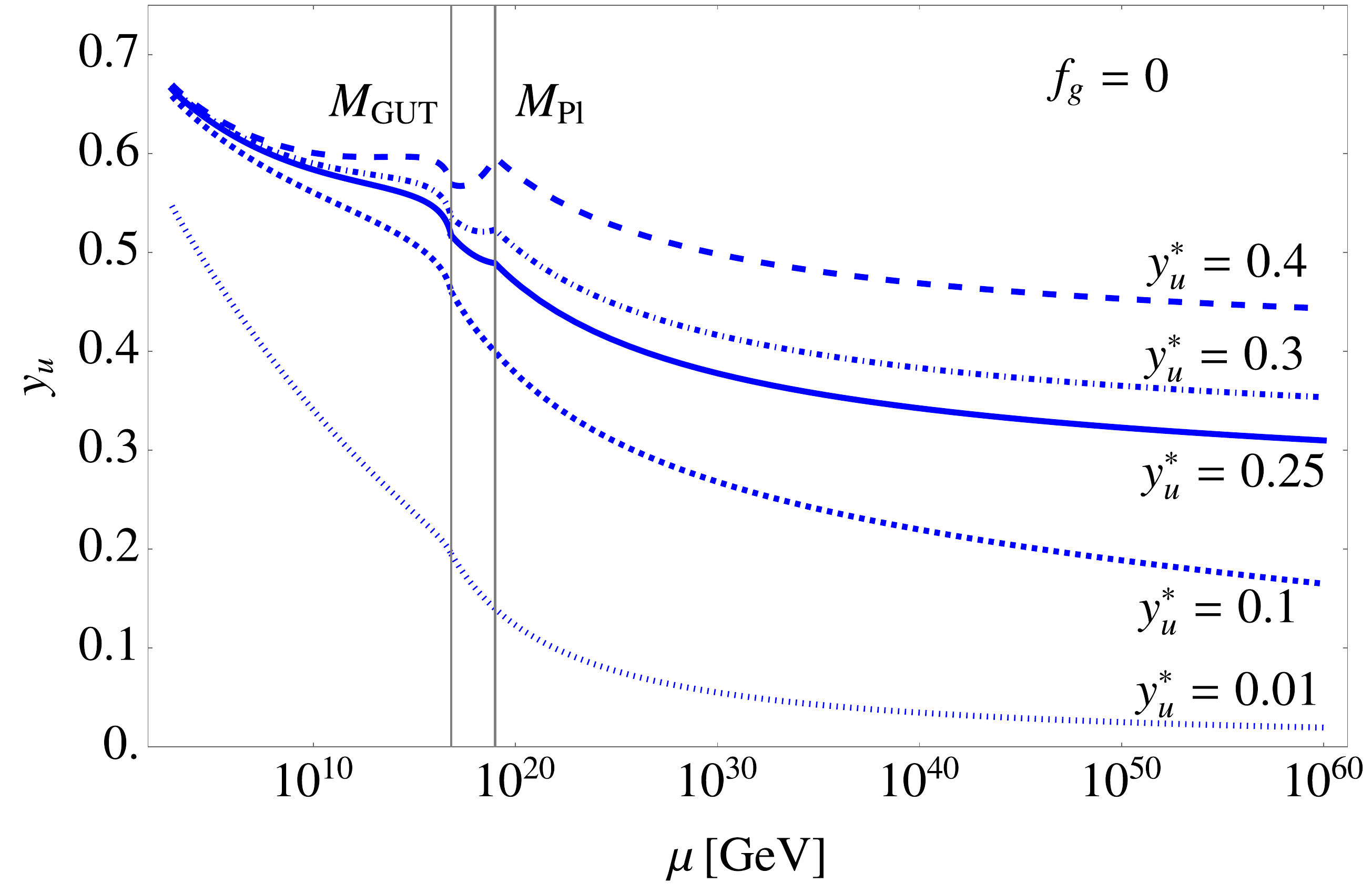}}
\caption{(a) The RG flow of the SU(6) Yukawa~(solid) and gauge~(dotted) couplings from the deep trans-Planckian UV to the EWSB scale for $f_g=0.05$ and $f_y=0.016$. We have selected $\tanb=1$ at the EWSB scale. (b) The flow of the SU(6) Yukawa~(solid) and gauge~(dotted) couplings from the deep trans-Planckian UV to the EWSB scale for $f_g=0.01$ and $f_y=0.016$. (c) The flow of the top-quark Yukawa coupling from the deep trans-Planckian UV to the EWSB scale for $f_g=0$ and different choices of the fixed point, corresponding to different $f_y$. }
\label{fig:fullrun}
\end{figure}

In \reffig{fig:fullrun}(a) we show the flow of the Yukawa couplings from the deep trans-Planckian UV down to the EWSB scale for chosen values 
of the parameters: $\tanb=1$, $f_y=0.016$, and $f_g=0.05$. Below the GUT scale, the RGEs of the low-energy model are computed with the public tool \texttt{RGBeta}\cite{Thomsen:2021ncy}. 

In \reffig{fig:fullrun}(b) we show the flow of the Yukawa couplings for 
$f_g=0.01$. A comparison with \reffig{fig:fullrun}(a) shows that the size of the Yukawa couplings at the fixed point is essentially independent of the value of the gravitational correction $f_g$, which can even be zero. 
This is because near the Planck scale the Yukawa-coupling flow is mostly driven by the (relevant) gauge coupling $g_6$, which is pinned in the IR by the measured value of the SM gauge couplings. This RG ``focusing'' behavior of irrelevant couplings driven by a relevant one is well known in BSM theories with UV fixed points (for a discussion see, \textit{e.g.}, Sec.~3.1 of Ref.\cite{Kotlarski:2023mmr}). 
Note also that the sub-GUT flow of the BSM abelian gauge coupling $g_X$ (dark dotted gray) is very steep, due to the large 
U(1)$_X$ charges of the low-energy fields. It leads to a small value of the coupling at the low scale: 
\be
g_X(\mu=1\tev)=0.07.
\ee
We do not show in 
\reffig{fig:fullrun} the running of the (abelian) kinetic mixing parameter
$g_{\epsilon}$. We have checked that the boundary condition $g_{\epsilon}(M_{\textrm{GUT}})=0$ leads to a negligible low-scale value, $g_{\epsilon}(\mu=1\tev)\lesssim 10^{-3}$. 

In \reffig{fig:fullrun}(c) we show the flow of $y_u$ for
$f_g=0$ and different choices of~$f_y$\,. While changing the $f_y$ value leads to different fixed points for $y_u$, 
the low-scale DM phenomenology does not, to a great extent, depend on the Yukawa size at the fixed point. 
This ceases to be true, however, if one takes a very
small $f_y$\,, which leads to a very small $y_u^{\ast}$. 
In that case, the pull of the gauge coupling is not
strong enough to drag the top quark Yukawa coupling to its measured low-scale value (inefficient focusing). 
All in all,  
the lesson we draw from \reffig{fig:fullrun} is that the low-scale Yukawa couplings are 
very mildly dependent on the outcome 
and details of a quantum-gravity calculation. This is true 
as long as the couplings' scaling behavior 
at the fixed point is consistent with the critical exponents presented in \reftable{tab:fixpoint}. 

The low-scale value of irrelevant Yukawa couplings are predicted uniquely. The only remaining free coupling is $y_{11}$~(relevant), 
which, as we shall see below, 
is adjusted to the mass of the bottom quark. Note also 
that there are two marginal couplings, thus needing higher-loop corrections to determine their fate.

One can see in \reftable{tab:fixpoint} that six Yukawa couplings are prevented from appearing in the Lagrangian by trans-Planckian scale symmetry, as their critical exponents are negative (we also include in this group the marginal couplings $y_{12}$ and $y_{21}$). Due to the chiral symmetry of the sub-Planckian gauge theory, they remain zero even when gravity decouples at $M_{\textrm{Pl}}$, and through the process of GUT symmetry-breaking.\footnote{For a comparison with the literature, we note that the couplings $y_{12}$, $y_{21}$, $\tilde{y}_{12}$ and $\hat{y}_{12}$ are forbidden in
Ref.\cite{Ma:2020hyy} by an imposed $\mathbb{Z}_2$ symmetry. However, the $\mathbb{Z}_2$ 
symmetry is not sufficient 
to forbid also the appearance of 
$\tilde{y}_{11}$ and $\tilde{y}_{22}$. It can be checked that, in order to prevent the DM particle from decaying too fast via the tree-level decays $\textrm{DM}\to h_1\, \nu_{\textrm{SM}}$, $\textrm{DM}\to Z\, \nu_{\textrm{SM}}$,  
and $\textrm{DM}\to W^{\pm}\, l^{\mp}$, an upper bound must be imposed, $\tilde{y}_{11}\approx \tilde{y}_{22} < 10^{-9}$. In the AS-inspired framework considered here, $\tilde{y}_{11}$ and $\tilde{y}_{22}$ are set to zero automatically due to the presence of an irrelevant trans-Planckian fixed point.} Thus, the Lagrangian (\ref{eq:yuk}) reduces at the DM mass scale to
\bea
\mathcal{L}_{\textrm{IR1}} &\supset &  2 y_u\, u H_u^{c\dag} Q + y_d\, d_1 H_d Q +  y_e\, e H_d L_1 +y_{\nu}\, L' H_d^{c\dag} \nu_1 \nonumber \\
& &+y_D\, d_2 d' s_6 +\, y_L\, L' L_2 s_6
+ y_{\nu_1}\, \nu_1 \nu_1 s_{21} + y_{\nu_2}\, \nu_2 \nu_2 s_{21}
+\textrm{H.c.}\,,\label{eq:le_lagr}
\eea
where spinor and weak isospin indices are contracted trivially following matrix multiplication rules, we have defined $H_{u,d}^c\equiv i\sigma_2 H_{u,d}^{\ast}$\,, and we restrict the analysis to the third SM generation only.
Couplings $y_d$, $y_e$, and $y_{\nu}$ in \refeq{eq:le_lagr} originate from the relevant UV parameter $y_{11}$, after following the RG flow down to the EWSB scale.
Couplings $y_D$ and $y_L$ originate from the irrelevant $y_{22}$ and are thus uniquely 
predicted at every scale of phenomenological interest. We show in \reftable{tab:le_values}, as an example, the predicted values of the irrelevant Yukawa couplings at the scale $\mu=1\tev$ for $\tanb=1$, $f_g=0.05$, and $f_y=0.016$ (we reiterate that the numerical values depend very mildly on the actual choice of the gravitational parameters). 
Similar considerations  allow one to derive from the RG flow the values of the irrelevant couplings that are important for the DM phenomenology at an arbitrary renormalization scale.

\begin{table}[t]
\centering
\begin{tabular}{|c|c|c|c|c|c||c|c|}
\hline
SU(6) & $y_u$ & \multicolumn{2}{c|}{$y_{22}$} & $\hat{y}_{11}$ & $\hat{y}_{22}$ & \multicolumn{2}{c|}{$y_{11}$} \\
\hline
 & $y_u$ & $y_{D}$ & $y_{L}$ & $y_{\nu_1}$ & $y_{\nu_2}$ & $y_{d}$ & $y_{\nu}$ \\
\hline
$\mu=1\tev$ & $0.69$ & $1.1$ & $0.55$ & $0.57$ & $0.51$ & $0.027$ & $0.014$ \\
\hline
\end{tabular}
\caption{(Left-hand side) The predicted values of the irrelevant Yukawa couplings at the scale $\mu=1\tev$ for $\tanb=1$, $f_g=0.05$, and $f_y=0.016$. 
The upper line indicates the corresponding SU(6) Yukawa couplings.
The numerical values depend very mildly on $f_g$ and $f_y$. (Right-hand side) Relevant couplings $y_d$ and $y_{\nu}$ are related to each other by the RG flow. Their value at the low scale 
is adjusted so that $y_d v_d/\sqrt{2}$ matches the bottom quark mass.}
\label{tab:le_values}
\end{table}

Because some of the Lagrangian couplings allowed by gauge invariance are forbidden by the quantum scale symmetry, low-energy particles 
that belong to GUT multiplets with the same quantum numbers do not mix with one another. 
In fact, the mass matrices for the bottom-like quarks and tau-like leptons read, in the basis $\langle d_1, d_2|,\,|d_{Q}, d'\rangle$, and $\langle e, e_{L'}|,\,|e_{L_1}, e_{L_2}\rangle$, respectively, 
\begin{equation}
M_b=\frac{1}{\sqrt{2}}\left(\begin{array}{cc}
y_d v_d & 0  \\
0 &  y_D v_{s_6}
\end{array}\right)\,, \quad 
M_{\tau}=\frac{1}{\sqrt{2}}\left(\begin{array}{cc}
y_e v_d & 0  \\
0 &  y_L v_{s_6}
\end{array}\right)\,,
\end{equation}
where $d_Q\in Q$, $e_{L'}\in L'$, $e_{L_1}\in L_1$, and  $e_{L_2}\in L_2$. 

The mass matrix for the neutral fermions, in the basis 
$\langle \nu_{L_1}, \nu_{L_2} ,  \nu_{L'}, \nu_1, \nu_2  |,\, |  \nu_{L_1}, 
\nu_{L_2} ,  \nu_{L'}, \nu_1, \nu_2 \rangle$ reads
\be\label{eq:massnu}
\frac{1}{2} M_{\nu}=\frac{1}{\sqrt{2}}\left(\begin{array}{ccccc}
0 & 0 & 0 & 0 & 0  \\
0 & 0  & y_L v_{s_6}  & 0 & 0\\
0 & y_L v_{s_6}  & 0 & y_{\nu} v_d &  0 \\
0 & 0 & y_{\nu} v_d   & y_{\nu_1} v_{s_{21}} & 0  \\
0  & 0   & 0   & 0 & y_{\nu_2} v_{s_{21}} 
\end{array}\right)\,,
\ee
where $ \nu_{L_1}, \nu_{L_2} ,  \nu_{L'}\in L_1, L_2, L'$\,, respectively. One can see that the lightest neutrino of one generation remains massless (this is still a phenomenologically allowed possibility\cite{Esteban:2024eli}), while the heavy Majorana neutrinos of the same generation 
comprise a dark sector prevented by the scale symmetry from decaying into the lightest neutrino. The lightest dark particle is thus a good candidate for WIMP DM (albeit metastable once GUT interactions mediated by heavy states are factored in, see Ref.\cite{Ma:2020hyy}). 

\subsection{Naturally small Yukawa couplings\label{sec:nsyc}}

As we have discussed above, 
quantum scale symmetry -- codified by the negative critical exponents of the fixed points of the SU(6) Yukawa couplings -- has the effect of both providing unique predictions for the interactions of the low-scale theory, and of preventing 
the couplings that correspond to a Gaussian fixed point from appearing in the Lagrangian. On the other hand, some of us have shown in a couple of recent papers\cite{Kowalska:2022ypk,Chikkaballi:2023cce} that the mere presence of an irrelevant Gaussian fixed point of the RG flow does not always lead to a zero coupling in the IR. In fact, if the RGE system admits also a relevant fixed point at much higher values of $\mu$, some couplings may flow from the relevant UV fixed point down to the basin of attraction of the Gaussian fixed point lying in the trans-Planckian IR, without ever reaching zero. The IR attractor induces an exponential suppression of the Yukawa coupling $y$, whose run becomes approximately parameterized as\cite{Kowalska:2022ypk} 
\be\label{eq:analytic}
y(t,\kappa)=\left[\frac{16 \pi^2 c_X \left(f_{\textrm{crit}}-f_y \right)}{e^{2 c_X \left(f_{\textrm{crit}}-f_y\right)\left(16\pi^2 \kappa-t\right)}+\alpha_Z}\right]^{1/2}\,,
\ee
where $f_{\textrm{crit}}$ is a critical $f_y$ value below which a Gaussian irrelevant fixed point for $y$ appears, $c_X$ is a linear combination of the coefficients of the one-loop beta functions of Yukawa couplings other than $y$, and $\alpha_Z$ is the $y^3$ coefficient of the one-loop 
beta function of $y$. Once the theory and the UV completion are set, the suppression depends on one single parameter $\kappa$, which measures, in e-folds, the distance between the UV and the IR fixed points.  

\begin{table}[t]
\centering
\begin{tabular}{|c|c|c|c||c|c|c|c|c|c|c|}
\hline
$y_u^{\ast}$ & $y_{22}^{\ast}$ & $\hat{y}_{11}^\ast$ & $\hat{y}_{22}^{\ast}$ & $y_{11}^{\ast}$ & $y_{12}^{\ast}$ & $y_{21}^{\ast}$ & $\tilde{y}_{11}^{\ast}$ & $\tilde{y}_{12}^{\ast}$ & $\tilde{y}_{22}^{\ast}$ & 
$\hat{y}_{12}^{\ast}$\\
\hline
$0.0$ & $0.54$ & $0.0$ & $0.0$ & $0.0$ & $0.0$ & $0.0$ & $0.0$ & $0.0$ & $0.0$ & $0.0$  \\
\hline
$\theta_t$ & $\theta_{22}$ & $\hat{\theta}_{11}$ & $\hat{\theta}_{22}$ & $\theta_{11}$ & $\theta_{12}$ & $\theta_{21}$ & $\tilde{\theta}_{11}$ & $\tilde{\theta}_{12}$ & $\tilde{\theta}_{22}$ & 
$\hat{\theta}_{12}$\\
\hline
$1.9$ & $-5.0$ & $2.5$ & $1.0$ & $2.2$ & $0$ & $0$ & $2.5$ & $1.8$ & $1.0$ & $1.8$  \\
\hline
\end{tabular}
\caption{SU(6) Yukawa couplings and their critical exponents times $16\pi^2$ at the extreme UV fixed point for $f_y=0.016$.}
\label{tab:fixpointUV}
\end{table}

The RGEs of the SU(6) theory do feature a relevant UV fixed point at high values of $\mu$. Its critical exponents are presented in \reftable{tab:fixpointUV}. As a consequence, some of the irrelevant 
Gaussian Yukawa couplings of \reftable{tab:fixpoint} are now allowed to take arbitrarily small values that are naturally suppressed according to \refeq{eq:analytic}. 

\begin{figure}[t]
	\centering%
		\includegraphics[width=0.5\textwidth]{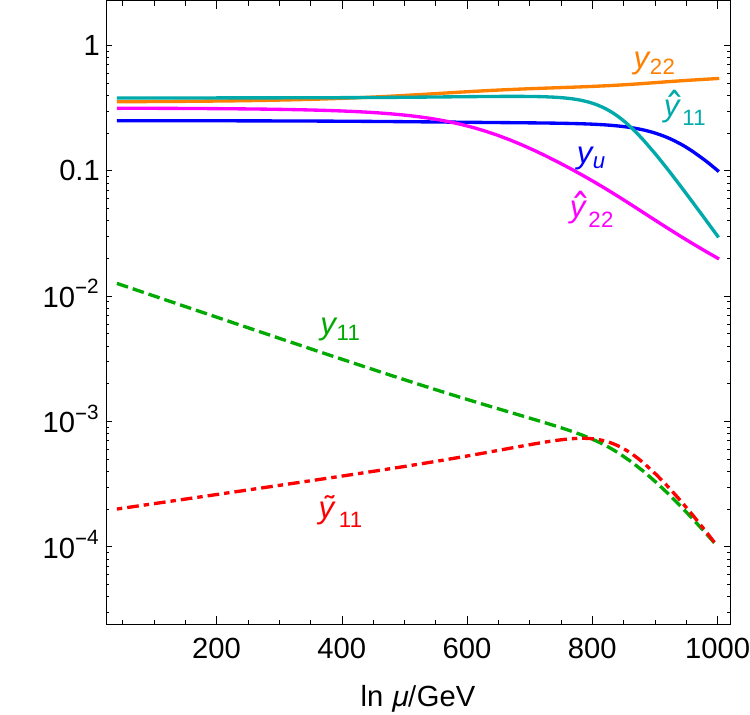}
\caption{Trans-Planckian RG flow of the Yukawa couplings from the deep UV fixed point of \reftable{tab:fixpointUV}. Solid lines show the couplings that are predictions of the theory, while dashed and dot-dashed are the couplings that can be freely adjusted.}
\label{fig:deepUV}
\end{figure}

In \reffig{fig:deepUV}, we show this effect for coupling~$\tilde{y}_{11}$ (in red, dot-dashed). The size of $\tilde{y}_{11}$ can be made arbitrarily small, depending on where in the flow couplings~$y_u$, $y_{22}$, $\hat{y}_{11}$, and $\hat{y}_{22}$ reach their interactive fixed point. One of the interesting properties of such a solution for $\tilde{y}_{11}$ 
is that its low-scale value can induce a small left-handed neutrino mass
via a light see-saw\cite{Lindner:2014oea}. We show for completeness, in \refeq{eq:fullmass} of Appendix~\ref{app:fer_mas}, the form of the 
full Majorana fermion mass matrix when all of the Gaussian irrelevant couplings of \reftable{tab:fixpoint} are allowed to appear. Assuming the lightest neutrino mass is $\mathcal{O}(10^{-10}\gev)$, for $y_{\nu_1} v_{s_{21}}\approx 1\tev$ one needs $\tilde{y}_{11}(\mu=1\tev)\approx 10^{-6}$ at $\tanb=1$.

\subsection{Scalar potential}\label{sec:scal}

As was mentioned in  \refsec{sec:mech_gi}, we will not investigate the fixed points of the scalar sector of the SU(6) model. Given the complexity of the scalar potential in \refeq{eq:su6pot} of Appendix~\ref{app:scap}, 
doing so would amount to a formidable task even at 
one loop. In this subsection, we limit ourselves to addressing 
qualitatively the expected scaling behavior of the scalar-potential parameters in relation to the current state of calculations in asymptotically safe quantum gravity.

Let us define the dimensionless running parameters of the scalar potential: mass parameters $\tilde{M}_i^2=M_i^2/\mu^2$, trilinear couplings  $\tilde{\kappa}_j=\kappa_j/\mu$, and quartic couplings $\alpha_k$, for $i,j,k$ spanning all the couplings given in \refeq{eq:su6pot} of Appendix~\ref{app:scap}.
The RGEs of the scalar potential are modified in the trans-Planckian regime with a ``correction'' due to the gravity fixed points, 
in analogy to Eqs.~(\ref{eq:betag}) and (\ref{eq:betay}). Following several studies in the literature\cite{Wetterich:2016uxm,Pawlowski:2018ixd,Wetterich:2019zdo} one can approximate the RGEs schematically, 
\bea
\frac{d\alpha_k}{d t}&= &\eta_{\alpha_k} \alpha_k + \beta_{\alpha_k,\textrm{add}} -f_{\lam} \alpha_k\,, \label{eq:alk}\\
\frac{d\tilde{\kappa}_j}{d t}&= &\left(-1+\eta_{\tilde{\kappa}_j}\right) \tilde{\kappa}_j + \beta_{\tilde{\kappa}_j,\textrm{add}} -f_{\lam} \tilde{\kappa}_j \,, \label{eq:kaj}\\
\frac{d\tilde{M}_i^2}{d t}&= &\left(-2+\eta_{\tilde{M}_i^2}\right) \tilde{M}_i^2 + \beta_{\tilde{M}_i^2,\textrm{add}} -f_{\lam} \tilde{M}_i^2\,, \label{eq:Mi}
\eea
where $\eta_{\alpha_k}$, $\eta_{\tilde{\kappa}_j}$, and $\eta_{\tilde{M}_i^2}$ indicate the matter anomalous dimensions. 
The terms with an ``add'' subscript parameterize additive contributions to the matter beta functions not included in the anomalous dimensions; they potentially depend on any of the couplings of the theory with the exception of the very one indicated in the subscript. $f_{\lam}$ is the universal multiplicative correction analogous to $f_g$ and $f_y$, which typically depends on the fixed points of the gravitational action. Note that for practicality  we neglect in Eqs.~(\ref{eq:alk})-(\ref{eq:Mi}) terms that parameterize additive contributions potentially arising from non-minimal direct couplings of the scalar potential to gravitational operators, see, \textit{e.g.}, the truncation introduced in Ref.\cite{Eichhorn:2020kca}. 

One may envision the outcome of the eventual UV calculation of $f_{\lam}$ broadly in three ways:\smallskip

\textbf{Case A.} $f_{\lam}\ll -2$\,. Under this condition, the set of Eqs.~(\ref{eq:alk})-(\ref{eq:Mi}) 
admit a fully irrelevant Gaussian fixed point. 
It was shown in 
Refs.\cite{Wetterich:2016uxm,Pawlowski:2018ixd,Wetterich:2019zdo} that
$f_{\lam}\ll -2$ may emerge in an FRG calculation of the Higgs potential and Einstein-Hilbert truncation of the gravitational action, for certain values of the running Planck mass. Under this assumption the scalar potential of the matter theory 
features complete quantum scale symmetry. Its dimensionful couplings may tend, 
when gravity decouples rapidly below the Planck scale, 
to specific predicted values that are naturally small\cite{Wetterich:2016uxm,Pawlowski:2018ixd,Wetterich:2019zdo}. 
For the same reason, this assumption is in strong tension with the hierarchical ladder of physical scales typical of GUT theories with a realistic phenomenology. \smallskip
  
\textbf{Case B.} $-1 \lesssim f_{\lam} < 0$. An eventual outcome of the UV calculation within this range 
would allow the dimensionful couplings of the scalar potential to remain relevant, while the
fixed points of the quartic couplings may be irrelevant \cite{Eichhorn:2019dhg}. While in this framework one can easily accommodate the
generation of physical scales below Planck, it is certainly hard to envision,
without a detailed fixed-point analysis, whether 
the requirement of doublet-triplet splitting, \refeq{eq:dt-spl} in Appendix~\ref{app:gsb}, and other scale separations could be implemented at all.   
\smallskip

\textbf{Case C.} $f_{\lam}\gg 0$. In this case the fixed points of the scalar potential are all relevant. 
Dimensionful and dimensionless couplings cannot be predicted, as any chosen IR value is eventually consistent with AS. It was shown in 
Ref.\cite{Pawlowski:2018ixd} that $f_{\lam}>0$ cannot be an
outcome of the FRG calculation with the action comprising the SM Higgs potential and gravity 
in the Einstein-Hilbert truncation, independently of the fixed-point value of the gravitational parameters and of the choice of regulator. If this negative outcome were to be confirmed by a detailed analysis of the fixed points of the gravitational and scalar potential of the SU(6) model presented here, it would place it squarely
outside of the ``landscape'' of realistic theories that can be derived from asymptotically safe quantum gravity~(see\cite{Basile:2021krr,deBrito:2021akp,Knorr:2024yiu,Eichhorn:2024rkc,Basile:2025zjc} 
for a discussion).

On the other hand, it was recently shown in Ref.\cite{Pastor-Gutierrez:2022nki} for the SM Higgs,
that the situation may be different if one keeps track 
in the FRG calculation of several 
higher-order operators arising from a Taylor expansion of the scalar potential. 
In the specific case of the SM coupled to Einstein-Hilbert gravity, 
one can observe the emergence
of a second relevant direction at the Gaussian fixed point, in addition to the one typically associated with the Higgs mass squared. This additional relevant direction, which remains hidden in calculation performed in a $\mathcal{O}(H^4)$ truncation, is indeed welcome as it allows a phenomenologically viable connection between the UV fixed point and the physical SM at low energies. 

A complete FRG analysis of the SU(6) scalar potential exceeds the purposes of this paper. Here we work under the assumption that $f_{\lam}\gg 0$, so that all the quartic couplings of \refeq{eq:su6pot} in Appendix~\ref{app:scap} are relevant at the Gaussian fixed point. An interesting exercise for future research would be to determine the minimal truncation of the SU(6)$+$gravity effective action that is consistent with the requirement that $f_{\lam}\gg 0$.


\section{Predictions for dark matter\label{sec:DM}}

\subsection{Two-component dark matter\label{sec:2DM}}

With the value of all BSM Yukawa couplings fixed by the UV completion at every scale of phenomenological interest (cf.~\reftable{tab:le_values}), BSM fermion masses become uniquely determined by two relevant parameters, the vevs $v_{s_6}$ and $v_{s_{21}}$. In particular, the masses of the heavy Majorana fermions, derived by diagonalizing \refeq{eq:massnu}, read, in the limit $v_d\ll v_{s_6},v_{s_{21}}$,
\bea
m_{N_1}&\simeq &\sqrt{2}\, y_{\nu_2}v_{s_{21}}\,,\nonumber\\
m_{N_{2}}&\simeq& \sqrt{2}\left(y_{L}v_{s_{6}}-\frac{1}{2}\frac{y^2_{\nu}v_d^2}{y_{\nu_1}v_{s_{21}}}\right)\,,\nonumber\\
m_{N_{3}}&\simeq& \sqrt{2}\left(y_{L}v_{s_{6}}+\frac{1}{2}\frac{y^2_{\nu}v_d^2}{y_{\nu_1}v_{s_{21}}}\right)\,,\nonumber\\ 
m_{N_4}&\simeq&\sqrt{2}\,y_{\nu_1}v_{s_{21}}\,.\label{eq:physmas}
\eea

Depending on the hierarchy between $v_{s_6}$ and $v_{s_{21}}$, different DM scenarios are possible.
\begin{itemize}
\item $v_{s_{21}}< v_{s_6}$ This case results in the mass ordering $m_{N_1}< m_{N_4}< m_{N_{2,3}}$\,. The DM candidate is in this case $N_1$. On the other hand, since $N_2$, $N_3$, and $N_4$ do not mix with $N_1$ at the tree level, the lightest of these three heavy neutrinos, $N_4$, is stable as well. As a consequence, this scenario features a \textit{two-component} DM sector, which is a SM-singlet.
\item {$v_{s_6}< v_{s_{21}}$}. This case leads to the mass ordering $m_{N_{2,3}}< m_{N_1}< m_{N_4}$. 
The stable particles are in this case $N_1$, $N_2$, and $N_3$, 
resulting in two-component Majorana 
DM with an SU(2)$_L$ doublet and a singlet as constituents. 
\end{itemize}

\begin{figure}[t]
	\centering
	\subfloat[]{%
		\includegraphics[width=0.4\textwidth]{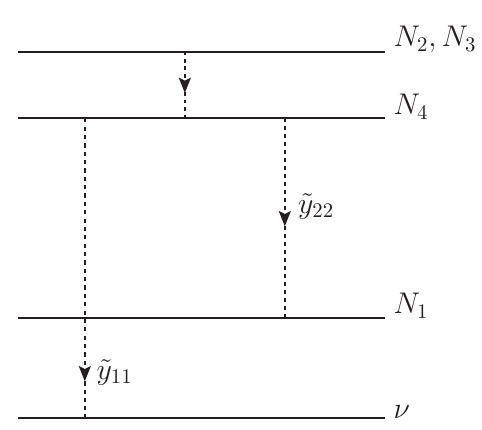}}
	\hspace{0.8cm}	
	\subfloat[]{%
		\includegraphics[width=0.4\textwidth]{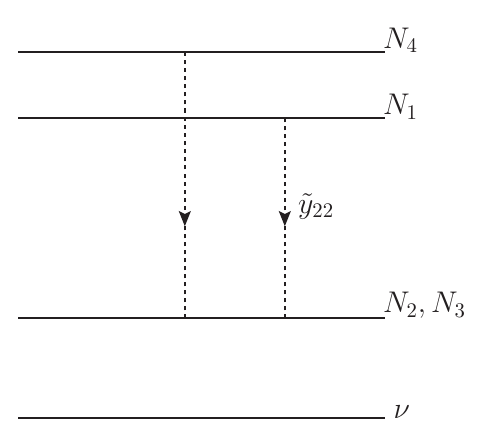}}
\caption{Decay chains of the heavy Majorana fermions $N_1$, $N_2$, $N_3$, and $N_4$ in the presence of non-zero (yet arbitrarily small) Yukawa couplings $\tilde{y}_{11}$ and $\tilde{y}_{22}$.}
\label{fig:onecomdm}
\end{figure}

The situation changes if one considers the scenario discussed in \refsec{sec:nsyc}. In that case very small, yet non-zero couplings~$\tilde{y}_{11}$, $\tilde{y}_{22}$ can be generated dynamically at the low-energy scale. The presence of these extra interactions facilitates the decay of the heaviest of the two-component DM particles into either the SM neutrinos (if $\tilde{y}_{11}$ is turned on), or the lightest BSM Majorana fermion (if $\tilde{y}_{22}$ is turned on). This is illustrated in \reffig{fig:onecomdm}(a) for $v_{s_{21}}< v_{s_6}$\,, and in \reffig{fig:onecomdm}(b) for $v_{s_6}< v_{s_{21}}$\,. 
As a result, DM is in this case one-component: either a SM-singlet (provided $v_{s_{21}}< v_{s_6}$), or a weak isospin doublet (provided $v_{s_6}< v_{s_{21}}$), reminiscent of a supersymmetric Higgsino-like neutralino. Note that, in the latter case, $\tilde{y}_{11}$ has to either remain zero or be very small in order to make the DM candidate stable over cosmological timescales.

Finally, one may consider the possibility that SU(6) coupling~$y_{21}$, which is marginal at the one loop order~(see \reftable{tab:fixpoint}) will eventually emerge as relevant once a higher-order analysis of the stability matrix is performed.
The DM sector would be then equivalent to the case of a non-zero $\tilde{y}_{22}$ illustrated in \reffig{fig:onecomdm}, where the role of $\tilde{y}_{22}$ is replaced by $y'_{\nu}$ (see \refeq{eq:extra_lagr} in Appendix~\ref{app:fer_mas}). Incidentally $y_{21}$ may also generate a small $y'_{d}$, which would in turn induce mixing of the heavy down-type quark with its SM counterpart, thus opening a two-body decay channel of the heavy quark which is otherwise absent.\footnote{Absent the two-body decay channel, one may have to adjust the mass of a color triplet component in $\mathbf{21}^{(S)}$ to allow a three-body decay of the heavy down-type quark into the bottom quark and two singlet BSM Majorana fermions\cite{Ma:2020hyy}.}

\subsection{Singlet fermion DM ($v_{s_{21}} < v_{s_6}$)}

Let us briefly discuss the current bounds on the single-particle DM cases discussed in the previous two paragraphs. In this subsection we consider the case where the DM candidate is a SM-singlet Majorana fermion with mass
\be\label{eq:mdms}
m_{\textrm{DM}}\simeq \sqrt{2}\, y_{\nu_2}v_{s_{21}}\,.
\ee
Its relic abundance at freeze-out is determined by pair annihilation into SM particles via an $s$-channel exchange of heavy BSM scalars, or an $s$-channel exchange of the $Z'$ vector boson (note that $t$-channel ``bulk-like'' annihilation would require mediators with masses of the order of~100\gev\cite{Drees:1992am,Baer:1995nc,Bai:2014osa}). Annihilation via the $s$-channel exchange of a scalar was discussed in Ref.\cite{Ma:2020hyy}.
Following our discussion in \refsec{sec:scal}, we assume in this work that the spectrum in the scalar sector is entirely controlled by relevant parameters in the Lagrangian (mass terms and trilinear couplings), which do not emerge as predictions of the fixed points. For this reason, we focus here on the $Z'$ resonance, which is instead entirely determined by gauge and Yukawa couplings that can be predicted from UV considerations. (In other words, we work under the assumption that 
the physical masses of all CP-even and odd
scalar fields of the model, bar the Higgs boson at 125\gev, are larger than $m_{Z'}$, cf.~Eqs.~(\ref{eq:MH}), (\ref{eq:MHa}) in Appendix~\ref{app:scap}.)

 \begin{figure}[t]
	\centering%
		\includegraphics[width=0.5\textwidth]{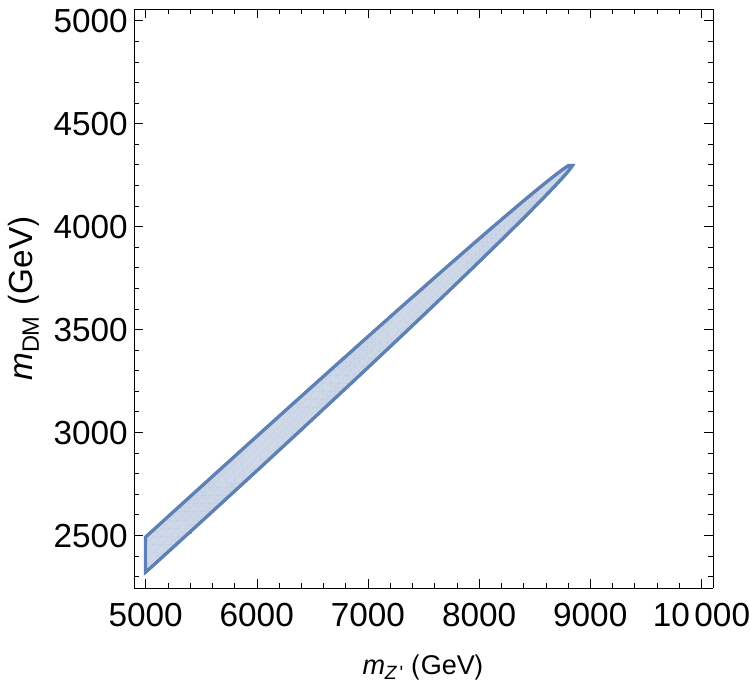}
\caption{The region of under-abundant relic density (in blue) in the SM-singlet DM case.}
\label{fig:Omegah2}
\end{figure}

Upon U(1)$_X$ symmetry breaking, the $Z'$ boson acquires a mass,
\be\label{eq:mzp}
m_{Z'}\simeq 5\, g_X\sqrt{v_{s_{6}}^2+4v^2_{s_{21}}}\,.
\ee
The resonance condition states $m_{Z'}\approx 2\,m_{\textrm{DM}}$, where the approximate sign accounts for the effects of a finite decay width.
We show in \reffig{fig:Omegah2} the region corresponding to the condition $\langle \sigma v\rangle\gsim  2.15 \times 10^{-26}\,\textrm{cm}^3/\textrm{s}$\cite{Cirelli:2024ssz} in the ($m_{Z'}$, $m_{\textrm{DM}}$) plane, where the thermally averaged pair annihilation cross section is defined, following, \textit{e.g.},  Ref.\cite{Okada:2018tgy,Okada:2020cue}, in \refeq{eq:ther_sig} of Appendix~\ref{app:Omegah2}. The lower bound on the $Z'$ mass, $m_{Z'}\gsim 5\tev$, is set by the null results of searches for narrow resonances with sequential SM couplings 
at $\sqrt{s}=13$\tev\ and a total integrated luminosity of up to 140~$\textrm{fb}^{-1}$ by CMS\cite{CMS:2021ctt}. The upper bound on the $Z'$ mass, $m_{Z'}\lesssim 9\tev$ emerges when the annihilation cross section starts to become kinematically suppressed by the mass of the mediator. Note that, since the Yukawa coupling $y_{\nu_2}$ and the gauge coupling $g_X(1\tev)\approx 0.07$ are entirely predicted by the UV completion, the relic-abundance condition indirectly constrains two of the relevant parameters of scalar potential~(\ref{eq:su6pot}) ($M_{62}^2$ and $M_{21}^2$, related to vevs $v_{s_6}$ and $v_{s_{21}}$, respectively) to a very narrow slice of the parameter space. 

Incidentally, the possibility of precisely constraining the physical mass scales of the theory via the measurement of an observable or an effective operator at low energy is one of the most attractive features of UV completions based on AS because it drastically increases the testability of the theory. For a discussion of the uncertainties involved in the process see, \textit{e.g.}, Sec.~3 of Ref.\cite{Kotlarski:2023mmr}.  

\subsection{Doublet DM ($v_{s_{6}}\ll v_{s_{21}}$)} 

The DM is the SU(2)$_L$ doublet Majorana fermion, similar to a supersymmetric Higgsino-like neutralino~(see, \textit{e.g.}, Refs.\cite{Roszkowski:2017nbc,Kowalska:2018toh,Cirelli:2024ssz} for reviews). 
As in the previous case, we work under the assumption that the masses of all CP-even and odd
scalar fields of the model, bar the Higgs boson at 125\gev, are much heavier than the DM particle. It is well known that condition $\langle \sigma v\rangle\approx 2.15 \times 10^{-26}\,\textrm{cm}^3/\textrm{s}$ requires in this case $m_{\textrm{DM}}=m_{N_2}\approx 1.1\tev$, see also \refeq{eq:higgsinosigv} in Appendix~\ref{app:Omegah2}.

As in the singlet DM case, 
all the Yukawa couplings and the gauge coupling $g_X$ are uniquely predicted. Indirect constraints can be derived on the heaviest vev of the DM sector, $v_{s_{21}}$, or, equivalently, on the mass $m_{Z'}$. The lower bound on the $Z'$ mass is established again by the null results of searches for narrow resonances with sequential SM couplings
at $\sqrt{s}=13\tev$ and a total integrated luminosity of up to 140~$\textrm{fb}^{-1}$ by CMS\cite{CMS:2021ctt}. 

The upper bound on the $Z'$ mass, $m_{Z'}\lesssim 50\tev$, 
is extracted from the inelastic scattering limit of the spin-independent DM-nucleon cross section. 
One can derive from 
\refeq{eq:physmas} the mass splitting of the neutral components of the doublet in the limit $v_{s_{21}}\gg v_{s_6}$\,:
\be
\delta m_{\textrm{DM}}\equiv m_{N_3}-m_{N_2} = \sqrt{2}\frac{y_{\nu}^2 v_d^2}{y_{\nu_1} v_{s_{21}}}\,.
\ee
In Ref.\cite{Nagata:2014wma} it was estimated that the lower bound on the mass splitting should be $\delta m_{\textrm{DM}}\geq 200\kev$. For smaller values of $\delta m_{\textrm{DM}}$ inelastic scattering $N_2 p \to N_3 p$, with exchange of a SM $Z$ boson in the $t$-channel, becomes extremely constrained in DM direct-detection experiments. 

 \begin{figure}[t]
	\centering%
		\includegraphics[width=0.5\textwidth]{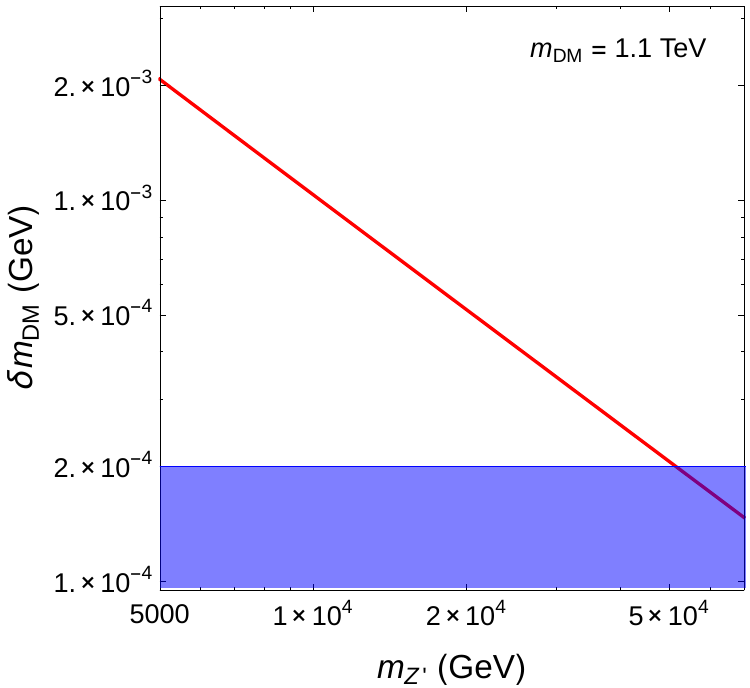}
\caption{The upper bound on $m_{Z'}$ from inelastic scattering limits on the spin-independent DM-nucleon cross section in the case of doublet DM. The exclusion bound on $\delta m_{\textrm{DM}}$ (in blue) is taken from Ref.\cite{Nagata:2014wma}.}
\label{fig:DD_inelastic}
\end{figure}

In \reffig{fig:DD_inelastic} we plot the mass splitting $\delta m_{\textrm{DM}}$ in red solid as a function of $m_{Z'}$. The region excluded by the inelastic scattering limit is shown in blue. In similar fashion to the case of singlet DM, relevant parameters 
$M_{62}^2$ and $M_{21}^2$ of the scalar potential, related to the vevs $v_{s_6}$ and $v_{s_{21}}$ are indirectly constrained to a narrow range of the parameter space. 
 
We conclude by emphasizing that our goal 
in this section is to highlight different ways in which UV boundary conditions based on quantum scale symmetry can constrain the DM sector of the SU(6) theory. It should thus be noted that we are not trying 
to constrain the free parameters of the low-energy theory by means of a very precise quantitative analysis of all relic-density channels and DM-detection bounds (direct and indirect). It goes without saying 
that a full numerical analysis of the DM sector (including two-component DM) would be required to improve on the approximate treatment used for Figs.~\ref{fig:Omegah2} 
and \ref{fig:DD_inelastic}, which is based on formulas
summarized in Appendix~\ref{app:Omegah2}. Such full analysis should also involve the details of the scalar potential, which may in turn open up additional mediator and final-state channels for the DM particle(s) to decay to. We leave these interesting aspects of investigation for future work. 

\section{Summary and conclusions}\label{sec:sum}

In this study, we have investigated the possibility of using trans-Planckian quantum scale symmetry, defined as the existence of Gaussian IR-attractive fixed points of the RG flow, 
to prevent the appearance in a gauge-Yukawa theory 
of couplings that would otherwise be allowed by the gauge symmetry. In particular, we have shown
that this feature can supply a BSM model with a dark sector hosting a DM candidate, and it thus provides a viable alternative to 
the introduction of global or discrete symmetries ``by hand.'' 

As an example, we have analyzed the gauge-Yukawa sector of an SU(6) GUT theory spontaneously broken to SU(5)$\times$U(1)$_X$. SU(5) is subsequently broken spontaneously to the SM group. Incidentally, once the minimal anomaly-free fermionic content is determined the Yukawa sector shows remarkable similarities in all SU($N$) GUTs. As a consequence, our discussion can be generalized straightforwardly to SU($N$) groups of rank higher than~6. The low-energy phenomenology, on the other hand, strongly depends on the details of the scalar sector and 
has to be investigated on a case-by-case basis.
In particular, in our example the SU(6) UV completion gives rise to 
an extended 
version of the 2HDM at scales about or slightly above EWSB.   
The gauge-Yukawa RGE system admits a trans-Planckian fixed point at which all but one Yukawa couplings are irrelevant (or marginal). Of those irrelevant couplings, the nonzero ones provide unique predictions for the strength of the interactions of the extended 2HDM, whereas the Gaussian ones prevent the appearance of interactions 
that can jeopardize the stability of DM.    
Predictivity is thus maximally
enhanced and the only parameters of the theory that remain unconstrained in the UV are the masses of the BSM particles. Importantly, however,  once the interaction strengths are known, physical mass scales can be constrained quite precisely measuring a low-scale observable, which in this case is the relic density of DM.  

An important feature of the asymptotically safe SU(6) GUT discussed in this paper is that the low-scale predictions for the gauge and Yukawa couplings depend only very mildly on the outcome of a quantum-gravity calculation of the effective parameters $f_g$ and $f_y$. While $f_y\neq 0$ is required to obtain the negative critical exponents enforcing the quantum scale symmetry dynamically, close to the Planck scale the RG flow of the irrelevant Yukawa couplings deviates from the scaling solution, driven predominantly by the relevant gauge coupling $g_6$, whose IR value is fixed by the measured values of the SM gauge couplings. As a result the DM phenomenology is much more sensitive to measured/measurable parameters like the strong coupling constant and $\tan\beta$, than it is to the actual value of UV parameters like $f_y$, whose computation is notoriously ridden with large uncertainties. 

The analysis presented in this work can be extended in several directions. First of all, the low-scale constraints on the DM sector were derived here under the assumption that the BSM scalar sector of the 2HDM decouples from the lightest (meta-)stable fermions (by acquiring an appropriately large mass). This is not automatically warranted and, even more importantly, it is not known whether this assumption is in itself consistent with an FRG calculation in AS (in the form of asymptotically safe quantum gravity). We leave it for future work to perform a full fixed-point analysis of the quartic couplings of the SU(6) scalar potential~(\ref{eq:su6pot}), either in itself or in relation to simple truncations of the gravitational action. 

It would also be instructive to perform a detailed numerical analysis of the parameter space of the DM sector of our model (including the quartic and trilinear couplings and the masses of the scalar potential) over several mass scales, in order to identify non-trivial regions and annihilation mechanisms for the correct relic abundance that go beyond the simple cases considered in \refsec{sec:DM}. 

In conclusion, in this work we demonstrated in yet another concrete example the promising role of trans-Planckian AS in shaping predictive UV-complete extensions of the SM, without relying on imposed symmetries to justify hard-to-achieve observational features like the extreme smallness of certain couplings or the stability of DM. This adds to the large body of work that has investigated potential measurable imprints of quantum gravity effects on particle phenomenology. 

\bigskip
\begin{center}
\textbf{ACKNOWLEDGMENTS} 
\end{center}
\noindent 
We would like to thank Nobuchika Okada for a helpful email exchange. RRLdS would like to thank Shiuli Chatterjee, Wojciech Kotlarski, and Daniele Rizzo for discussions and support with numerical codes. AC, RRLdS, and EMS are supported in part by the National Science Centre (Poland) under the research Grant No.~2020/38/E/ST2/00126. KK is supported in part by the National Science Centre (Poland) under the research Grant No.~2017/26/E/ST2/00470. The use of the CIS computer cluster at the National Centre for Nuclear Research in Warsaw is gratefully acknowledged.

\newpage

\appendix
\addcontentsline{toc}{section}{Appendices}


\section{SU(6) scalar potential}
\label{app:scap}

The model introduced in \refsec{sec:su6} contains four scalar multiplets which mediate the Yukawa interactions of \refeq{eq:yuk}: $\mathbf{6_1}^{(S)}$, $\mathbf{6_2}^{(S)}$, $\mathbf{15}^{(S)}$, $\mathbf{21}^{(S)}$, as well as an adjoint $\mathbf{35}^{(S)}$ whose vev breaks the SU(6) symmetry.
The full SU(6)-symmetric scalar potential reads (the superscript $(S)$ is omitted):
\bea\label{eq:su6pot}
V_{\textrm{SU(6)}}&=&-\frac{1}{2}M^2_{35}\textrm{Tr}({\bf 35}^2)+\frac{1}{4}A_{35}[\textrm{Tr}({\bf 35}^2)]^2+\frac{1}{2}B_{35}\textrm{Tr}({\bf 35}^4)\nonumber\\
&&-M^2_{21}\textrm{Tr}({\bf 21^\dagger}{\bf 21})+A_{21}[\textrm{Tr}({\bf 21^\dagger}{\bf 21})]^2+B_{21}\textrm{Tr}[({\bf 21^\dagger}{\bf 21})^2]\nonumber\\[5pt]
&&-M^2_{15}\textrm{Tr}({\bf 15^\dagger}{\bf 15})+A_{15}[\textrm{Tr}({\bf 15^\dagger}{\bf 15})]^2+B_{15}\textrm{Tr}[({\bf 15^\dagger}{\bf 15})^2]\nonumber\\[5pt]
&&-M^2_{61}({\bf 6_{1}^\dagger}{\bf 6_{1}})-M^2_{62}({\bf 6_{2}^\dagger}{\bf 6_{2}})-M^2_{12}({\bf 6_{1}^\dagger}{\bf 6_{2}}+\textrm{H.c.})+B_{61}({\bf 6^\dagger_{1}}{\bf 6_{1}})^2+B_{62}({\bf 6^\dagger_{2}}{\bf 6_{2}})^2\nonumber\\[5pt]
&&+C_3({\bf 6^\dagger_{1}}{\bf 6_{1}})({\bf 6^\dagger_{2}}{\bf 6_{2}})+C_4({\bf 6^\dagger_{1}}{\bf 6_{2}})({\bf 6^\dagger_{2}}{\bf 6_{1}})\nonumber\\[5pt]
 & &+\big(C_5({\bf 6^\dagger_{1}}{\bf 6_{2}})^2+C_6({\bf 6^\dagger_{1}}{\bf 6_{1}})({\bf 6^\dagger_{1}}{\bf 6_{2}})+C_7({\bf 6^\dagger_{2}}{\bf 6_{2}})({\bf 6^\dagger_{1}}{\bf 6_{2}})+\textrm{H.c.}\big)\nonumber\\[5pt]
&&+\alpha_1({\bf 6^\dagger_{1}}{\bf 6_{1}})\textrm{Tr}({\bf 35}^2)+\beta_1{\bf 6^\dagger_{1}}({\bf 35}^2){\bf 6_{1}}+\alpha_2({\bf 6^\dagger_{2}}{\bf 6_{2}})\textrm{Tr}({\bf 35}^2)+\beta_2{\bf 6^\dagger_{2}}({\bf 35}^2){\bf 6_{2}}\nonumber\\[5pt]
&&+\alpha_{12}({\bf 6^\dagger_{1}}{\bf 6_{2}})\textrm{Tr}({\bf 35}^2)+\beta_{12}{\bf 6^\dagger_{1}}({\bf 35}^2){\bf 6_{2}}+\alpha_{21}({\bf 6^\dagger_{2}}{\bf 6_{1}})\textrm{Tr}({\bf 35}^2)+\beta_{21}{\bf 6^\dagger_{2}}({\bf 35}^2){\bf 6_{1}}\nonumber\\[5pt]
&&+\gamma_1({\bf 6^\dagger_{1}}{\bf 6_{1}})\textrm{Tr}({\bf 21^\dagger}{\bf 21})+\delta_1{\bf 6^\dagger_{1}}({\bf 21^\dagger}{\bf 21}){\bf 6_{1}}+\gamma_2({\bf 6^\dagger_{2}}{\bf 6_{2}})\textrm{Tr}({\bf 21^\dagger}{\bf 21})+\delta_2{\bf 6^\dagger_{2}}({\bf 21^\dagger}{\bf 21}){\bf 6_{2}}\nonumber\\[5pt]
&&+\gamma_{12}({\bf 6^\dagger_{1}}{\bf 6_{2}})\textrm{Tr}({\bf 21^\dagger}{\bf 21})+\delta_{12}{\bf 6^\dagger_{1}}({\bf 21^\dagger}{\bf 21}){\bf 6_{2}}+\gamma_{21}({\bf 6^\dagger_{2}}{\bf 6_{1}})\textrm{Tr}({\bf 21^\dagger}{\bf 21})+\delta_{21}{\bf 6^\dagger_{2}}({\bf 21^\dagger}{\bf 21}){\bf 6_{1}}\nonumber\\[5pt]
&&+\rho_1({\bf 6^\dagger_{1}}{\bf 6_{1}})\textrm{Tr}({\bf 15^\dagger}{\bf 15})+\sigma_1{\bf 6^\dagger_{1}}({\bf 15^\dagger}{\bf 15}){\bf 6_{1}}+\rho_2({\bf 6^\dagger_{2}}{\bf 6_{2}})\textrm{Tr}({\bf 15^\dagger}{\bf 15})+\sigma_2{\bf 6^\dagger_{2}}({\bf 15^\dagger}{\bf 15}){\bf 6_{2}}\nonumber\\[5pt]
&&+\rho_{12}({\bf 6^\dagger_{1}}{\bf 6_{2}})\textrm{Tr}({\bf 15^\dagger}{\bf 15})+\sigma_{12}{\bf 6^\dagger_{1}}({\bf 15^\dagger}{\bf 15}){\bf 6_{2}}+\rho_{21}({\bf 6^\dagger_{2}}{\bf 6_{1}})\textrm{Tr}({\bf 15^\dagger}{\bf 15})+\sigma_{21}{\bf 6^\dagger_{2}}({\bf 15^\dagger}{\bf 15}){\bf 6_{1}}\nonumber\\[5pt]
&&+a_1\textrm{Tr}({\bf 15^\dagger}{\bf 15})\textrm{Tr}({\bf 21^\dagger}{\bf 21})+a_2\textrm{Tr}({\bf 15^\dagger}{\bf 15})\textrm{Tr}({\bf 35}^2)+a_3\textrm{Tr}({\bf 35}^2)\textrm{Tr}({\bf 21^\dagger}{\bf 21})\nonumber\\[5pt]
&&+b_1\textrm{Tr}({\bf 15^\dagger}{\bf 15}\,{\bf 21^\dagger}{\bf 21})+b_2\textrm{Tr}({\bf 15^\dagger}{\bf 15}\,{\bf 35}^2)+b_3\textrm{Tr}({\bf 35}^2\,{\bf 21^\dagger}{\bf 21})\nonumber\\[5pt]
&&+\big[\kappa_1({\bf 6_{1}}{\bf 15^\dagger}{\bf 6_{1}})+\kappa_2({\bf 6_{2}}{\bf 15^\dagger}{\bf 6_{2}})+\kappa_3({\bf 6_{1}}{\bf 15^\dagger}{\bf 6_{2}})\nonumber\\[5pt]
&&+\kappa_4({\bf 6_{1}}{\bf 21^\dagger}{\bf 6_{1}})+\kappa_5({\bf 6_{2}}\,{\bf 21^\dagger}{\bf 6_{2}})+\kappa_6({\bf 6_{1}}{\bf 21^\dagger}{\bf 6_{2}})\nonumber\\[5pt]
&&+\lambda_1({\bf 6_{1}^\dagger}{\bf 35}\,{\bf 6_{1}})+\lambda_2({\bf 6_{2}^\dagger}{\bf 35}\,{\bf 6_{2}})+\lambda_3({\bf 6_{1}^\dagger}{\bf 35}\,{\bf 6_{2}})+\lambda_4({\bf 6_{2}^\dagger}{\bf 35}\,{\bf 6_{1}})\nonumber\\[5pt]
&&+\eta_1({\bf 15^\dagger}\,{\bf 35}\,{\bf 15})+\eta_2({\bf 15}\,{\bf 15}\,{\bf 15})+\eta_3({\bf 21^\dagger}{\bf 35}\,{\bf 21})+\eta_4({\bf 21^\dagger}{\bf 35}\,{\bf 15})+\textrm{H.c.}\big]\nonumber\\[5pt]
&&+\xi_1\textrm{Tr}({\bf 21}\,{\bf 21^\dagger}{\bf 21}\,{\bf 15^\dagger})+\xi_2\textrm{Tr}({\bf 15}\,{\bf 15^\dagger}{\bf 15}\,{\bf 21^\dagger})+\xi_3\textrm{Tr}({\bf 15}\,{\bf 15}\,{\bf 15}\,{\bf 35})\nonumber\\[5pt]
&&+\big(\xi_4({\bf 6_1^\dagger}{\bf 15}\,{\bf 21^\dagger}{\bf 6_1})+\xi_5({\bf 6_2^\dagger}{\bf 15}\,{\bf 21^\dagger}{\bf 6_2})+\xi_6({\bf 6_1^\dagger}{\bf 15}\,{\bf 21^\dagger}{\bf 6_2})+\textrm{H.c.}\big)\nonumber\\[5pt]
&&+\xi_7({\bf 6_1}{\bf 15^\dagger}{\bf 35}\,{\bf 6_1})+\xi_8({\bf 6_2}{\bf 15^\dagger}{\bf 35}\,{\bf 6_2})+\xi_9({\bf 6_1}{\bf 15^\dagger}{\bf 35}\,{\bf 6_2})+\xi_{10}({\bf 6_2}{\bf 15^\dagger}{\bf 35}\,{\bf 6_1})\nonumber\\[5pt]
&&+\xi_{11}({\bf 6_1}{\bf 21^\dagger}{\bf 35}\,{\bf 6_1})+\xi_{12}({\bf 6_2}{\bf 21^\dagger}{\bf 35}\,{\bf 6_2})+\xi_{13}({\bf 6_1}{\bf 21^\dagger}{\bf 35}\,{\bf 6_2})+\xi_{14}({\bf 6_2}{\bf 21^\dagger}{\bf 35}\,{\bf 6_1})
\,.
\eea
In \refeq{eq:su6pot} the parameters $M_{i}$ have dimension mass$^2$, the trilinear couplings $\kappa_{i}$, $\lambda_i$ and $\eta_i$ have dimension mass$^1$, and all the quartic couplings $(A_i,B_i,C_i,\alpha_i,\beta_i,\gamma_i,\delta_i,\rho_i,\sigma_i,a_i,b_i,\xi_i)$ are dimensionless. 

\subsection{GUT symmetry breaking\label{app:gsb}}
The SU(6) symmetry is spontaneously broken to SU(5)$\times$U(1)$_X$
by $\langle {\bf 35}\rangle=v_6\,\textrm{diag}(-5,1,1,1,1,1)$. 10 gauge bosons acquire 
masses of the order of $v_6$. The vev is given by 
\be\label{eq:v6}
v_6^2=\frac{M_{35}^2}{30A_{35}+42B_{35}}\,,
\ee
where $M_{35}^2>0$, $A_{35}>0$, and $B_{35}>0$. We decompose the $\mathbf{35}$ in the following way:
\be\label{eq:35matr}
\mathbf{35}-\langle \mathbf{35}\rangle=
  \begin{pmatrix}
        -\frac{5}{\sqrt{30}}A_0 & 0\\
        0 & \mathbf{24}_0+\frac{A_0}{\sqrt{30}}\,\delta_{ij}
        \end{pmatrix},
\ee
where $A_0$ is a singlet of SU(5), $i,j=1,...,5$, and we neglect to write the massless Goldstone bosons in the off-diagonal terms. Given \refeq{eq:su6pot} and \refeq{eq:35matr}, the potential for the $\mathbf{24}_0$ becomes
\be
V_{\textrm{SU(5)}\times\textrm{U(1)}}=
-\frac{1}{2}M_{35}^2\left(\frac{6 B_{35}}{5 A_{35}+7 B_{35}} \right)\textrm{Tr}(\mathbf{24}_0^2)+\frac{1}{4} A_{35}\left[ \textrm{Tr}(\mathbf{24}_0^2)\right]^2+\frac{1}{2} B_{35}\textrm{Tr}(\mathbf{24}_0^4)\,.
\ee

The SU(5) symmetry is spontaneously broken by $\langle {\bf 24}_0\rangle=v_5\,\textrm{diag}(1,1,1,-3/2, -3/2)$,
where 
\be\label{eq:v5}
v_5^2=\frac{12 B_{35} M_{35}^2}{\left(5 A_{35}+7 B_{35}\right)\left(15 A_{35}+7 B_{35} \right)}\,.
\ee
12 gauge bosons acquire masses of the order of $v_5$. We decompose the $\mathbf{24}_0$ in the standard way:
\be\label{eq:24matr}
\mathbf{24}_0-\langle \mathbf{24}_0\rangle=
  \begin{pmatrix}
        \mathbf{8}+\frac{2}{\sqrt{30}} P_0\, \delta_{ij} & 0\\
        0 & \mathbf{3}-\frac{3}{\sqrt{30}}P_0\, \delta_{lm}
        \end{pmatrix},
\ee
where we isolate an octet of color, $\mathbf{8}$, and a weak isospin triplet, $\mathbf{3}$; $P_0$ is a singlet of the SM group and $i,j=1,...,3$, $l,m=1,2$. As before, we neglect to write the massless Goldstone bosons in the off-diagonal terms.  
Note that $M_{35}^2>0$, $A_{35}>0$, and $B_{35}>0$ imply
that the octet, triplet, and singlet fields have positive mass terms.

The model features two scalar multiplets in the fundamental representation, $\mathbf{6_{1,2}}\supset \mathbf{1_{1,2}}(-5)+\mathbf{5_{1,2}}(1)$, 
where we have indicated the U(1)$_X$ charge in parentheses. The model also features scalars $\mathbf{15}$ and $\mathbf{21}$.
After the breaking of SU(6), we can extract from the fourth and seventh lines 
in \refeq{eq:su6pot} the mass terms for the scalar components,
\be
V_{\textrm{SU(5)}\times\textrm{U(1)}}\supset \left[\left(30\alpha_2 +25 \beta_2 \right)v_6^2-M^2_{62}\right] \left|\mathbf{1_2}\right|^2+ \left[\left(30\alpha_2 +\beta_2 \right)v_6^2-M^2_{62}\right]
 (\mathbf{5_2})^{\dag}\mathbf{5_2}\,.
\ee
If the relation $30\alpha_2+25\beta_2=0$ holds, $v_6$ will not contribute to the mass of the SM singlet field $s_6\equiv \mathbf{1_2}(-5)$, which, if $M_{62}^2\ll v_6^2$, will remain light as required in \refsec{sec:mech}.

Equivalently, after the breaking of SU(5) we can extract from the fourth and seventh lines in \refeq{eq:su6pot} the mass terms 
\begin{multline}
V_{\textrm{SU(3)}\times\textrm{SU(2)}\times\textrm{U(1)}\times \textrm{U(1)}}\supset  \left[\left(30 \alpha_1 +25 \beta_1\right)v_6^2-M_{61}^2  \right]  \left|\mathbf{1_1}\right|^2\\
+  \left[\left(30 v_6^2+\frac{15}{2}v_5^2 \right)\alpha_1 +\left(v_6^2+v_5^2 \right) \beta_1-M_{61}^2  \right] \textrm{Tr}(T^{\dag}T)\\
+  \left[\left(30 v_6^2+\frac{15}{2}v_5^2 \right)\alpha_1 +\left(v_6^2+\frac{9}{4}v_5^2 \right) \beta_1-M_{61}^2  \right] H_d H_d^{\dag}\,,
\end{multline}
where doublet $H_d$ was defined in \refeq{eq:lightf} and we have introduced a color triplet $T\equiv(\mathbf{\bar{3}},\mathbf{1},1/3;-1)$.
If the relation
\be\label{eq:dt-spl}
\left(30 v_6^2+\frac{15}{2}v_5^2 \right)\alpha_1 +\left(v_6^2+\frac{9}{4}v_5^2 \right) \beta_1=0
\ee
holds, neither $v_5$ nor $v_6$ contribute to the mass term of the weak isospin doublet. 

Similar relations can be derived in order to make sure that the other isospin doublet, $H_u$, is the only multiplet 
remaining light inside the $\mathbf{15}$, and that the singlet $s_{21}$ is the only field remaining light inside the $\mathbf{21}$.

\subsection{Scalar mass matrices}

After the SU(6) and SU(5) GUT symmetries are broken, we are left with the following potential for the SU(2)$_L$ doublets $H_u$, $H_d$ and the singlets $s_6$, $s_{21}$:
\bea\label{eq:lowpot}
V_{\textrm{low}}&= &\mu_d^2\, H_d H_d^{\dag}+\mu_u^2\, H_u^\dagger H_u+\mu_6^2|s_6|^2+\mu_{21}^2 |s_{21}|^2\nonumber\\
&&+\,z_d\left(H_d H_d^{\dag}\right)^2+z_u\left(H_u^\dagger H_u\right)^2+z_6 |s_6|^4+z_{21} |s_{21}|^4\nonumber\\
&&+\,z_{d6}H_d H_d^{\dag} |s_6|^2+z_{d21} H_d H_d^{\dag} |s_{21}|^2+z_{u6} H_u^\dagger H_u |s_6|^2+z_{u21}H_u^\dagger H_u |s_{21}|^2\nonumber\\
&&+\,z_{du}\left(H_d H_d^{\dag}\right)\left( H_u^\dagger H_u\right)
+z_{dudu}H_d H_u\left(H_d H_u\right)^{\dag}+z_{621}|s_6|^2 |s_{21}|^2\nonumber\\
&&+\left(z_{ud621} H_d H_u s_6^{\dag} s_{21}^{\dag}+w_{du6}H_d H_u s_6+w_{2166}\, s_{21} s_6^2+\textrm{H.c.}\right),
\eea
where the couplings $z_i, w_j$ should be expressed in terms of the parameters of the potential in \refeq{eq:su6pot}. The fields $H_u$, $H_d$, $s_6$ and $s_{21}$ are expanded around their vacuum states, 
\begin{eqnarray}\label{eq:higgses}
H_{u} = 
\begin{pmatrix}
h_{u}^{+} \\
\frac{1}{\sqrt{2}} \left( v_{u} + h_{u}^{0} + i\sigma_u \right)
\end{pmatrix},\qquad\nonumber
H_{d} = 
\begin{pmatrix}
h_{d}^{-}\\
\frac{1}{\sqrt{2}} \left( v_{d} + h_{d}^{0} + i\sigma_d \right)
\end{pmatrix}^T,\\
s_6 = \frac{1}{\sqrt{2}} \left(v_{s_6} + s_6^0 + i\sigma_6 \right),\qquad s_{21} = \frac{1}{\sqrt{2}} \left(v_{s_{21}} + s_{21}^0 + i\sigma_{21} \right)\,.
\end{eqnarray}

The CP-even scalar mass matrix in the basis $(h_{u}^{0},\, h_{d}^{0},\,s_6^0,\, s^0_{21})$ -- $\mathbf{M}_{\textrm{CP-even}}^{2}$ -- can be diagonalized by an orthogonal matrix $R_h$,
\be
\textrm{diag}\{M_{h_1}^2,M_{h_2}^2,M_{h_3}^2,M_{h_4}^2\}=R_h (\mathbf{M}_{\textrm{CP-even}}^{2}) R_h^T\,.
\ee
The lightest eigenvalue, $h_1$, corresponds to the SM Higgs boson.
The masses of the remaining neutral scalars, in the limit $v_{s_6},v_{s_{21}}\gg v_u,v_d$, are approximately given by 
\bea\label{eq:MH}
M_{h_2}^2&\simeq&-z_{ud621}v_{s_{21}} v_{s_6} - \sqrt{2}w_{du6} v_{s_6}\,,\nonumber\\
M_{h_3}^2&\simeq&z_{21}v_{s_{21}}^2 + z_6v_{s_6}^2  -w_{2166} \frac{v_{s_6}^2}{\sqrt{2} v_{s_{21}}}-\sqrt{A}\,,\nonumber\\
M_{h_4}^2&\simeq&z_{21}v_{s_{21}}^2 + z_6v_{s_6}^2  -w_{2166} \frac{v_{s_6}^2}{\sqrt{2} v_{s_{21}}}+\sqrt{A}\,,
\eea
where the factor under square root reads
\begin{multline}
A=\left(z_{21}v^2_{s_{21}}-z_6v^2_{s_6}\right)^2+v^2_{s_6}\left(\sqrt{2}w_{2166}+z_{621}v_{s_{21}}\right)^2\\
-z_{21} w_{2166}\frac{v_{s_{21}} v_{s_6}^2 }{\sqrt{2}}
 + w_{2166}^2\frac{v_{s_6}^4}{8\, v_{s_{21}}^2} + w_{2166} z_6\frac{v_{s_6}^4}{\sqrt{2}\, v_{s_{21}}}\,.
\end{multline}

The CP-odd scalar mass matrix in the basis $(\sigma_u,\, \sigma_d,\,\sigma_6,\, \sigma_{21})$ -- $\mathbf{M}_{\textrm{CP-odd}}^{2}$ -- after diagonalization with an orthogonal matrix $R_a$, leads to the physical spectrum with two massless Goldstone bosons and two massive pseudoscalars, $a_1$ and $a_2$,
\be
{\rm diag}\{0,0,M_{a_1}^2,M_{a_2}^2\}=R_a (\mathbf{M}_{\textrm{CP-odd}}^{2}) R_a^T\,,
\ee
where, in the limit $v_{s_6},v_{s_{21}}\gg v_u,v_d$, one finds
\bea\label{eq:MHa}
M_{a_1}^2&\simeq&-z_{du621}v_{s_{21}} v_{s_6} - \sqrt{2}w_{du6} v_{s_6}\,, \nonumber\\
M_{a_2}^2&\simeq&-\sqrt{2}w_{2166}\frac{4v_{s_{21}}^2 + v_{s_6}^2}{2 v_{s_{21}}}\,.
\eea

Finally, the charged scalar mass matrix -- $\mathbf{M}_{\textrm{Charged}}^{2}$ -- can be diagonalized with a mixing matrix $R_\beta$, 
\be
{\rm diag}\{0,M_{h^\pm}^2\}=R_\beta (\mathbf{M}_{\textrm{Charged}}^{2}) R_\beta^T\,.
\ee
In the physical basis, one is then left with a massless charged Goldstone boson and a charged Higgs boson, whose mass squared reads
\be\label{eq:MHpm}
M^2_{h^\pm}= z_{dudu}  \frac{v_u^2+v_d^2}{2} - \left(v_{s_{21}} z_{du621} + \sqrt{2} w_{du6}\right)\frac{v_{s_6}\left(v_u^2+v_d^2\right)}{2v_dv_u}\,.
\ee
One should notice that, had we not relied on AS to seclude different sectors of the theory, we would need to impose a $\mathbb{Z}_2$ or other symmetry to ensure the (meta-)stability of DM\cite{Ma:2020hyy}. The quartic and trilinear terms in parentheses at the bottom of~\refeq{eq:lowpot}, which are necessary to give the desired masses to the two pseudoscalars and the charged Higgs boson, would then break the $\mathbb{Z}_2$ symmetry. Thus, with quantum scale invariance we also avoid soft-breaking terms.

\section{Renormalization group equations}
\label{app:rges}

In this appendix, we present the leading-order RGEs of the SU(6) gauge-Yukawa sector described in the main text. 
For a generic coupling $c_i$ they take the following form in the trans-Planckian regime:
\be
\frac{d c_i}{dt}=\frac{1}{16 \pi^2}\beta^{(1)}(c_i)-f_{c}\,c_i\,,
\ee
where $f_c=f_g$ (for the gauge coupling) and $f_c=f_y$ (for all Yukawa couplings) are gravitational corrections at the UV fixed point, which rapidly tend to zero below the Planck scale (see Sec.~3.2 of Ref.\cite{Kotlarski:2023mmr} for a discussion of the impact of the position of the Planck scale on the predictions of AS). 
The superscript~(1) means that we work at the 1-loop order. All matter RGEs 
are derived with the public tool~\texttt{PyR@TE 3}\cite{Poole:2019kcm,Sartore:2020gou}. 
{\allowdisplaybreaks
\begin{align*}
\beta^{(1)}(g_6) =- \frac{38}{3} g_6^{3}
\end{align*}
}
{\allowdisplaybreaks
\begin{align*}
\beta^{(1)}(y_{11}) =\left(\frac{17}{2} y_{11}^{2}
+ \frac{17}{2}  y_{12}^{2}
+ \frac{17}{2} y_{21}^{2}
+ y_{22}^{2}
- 10 \tilde{y}_{11}^{2} 
+ \frac{5}{2} \tilde{y}_{12}^{2} 
+ 7 \hat{y}_{11}^{2} + \frac{7}{4} \hat{y}_{12}^{2} + 12 y_{u}^{2} -  \frac{91}{4} g_6^{2}  \right) y_{11}\\
+ \frac{15}{2} y_{12} y_{21} y_{22}
- 5 \tilde{y}_{11} \tilde{y}_{12} y_{21}
- 5 \tilde{y}_{12} \tilde{y}_{22} y_{21}
+ \frac{7}{2} \hat{y}_{11} \hat{y}_{12} y_{21}
+ \frac{7}{2} \hat{y}_{12} \hat{y}_{22} y_{21}
\end{align*}
\begin{align*}
\beta^{(1)}(y_{12}) =\left(
\frac{17}{2} y_{11}^{2}
+ \frac{17}{2} y_{12}^{2}
+ y_{21}^{2}
+ \frac{17}{2} y_{22}^{2}
- 10 \tilde{y}_{11}^{2} 
+ \frac{5}{2} \tilde{y}_{12}^{2} 
+ 7 \hat{y}_{11}^{2} 
+ \frac{7}{4} \hat{y}_{12}^{2} 
+ 12 y_{u}^{2} -  \frac{91}{4} g_6^{2}  \right) y_{12}\\
+\frac{15}{2} y_{11} y_{21} y_{22}
- 5 \tilde{y}_{11}\tilde{y}_{12} y_{22}
- 5 \tilde{y}_{12} \tilde{y}_{22} y_{22}
+ \frac{7}{2} \hat{y}_{11} \hat{y}_{12} y_{22}
+ \frac{7}{2} \hat{y}_{12}\hat{y}_{22} y_{22}
\end{align*}
\begin{align*}
\beta^{(1)}(y_{21}) =\left(
\frac{17}{2} y_{11}^{2} 
+ y_{12}^{2} 
+ \frac{17}{2} y_{21}^{2}
+ \frac{17}{2}  y_{22}^{2}
+ \frac{5}{2} \tilde{y}_{12}^{2}
- 10 \tilde{y}_{22}^{2} 
+ \frac{7}{4} \hat{y}_{12}^{2} 
+ 7 \hat{y}_{22}^{2} 
+ 12 y_{u}^{2} -  \frac{91}{4} g_6^{2}  \right) y_{21}\\
+ \frac{15}{2} y_{11} y_{12} y_{22}
+ 5 \tilde{y}_{11} \tilde{y}_{12} y_{11}
+ 5 \tilde{y}_{12} \tilde{y}_{22} y_{11}
+ \frac{7}{2} \hat{y}_{11} \hat{y}_{12} y_{11}
+ \frac{7}{2} \hat{y}_{12} \hat{y}_{22} y_{11}
\end{align*}
\begin{align*}
\beta^{(1)}(y_{22}) =\left(
y_{11}^{2}
+ \frac{17}{2} y_{12}^{2} 
+ \frac{17}{2} y_{21}^{2} 
+ \frac{17}{2} y_{22}^{2}
+ \frac{5}{2} \tilde{y}_{12}^{2} 
- 10 \tilde{y}_{22}^{2} 
+ \frac{7}{4} \hat{y}_{12}^{2} 
+ 7 \hat{y}_{22}^{2} + 12 y_{u}^{2}  -  \frac{91}{4} g_6^{2} \right) y_{22}\\
+ \frac{15}{2} y_{11} y_{12} y_{21} + 5 \tilde{y}_{11} \tilde{y}_{12} y_{12}
+ 5 \tilde{y}_{12} \tilde{y}_{22} y_{12}
+ \frac{7}{2} \hat{y}_{11} \hat{y}_{12} y_{12}
+ \frac{7}{2} \hat{y}_{12} \hat{y}_{22} y_{12}
\end{align*}
\begin{align*}
\beta^{(1)}(\tilde{y}_{11}) =
\left( 5 y_{11}^{2}
+ 5 y_{12}^{2}
- 24 \tilde{y}_{11}^{2}
+ 12 \tilde{y}_{12}^{2}
- 4 \tilde{y}_{22}^{2}
+ 14 \hat{y}_{11}^{2}
+ \frac{7}{2} \hat{y}_{12}^{2}
+ 12 y_{u}^{2}
-  \frac{35}{2} g_6^{2}
\right)\tilde{y}_{11}
+ 5 \tilde{y}_{12}^{2} \tilde{y}_{22}
\end{align*}
\begin{align*}
\beta^{(1)}(\tilde{y}_{12}) =
\left( \frac{5}{2} y_{11}^{2}
+ \frac{5}{2} y_{12}^{2}
+ \frac{5}{2} y_{21}^{2}
+ \frac{5}{2} y_{22}^{2}
- 24 \tilde{y}_{11}^{2}
+ 7 \tilde{y}_{12}^{2}
- 24 \tilde{y}_{22}^{2}
- 20 \tilde{y}_{11} \tilde{y}_{22} 
+ 7 \hat{y}_{11}^{2}
+ \frac{7}{2} \hat{y}_{12}^{2}
+ 7 \hat{y}_{22}^{2}
 \right.\\
\left. + 12 y_{u}^{2}
-  \frac{35}{2} g_6^{2}
\right) \tilde{y}_{12}
+ 5 \tilde{y}_{11} y_{11} y_{21}
+ 5 \tilde{y}_{11} y_{12} y_{22}
+ 7 \tilde{y}_{11} \hat{y}_{11} \hat{y}_{12}
+ 7 \tilde{y}_{11} \hat{y}_{12} \hat{y}_{22}
\\ + 5 \tilde{y}_{22} y_{11} y_{21}
+ 5 \tilde{y}_{22} y_{12} y_{22}
+ 7 \tilde{y}_{22} \hat{y}_{11} \hat{y}_{12}
+ 7 \tilde{y}_{22} \hat{y}_{12} \hat{y}_{22}
\end{align*}
\begin{align*}
\beta^{(1)}(\tilde{y}_{22}) =
\left( 5 y_{21}^{2}
+ 5 y_{22}^{2}
- 4 \tilde{y}_{11}^{2}
+ 12 \tilde{y}_{12}^{2} 
- 24 \tilde{y}_{22}^{2}
+ \frac{7}{2} \hat{y}_{12}^{2}
+ 14\hat{y}_{22}^{2}
+ 12 y_{u}^{2}
-  \frac{35}{2} g_6^{2} 
\right)\tilde{y}_{22}
+ 5 \tilde{y}_{11} \tilde{y}_{12}^{2}
\end{align*}
\begin{align*}
\beta^{(1)}(\hat{y}_{11}) =\left( 
5 y_{11}^{2}
+ 5 y_{12}^{2}
- 20 \tilde{y}_{11}^{2}
+ 5 \tilde{y}_{12}^{2} 
+ 16 \hat{y}_{11}^{2}
+ 8 \hat{y}_{12}^{2}
+ 2 \hat{y}_{22}^{2}
-  \frac{35}{2} g_6^{2}
\right)\hat{y}_{11}\\
+ \frac{5}{2} \hat{y}_{12} y_{11} y_{21}
+ \frac{5}{2} \hat{y}_{12} y_{12} y_{22}
+ \frac{7}{2} \hat{y}_{12}^{2} \hat{y}_{22}
\end{align*}
\begin{align*}
\beta^{(1)}(\hat{y}_{12}) =\left(
14\hat{y}_{11} \hat{y}_{22}
+ \frac{5}{2}  y_{11}^{2}
+ \frac{5}{2} y_{12}^{2}
+ \frac{5}{2} y_{21}^{2}
+ \frac{5}{2} y_{22}^{2}
- 10 \tilde{y}_{11}^{2}
+ 5 \tilde{y}_{12}^{2} 
- 10 \tilde{y}_{22}^{2} 
+ 16 \hat{y}_{11}^{2} 
+ \frac{9}{2} \hat{y}_{12}^{2}
\right.\\
\left.
+ 16 \hat{y}_{22}^{2}
-  \frac{35}{2} g_6^{2} 
\right)\hat{y}_{12}
- 10 \tilde{y}_{11} \tilde{y}_{12} \hat{y}_{11}
- 10 \tilde{y}_{11} \tilde{y}_{12} \hat{y}_{22}
- 10 \tilde{y}_{12} \tilde{y}_{22} \hat{y}_{11}
- 10 \tilde{y}_{12} \tilde{y}_{22} \hat{y}_{22}\\
+ 5 \hat{y}_{11} y_{11} y_{21}
+ 5 \hat{y}_{11} y_{12} y_{22}
+ 5 \hat{y}_{22} y_{11} y_{21}
+ 5 \hat{y}_{22} y_{12} y_{22}
\end{align*}
\begin{align*}
\beta^{(1)}(\hat{y}_{22}) =\left( 
5 y_{21}^{2}
+ 5 y_{22}^{2}
+ 5 \tilde{y}_{12}^{2}
- 20 \tilde{y}_{22}^{2} 
+ 2 \hat{y}_{11}^{2} 
+ 8 \hat{y}_{12}^{2} 
+ 16 \hat{y}_{22}^{2}
-  \frac{35}{2} g_6^{2}
\right)\hat{y}_{22}\\
+ \frac{7}{2} \hat{y}_{11} \hat{y}_{12}^{2}
+ \frac{5}{2} \hat{y}_{12} y_{11} y_{21}
+ \frac{5}{2} \hat{y}_{12} y_{12} y_{22}
\end{align*}
\begin{align*}
\beta^{(1)}(y_{u}) =\left(
2 y_{11}^{2} 
+ 2 y_{12}^{2} 
+ 2 y_{21}^{2} 
+ 2 y_{22}^{2} 
- 4 \tilde{y}_{11}^{2}
+ 2 \tilde{y}_{12}^{2} 
- 4 \tilde{y}_{22}^{2} 
+ 36 y_{u}^{2}
- 28 g_6^{2} 
\right)y_u
\end{align*}
}


\section{Alternative fixed-point solutions\label{app:new_fp}}


For the sake of completeness, we present in this appendix a few 
 fixed-point solutions to the trans-Planckian RGE system, 
 alternative to the one selected in \reftable{tab:fixpoint} of \refsec{sec:z2}.
They are reported, together with their critical exponents, in \reftable{tab:z2_ma}. Each of the featured fixed points corresponds to a different discrete symmetry of the Yukawa sector in the deep UV. Note that FP2-FP5 are all characterized by a larger number of relevant directions than the fixed point in \reftable{tab:fixpoint}, and are thus less predictive. In contrast, FP1 features only one extra marginal direction with respect to the point selected in \refsec{sec:z2}. However, it is typically excluded on phenomenological grounds, since $y_{11}^{\ast}=y_{22}^{\ast}=0.33$ predicts a low-scale value of the the bottom quark Yukawa larger than the top Yukawa value.

\begin{table}[t]
\centering
\begin{tabular*}{.9\linewidth}{@{\extracolsep{\fill}}|c c c c c c c || c c c c|}
\hline
\multicolumn{11}{|c|}{FP1}\\
\hline
$y_u^{\ast}$ & $\hat{y}_{11}^\ast$ & $\hat{y}_{22}^{\ast}$ & $\tilde{y}_{11}^{\ast}$ & $\tilde{y}_{22}^{\ast}$ & $y_{11}^{\ast}$ & $y_{22}^{\ast}$  & $y_{12}^{\ast}$ & $y_{21}^{\ast}$ & $\tilde{y}_{12}^{\ast}$ & 
$\hat{y}_{12}^{\ast}$  \\
\hline
$0.25$ & $0.37$ & $0.38$  &  $0.15$ & $0.17$ & $0.33$ & $0.33$ & $0.0$ & $0.0$ & $0.0$ & $0.0$  \\
\hline
$\theta_u$ & $\hat{\theta}_{11}$ & $\hat{\theta}_{22}$ & $\tilde{\theta}_{11}$ & $\tilde{\theta}_{22}$ & $\theta_{11}$ & $\theta_{22}$ & $\theta_{12}$ & $\theta_{21}$ &   $\tilde{\theta}_{12}$ & $\hat{\theta}_{12}$\\
\hline
$-5.1$ & $-4.7$ & $-3.6$  & $0$ & $0.62$ & $-3.6$ & $-1.1$ & $0$ & $0$ & $-0.87$ &  $-3.8$  \\
\hline
\end{tabular*}
\begin{tabular*}{.9\linewidth}{@{\extracolsep{\fill}}|c c c c c c c || c c c c|}
\hline
\multicolumn{11}{|c|}{FP2}\\
\hline
$y_u^{\ast}$ & $\hat{y}_{11}^\ast$ & $\hat{y}_{22}^{\ast}$ & $\tilde{y}_{11}^{\ast}$ & $\tilde{y}_{22}^{\ast}$ & $y_{11}^{\ast}$ & $y_{12}^{\ast}$  & $y_{22}^{\ast}$ & $y_{21}^{\ast}$ & $\tilde{y}_{12}^{\ast}$ & 
$\hat{y}_{12}^{\ast}$  \\
\hline
$0.26$ & $0.37$ & $0.41$  &  $0.17$ & $0.15$ & $0.19$ & $0.29$ & $0.0$ & $0.0$ & $0.0$ & $0.0$  \\
\hline
$\theta_u$ & $\hat{\theta}_{11}$ & $\hat{\theta}_{22}$ & $\tilde{\theta}_{11}$ & $\tilde{\theta}_{22}$ & $\theta_{11}$ & $\theta_{12}$ & $\theta_{22}$ & $\theta_{21}$ &   $\tilde{\theta}_{12}$ & $\hat{\theta}_{12}$\\
\hline
$-5.1$ & $-3.6$ & $-5.2$  & $0.64$ & $0$ & $0$ & $-1.0$ & $0$ & $0.61$ & $0.18$ &  $-3.7$  \\
\hline
\end{tabular*}
\begin{tabular*}{.9\linewidth}{@{\extracolsep{\fill}}|c c c c c || c | c | c | c |c c|}
\hline
\multicolumn{11}{|c|}{FP3}\\
\hline
$y_u^{\ast}$ & $\hat{y}_{11}^\ast$ & $\hat{y}_{22}^{\ast}$  & $\tilde{y}_{11}^{\ast}$ & $\tilde{y}_{22}^{\ast}$ & $y_{11}^{\ast}$ & $y_{12}^{\ast}$ & $y_{22}^{\ast}$ & $y_{21}^{\ast}$ & $\tilde{y}_{12}^{\ast}$ & 
$\hat{y}_{12}^{\ast}$  \\
\hline
$0.28$ & $0.41$ & $0.41$ & $0.17$ & $0.17$  &  $0.0$ & $0.0$ & $0.0$ & $0.0$ & $0.0$ & $0.0$  \\
\hline
$\theta_u$ & $\hat{\theta}_{11}$ & $\hat{\theta}_{22}$ & $\tilde{\theta}_{11}$ & $\tilde{\theta}_{22}$ & $\theta_{11}$ & $\theta_{12}$ & $\theta_{22}$ & $\theta_{21}$ &   $\tilde{\theta}_{12}$ & $\hat{\theta}_{12}$\\
\hline
$-5.7$ & $-3.7$ & $-5.1$ & $0$ & $0.66$ & $0.69$ & $0.69$ & $0.69$ & $0.69$ & $0$ &  $-3.7$  \\
\hline
\end{tabular*}
\begin{tabular*}{.9\linewidth}{@{\extracolsep{\fill}}|c c c c c c c || c c | c c|}
\hline
\multicolumn{11}{|c|}{FP4}\\
\hline
$y_u^{\ast}$ & $\hat{y}_{11}^\ast$ & $\hat{y}_{22}^{\ast}$ & $\tilde{y}_{11}^{\ast}$ & $\tilde{y}_{22}^{\ast}$ & $y_{21}^{\ast}$ & $y_{22}^{\ast}$  & $y_{11}^{\ast}$ & $y_{12}^{\ast}$ & $\tilde{y}_{22}^{\ast}$ & 
$\hat{y}_{12}^{\ast}$  \\
\hline
$0.26$ & $0.41$ & $0.37$  &  $0.15$ & $0.17$ & $0.29$ & $0.19$ & $0.0$ & $0.0$ & $0.0$ & $0.0$  \\
\hline
$\theta_u$ & $\hat{\theta}_{11}$ & $\hat{\theta}_{22}$ & $\tilde{\theta}_{11}$ & $\tilde{\theta}_{22}$ & $\theta_{21}$ & $\theta_{22}$ & $\theta_{11}$ & $\theta_{12}$ &   $\tilde{\theta}_{22}$ & $\hat{\theta}_{12}$\\
\hline
$-5.1$ & $-5.2$ & $-3.6$  & $0$ & $0.64$ & $0$ & $-1.0$ & $0$ & $0.61$ & $0$ &  $-3.5$  \\
\hline
\end{tabular*}
\begin{tabular*}{.9\linewidth}{@{\extracolsep{\fill}}|c c c c c c c || c c c c|}
\hline
\multicolumn{11}{|c|}{FP5}\\
\hline
$y_u^{\ast}$ & $\hat{y}_{11}^\ast$ & $\hat{y}_{22}^{\ast}$ & $\tilde{y}_{11}^{\ast}$ & $\tilde{y}_{22}^{\ast}$ & $\tilde{y}_{12}^{\ast}$ & 
$\hat{y}_{12}^{\ast}$ & $y_{11}^{\ast}$ & $y_{12}^{\ast}$  & $y_{21}^{\ast}$ & $y_{22}^{\ast}$  \\
\hline
$0.26$ & $0.37$ & $0.41$  &  $0.17$ & $0.15$ & $0.19$ & $0.29$ & $0.0$ & $0.0$ & $0.0$ & $0.0$  \\
\hline
$\theta_u$ & $\hat{\theta}_{11}$ & $\hat{\theta}_{22}$ & $\tilde{\theta}_{11}$ & $\tilde{\theta}_{22}$ & $\tilde{\theta}_{12}$ & $\hat{\theta}_{12}$ & $\theta_{11}$ & $\theta_{12}$ &   $\theta_{21}$ & $\theta_{22}$\\
\hline
$-5.1$ & $0.86$ & $0.86$  & $0.35$ & $0$ & $-6.4$ & $1.5$ & $1.5$ & $0.48$ & $1.5$ &  $0.48$  \\
\hline
\end{tabular*}
\caption{A sample of trans-Planckian fixed points alternative to the one given in  \reftable{tab:fixpoint}, together with their critical exponents.}
\label{tab:z2_ma}
\end{table}

\section{Relic abundance}
\label{app:Omegah2}

In this appendix, we collect useful formulas that pertain to \reffig{fig:Omegah2} and \reffig{fig:DD_inelastic}.

\paragraph{Vector resonance} Following Ref.\cite{Okada:2018tgy,Okada:2020cue}, we write down a solution for the Boltzmann equation in the case of SM-singlet Majorana fermion pairs annihilating to the SM d.o.f.'s via the $Z'$ resonance, see \refsec{sec:DM}. The thermally averaged annihilation cross section reads\cite{Gondolo:1990dk} 
\be\label{eq:ther_sig}
\langle \sigma v \rangle= \frac{1}{16\pi^4}\left(\frac{m_{\textrm{DM}}}{x_f} \right)\frac{1}{n^2_{\textrm{eq}}}\int_{4 m_{\textrm{DM}}^2}^{\infty} ds\, \hat{\sigma}(s)\sqrt{s}K_1\left(\frac{x_f \sqrt{s}}{m_{\textrm{DM}}} \right),
\ee
where $x_f=m_{\textrm{DM}}/T\approx 30$ and $K_1(z)$ is the modified Bessel function of the second kind. The number density of DM, $n_{\textrm{eq}}$, the entropy density of the thermal plasma, $S(m_{\textrm{DM}})$, and the thermal equilibrium DM yield, $Y(m_{\textrm{DM}})$, are given by
\bea
n_{\textrm{eq}}&=&S(m_{\textrm{DM}})Y(m_{\textrm{DM}})/x_f^3\,,\\
S(m_{\textrm{DM}})&=&\frac{2 \pi^2}{45}g_{\ast}\, m_{\textrm{DM}}^3\,,\\
Y(m_{\textrm{DM}})&=&\frac{1}{\pi^2}\frac{x_f^2\, m_{\textrm{DM}}^3}{S(m_{\textrm{DM}})}K_2(x_f)\,,
\eea
where $g_{\ast}$ is the effective total number of degrees of freedom for the particles in thermal equilibrium. $g_{\ast}=106.75$ is the number of d.o.f.'s of the SM,
and $K_{2}(z)$ is the modified Bessel function of the second kind. The reduced cross section $\hat{\sigma}(s)$ is defined as 
\be
\hat{\sigma}(s)=2\left(s-4 m_{\textrm{DM}}^2 \right) \sigma_{\textrm{SM}}(s)\,,
\ee
which in our model leads to the following expression,
\be
\sigma_{\textrm{SM}}(s) = \frac{25\cdot 135\,\pi}{3}\frac{g_X^4}{16\pi^2}\frac{\sqrt{s\left(s-4 m_{\textrm{DM}}^2 \right)}}{\left(s-m_{Z'}^2 \right)^2+m_{Z'}^2 \Gamma_{Z'}^2}\,,
\label{eq:sigmaZp}
\ee
with the $Z'$ boson decay width given by
\be
\Gamma_{Z'}=\frac{135}{6}\frac{g_X^2}{4\pi} m_{Z'}\,.
\ee

\paragraph{Weak doublet DM} In the limit of all four components (neutral and charged) of the two weak isospin doublets $L_2$ and $L'$ being degenerate in mass, the effective thermally averaged cross section takes into account their co-annihilation into final states involving the gauge bosons of the SM. As long as the scalars of the theory are much heavier than the DM, it is well approximated by (see, \textit{e.g.}, Ref.\cite{ArkaniHamed:2006mb})
\be\label{eq:higgsinosigv}
\langle\sigma v \rangle_{\tilde{H}}^{(\textrm{eff})} \approx \frac{21\,g_2^4+3\,g_2^2 g_Y^2+11\,g_Y^2}{512\,\pi\,m_{\textrm{DM}}^2}\,,
\ee
where  $g_Y(\mu=1\tev)\approx 0.33$ is the U(1)$_Y$ gauge coupling 
and $g_2(\mu=1\tev)\approx 0.65$ is the SU(2)$_L$ gauge coupling.  
$\langle\sigma v \rangle_{\tilde{H}}^{(\textrm{eff})}= 2.15 \times 10^{-26}\,\textrm{cm}^3/\textrm{s} $ implies $m_{\textrm{DM}}=1.1\tev$. 

\section{Full Yukawa Lagrangian and fermion masses\label{app:fer_mas}}

When the couplings that in \reftable{tab:fixpoint} correspond to Gaussian irrelevant directions of the RG flow are allowed to appear via the mechanism discussed in \refsec{sec:nsyc}, more terms are featured in the low-scale Yukawa Lagrangian,
\bea
\mathcal{L}_{\textrm{IR2}}& \supset & \mathcal{L}_{\textrm{IR1}}+\left[ y'_d\, d_2 H_d Q +y'_e\, e H_d L_2 + y'_{\nu}\, L' H_d^{c\dag} \nu_2 \right. \nonumber \\
& &  +\,y'_D\, d_1 d' s_6 + y'_L\, L' L_1 s_6+2\tilde{y}_{11}\,\nu_1 H_u^{c\dag} L_1+2 \tilde{y}_{22}\, \nu_2 H_u^{c\dag} L_2\nonumber\\
& &\left. +\,\tilde{y}_{12}\left(\nu_1 H_u^{c\dag} L_2+\nu_2 H_u^{c\dag} L_1 \right)+\hat{y}_{12}\,\nu_1 \nu_2 s_{21}+\textrm{H.c.} \right],\label{eq:extra_lagr}
\eea
where $\mathcal{L}_{\textrm{IR1}}$ was defined in \refeq{eq:le_lagr}.
Couplings $y'_d$, $y'_e$, $y'_{\nu}$ depart from the common value, $y_{21}$, after the breaking of SU(6). Couplings $y'_D$, $y'_L$ stem from~$y_{12}$\,.  

The mass matrices of the bottom quark and tau lepton read,
\begin{equation}
M_b=\frac{1}{\sqrt{2}}\left(\begin{array}{cc}
y_d v_d & y'_D v_{s_6}  \\
y'_d v_d &  y_D v_{s_6}
\end{array}\right)\,, \quad 
M_{\tau}=\frac{1}{\sqrt{2}}\left(\begin{array}{cc}
y_e v_d & y'_e v_d  \\
y'_L v_{s_6} &  y_L v_{s_6}
\end{array}\right)\,.
\end{equation}

The mass matrix for the neutral fermions, 
in the basis 
$\langle \nu_{L_1}, \nu_{L_2} ,  \nu_{L'}, \nu_1, \nu_2  |,\, 
|  \nu_{L_1}, \nu_{L_2} ,  \nu_{L'}, \nu_1, \nu_2 \rangle$ reads
\be\label{eq:fullmass}
\frac{1}{2} M_{\nu}=\frac{1}{\sqrt{2}}\left(\begin{array}{ccccc}
0 & 0 & y'_L v_{s_6} & 2 \tilde{y}_{11} v_u & \tilde{y}_{12} v_u  \\
0 & 0  & y_L v_{s_6}  & \tilde{y}_{12} v_u & 2 \tilde{y}_{22} v_u\\
y'_L v_{s_6} & y_L v_{s_6}  & 0 & y_{\nu} v_d &  y'_{\nu} v_d \\
2 \tilde{y}_{11} v_u & \tilde{y}_{12} v_u & y_{\nu} v_d   & y_{\nu_1} v_{s_{21}} & \hat{y}_{12} v_{s_{21}}  \\
\tilde{y}_{12} v_u  & 2 \tilde{y}_{22} v_u   & y'_{\nu} v_d   & \hat{y}_{12} v_{s_{21}} & y_{\nu_2} v_{s_{21}} 
\end{array}\right)\,.
\ee



\bibliographystyle{utphysmcite}
\bibliography{mybib}

\ifx\mcitethebibliography\mciteundefinedmacro
\PackageError{unsrtM.bst}{mciteplus.sty has not been loaded}
{This bibstyle requires the use of the mciteplus package.}\fi
\begin{mcitethebibliography}{100}

\bibitem{inbookWS}
S.~Weinberg, {\em General Relativity}, pp.~790--831.
\newblock S.W.Hawking, W.Israel (Eds.), Cambridge Univ. Press, 1980\relax
\mciteBstWouldAddEndPunctfalse
\mciteSetBstMidEndSepPunct{\mcitedefaultmidpunct}
{}{\mcitedefaultseppunct}\relax
\EndOfBibitem
\bibitem{WETTERICH199390}
C.~Wetterich, ``Exact evolution equation for the effective potential,''
  \href{http://dx.doi.org/https://doi.org/10.1016/0370-2693(93)90726-X}{{\em
  Physics Letters B} {\bfseries 301} no.~1, (1993) 90 -- 94}\relax
\mciteBstWouldAddEndPunctfalse
\mciteSetBstMidEndSepPunct{\mcitedefaultmidpunct}
{}{\mcitedefaultseppunct}\relax
\EndOfBibitem
\bibitem{Morris:1993qb}
T.~R. Morris, ``{The Exact renormalization group and approximate solutions},''
  \href{http://dx.doi.org/10.1142/S0217751X94000972}{{\em Int. J. Mod. Phys. A}
  {\bfseries 9} (1994) 2411--2450},
  \href{http://arxiv.org/abs/hep-ph/9308265}{{\ttfamily
  arXiv:hep-ph/9308265}}\relax
\mciteBstWouldAddEndPunctfalse
\mciteSetBstMidEndSepPunct{\mcitedefaultmidpunct}
{}{\mcitedefaultseppunct}\relax
\EndOfBibitem
\bibitem{Reuter:1996cp}
M.~Reuter, ``{Nonperturbative evolution equation for quantum gravity},''
  \href{http://dx.doi.org/10.1103/PhysRevD.57.971}{{\em Phys. Rev. D}
  {\bfseries 57} (1998) 971--985},
  \href{http://arxiv.org/abs/hep-th/9605030}{{\ttfamily
  arXiv:hep-th/9605030}}\relax
\mciteBstWouldAddEndPunctfalse
\mciteSetBstMidEndSepPunct{\mcitedefaultmidpunct}
{}{\mcitedefaultseppunct}\relax
\EndOfBibitem
\bibitem{Lauscher:2001ya}
O.~Lauscher and M.~Reuter, ``{Ultraviolet fixed point and generalized flow
  equation of quantum gravity},''
  \href{http://dx.doi.org/10.1103/PhysRevD.65.025013}{{\em Phys. Rev. D}
  {\bfseries 65} (2002) 025013},
  \href{http://arxiv.org/abs/hep-th/0108040}{{\ttfamily
  arXiv:hep-th/0108040}}\relax
\mciteBstWouldAddEndPunctfalse
\mciteSetBstMidEndSepPunct{\mcitedefaultmidpunct}
{}{\mcitedefaultseppunct}\relax
\EndOfBibitem
\bibitem{Reuter:2001ag}
M.~Reuter and F.~Saueressig, ``{Renormalization group flow of quantum gravity
  in the Einstein-Hilbert truncation},''
  \href{http://dx.doi.org/10.1103/PhysRevD.65.065016}{{\em Phys. Rev. D}
  {\bfseries 65} (2002) 065016},
  \href{http://arxiv.org/abs/hep-th/0110054}{{\ttfamily
  arXiv:hep-th/0110054}}\relax
\mciteBstWouldAddEndPunctfalse
\mciteSetBstMidEndSepPunct{\mcitedefaultmidpunct}
{}{\mcitedefaultseppunct}\relax
\EndOfBibitem
\bibitem{Lauscher:2002sq}
O.~Lauscher and M.~Reuter, ``{Flow equation of quantum Einstein gravity in a
  higher derivative truncation},''
  \href{http://dx.doi.org/10.1103/PhysRevD.66.025026}{{\em Phys. Rev. D}
  {\bfseries 66} (2002) 025026},
  \href{http://arxiv.org/abs/hep-th/0205062}{{\ttfamily
  arXiv:hep-th/0205062}}\relax
\mciteBstWouldAddEndPunctfalse
\mciteSetBstMidEndSepPunct{\mcitedefaultmidpunct}
{}{\mcitedefaultseppunct}\relax
\EndOfBibitem
\bibitem{Litim:2003vp}
D.~F. Litim, ``{Fixed points of quantum gravity},''
  \href{http://dx.doi.org/10.1103/PhysRevLett.92.201301}{{\em Phys. Rev. Lett.}
  {\bfseries 92} (2004) 201301},
  \href{http://arxiv.org/abs/hep-th/0312114}{{\ttfamily
  arXiv:hep-th/0312114}}\relax
\mciteBstWouldAddEndPunctfalse
\mciteSetBstMidEndSepPunct{\mcitedefaultmidpunct}
{}{\mcitedefaultseppunct}\relax
\EndOfBibitem
\bibitem{Codello:2006in}
A.~Codello and R.~Percacci, ``{Fixed points of higher derivative gravity},''
  \href{http://dx.doi.org/10.1103/PhysRevLett.97.221301}{{\em Phys. Rev. Lett.}
  {\bfseries 97} (2006) 221301},
  \href{http://arxiv.org/abs/hep-th/0607128}{{\ttfamily
  arXiv:hep-th/0607128}}\relax
\mciteBstWouldAddEndPunctfalse
\mciteSetBstMidEndSepPunct{\mcitedefaultmidpunct}
{}{\mcitedefaultseppunct}\relax
\EndOfBibitem
\bibitem{Machado:2007ea}
P.~F. Machado and F.~Saueressig, ``{On the renormalization group flow of
  f(R)-gravity},'' \href{http://dx.doi.org/10.1103/PhysRevD.77.124045}{{\em
  Phys. Rev. D} {\bfseries 77} (2008) 124045},
  \href{http://arxiv.org/abs/0712.0445}{{\ttfamily arXiv:0712.0445
  [hep-th]}}\relax
\mciteBstWouldAddEndPunctfalse
\mciteSetBstMidEndSepPunct{\mcitedefaultmidpunct}
{}{\mcitedefaultseppunct}\relax
\EndOfBibitem
\bibitem{Codello:2008vh}
A.~Codello, R.~Percacci, and C.~Rahmede, ``{Investigating the Ultraviolet
  Properties of Gravity with a Wilsonian Renormalization Group Equation},''
  \href{http://dx.doi.org/10.1016/j.aop.2008.08.008}{{\em Annals Phys.}
  {\bfseries 324} (2009) 414--469},
  \href{http://arxiv.org/abs/0805.2909}{{\ttfamily arXiv:0805.2909
  [hep-th]}}\relax
\mciteBstWouldAddEndPunctfalse
\mciteSetBstMidEndSepPunct{\mcitedefaultmidpunct}
{}{\mcitedefaultseppunct}\relax
\EndOfBibitem
\bibitem{Benedetti:2009rx}
D.~Benedetti, P.~F. Machado, and F.~Saueressig, ``{Asymptotic safety in
  higher-derivative gravity},''
  \href{http://dx.doi.org/10.1142/S0217732309031521}{{\em Mod. Phys. Lett. A}
  {\bfseries 24} (2009) 2233--2241},
  \href{http://arxiv.org/abs/0901.2984}{{\ttfamily arXiv:0901.2984
  [hep-th]}}\relax
\mciteBstWouldAddEndPunctfalse
\mciteSetBstMidEndSepPunct{\mcitedefaultmidpunct}
{}{\mcitedefaultseppunct}\relax
\EndOfBibitem
\bibitem{Dietz:2012ic}
J.~A. Dietz and T.~R. Morris, ``{Asymptotic safety in the f(R)
  approximation},'' \href{http://dx.doi.org/10.1007/JHEP01(2013)108}{{\em JHEP}
  {\bfseries 01} (2013) 108}, \href{http://arxiv.org/abs/1211.0955}{{\ttfamily
  arXiv:1211.0955 [hep-th]}}\relax
\mciteBstWouldAddEndPunctfalse
\mciteSetBstMidEndSepPunct{\mcitedefaultmidpunct}
{}{\mcitedefaultseppunct}\relax
\EndOfBibitem
\bibitem{Falls:2013bv}
K.~Falls, D.~Litim, K.~Nikolakopoulos, and C.~Rahmede, ``{A bootstrap towards
  asymptotic safety},'' \href{http://arxiv.org/abs/1301.4191}{{\ttfamily
  arXiv:1301.4191 [hep-th]}}\relax
\mciteBstWouldAddEndPunctfalse
\mciteSetBstMidEndSepPunct{\mcitedefaultmidpunct}
{}{\mcitedefaultseppunct}\relax
\EndOfBibitem
\bibitem{Falls:2014tra}
K.~Falls, D.~F. Litim, K.~Nikolakopoulos, and C.~Rahmede, ``{Further evidence
  for asymptotic safety of quantum gravity},''
  \href{http://dx.doi.org/10.1103/PhysRevD.93.104022}{{\em Phys. Rev. D}
  {\bfseries 93} no.~10, (2016) 104022},
  \href{http://arxiv.org/abs/1410.4815}{{\ttfamily arXiv:1410.4815
  [hep-th]}}\relax
\mciteBstWouldAddEndPunctfalse
\mciteSetBstMidEndSepPunct{\mcitedefaultmidpunct}
{}{\mcitedefaultseppunct}\relax
\EndOfBibitem
\bibitem{Percacci:2002ie}
R.~Percacci and D.~Perini, ``{Constraints on matter from asymptotic safety},''
  \href{http://dx.doi.org/10.1103/PhysRevD.67.081503}{{\em Phys. Rev. D}
  {\bfseries 67} (2003) 081503},
  \href{http://arxiv.org/abs/hep-th/0207033}{{\ttfamily
  arXiv:hep-th/0207033}}\relax
\mciteBstWouldAddEndPunctfalse
\mciteSetBstMidEndSepPunct{\mcitedefaultmidpunct}
{}{\mcitedefaultseppunct}\relax
\EndOfBibitem
\bibitem{Percacci:2003jz}
R.~Percacci and D.~Perini, ``{Asymptotic safety of gravity coupled to
  matter},'' \href{http://dx.doi.org/10.1103/PhysRevD.68.044018}{{\em Phys.
  Rev. D} {\bfseries 68} (2003) 044018},
  \href{http://arxiv.org/abs/hep-th/0304222}{{\ttfamily
  arXiv:hep-th/0304222}}\relax
\mciteBstWouldAddEndPunctfalse
\mciteSetBstMidEndSepPunct{\mcitedefaultmidpunct}
{}{\mcitedefaultseppunct}\relax
\EndOfBibitem
\bibitem{Zanusso:2009bs}
O.~Zanusso, L.~Zambelli, G.~Vacca, and R.~Percacci, ``{Gravitational
  corrections to Yukawa systems},''
  \href{http://dx.doi.org/10.1016/j.physletb.2010.04.043}{{\em Phys. Lett. B}
  {\bfseries 689} (2010) 90--94},
  \href{http://arxiv.org/abs/0904.0938}{{\ttfamily arXiv:0904.0938
  [hep-th]}}\relax
\mciteBstWouldAddEndPunctfalse
\mciteSetBstMidEndSepPunct{\mcitedefaultmidpunct}
{}{\mcitedefaultseppunct}\relax
\EndOfBibitem
\bibitem{Daum:2009dn}
J.-E. Daum, U.~Harst, and M.~Reuter, ``{Running Gauge Coupling in
  Asymptotically Safe Quantum Gravity},''
  \href{http://dx.doi.org/10.1007/JHEP01(2010)084}{{\em JHEP} {\bfseries 01}
  (2010) 084}, \href{http://arxiv.org/abs/0910.4938}{{\ttfamily arXiv:0910.4938
  [hep-th]}}\relax
\mciteBstWouldAddEndPunctfalse
\mciteSetBstMidEndSepPunct{\mcitedefaultmidpunct}
{}{\mcitedefaultseppunct}\relax
\EndOfBibitem
\bibitem{Daum:2010bc}
J.-E. Daum, U.~Harst, and M.~Reuter, ``{Non-perturbative QEG Corrections to the
  Yang-Mills Beta Function},''
  \href{http://dx.doi.org/10.1007/s10714-010-1032-2}{{\em Gen. Rel. Grav.}
  {\bfseries 43} (2011) 2393}, \href{http://arxiv.org/abs/1005.1488}{{\ttfamily
  arXiv:1005.1488 [hep-th]}}\relax
\mciteBstWouldAddEndPunctfalse
\mciteSetBstMidEndSepPunct{\mcitedefaultmidpunct}
{}{\mcitedefaultseppunct}\relax
\EndOfBibitem
\bibitem{Folkerts:2011jz}
S.~Folkerts, D.~F. Litim, and J.~M. Pawlowski, ``{Asymptotic freedom of
  Yang-Mills theory with gravity},''
  \href{http://dx.doi.org/10.1016/j.physletb.2012.02.002}{{\em Phys. Lett. B}
  {\bfseries 709} (2012) 234--241},
  \href{http://arxiv.org/abs/1101.5552}{{\ttfamily arXiv:1101.5552
  [hep-th]}}\relax
\mciteBstWouldAddEndPunctfalse
\mciteSetBstMidEndSepPunct{\mcitedefaultmidpunct}
{}{\mcitedefaultseppunct}\relax
\EndOfBibitem
\bibitem{Dona:2013qba}
P.~Donà, A.~Eichhorn, and R.~Percacci, ``{Matter matters in asymptotically
  safe quantum gravity},''
  \href{http://dx.doi.org/10.1103/PhysRevD.89.084035}{{\em Phys. Rev. D}
  {\bfseries 89} no.~8, (2014) 084035},
  \href{http://arxiv.org/abs/1311.2898}{{\ttfamily arXiv:1311.2898
  [hep-th]}}\relax
\mciteBstWouldAddEndPunctfalse
\mciteSetBstMidEndSepPunct{\mcitedefaultmidpunct}
{}{\mcitedefaultseppunct}\relax
\EndOfBibitem
\bibitem{Meibohm:2015twa}
J.~Meibohm, J.~M. Pawlowski, and M.~Reichert, ``{Asymptotic safety of
  gravity-matter systems},''
  \href{http://dx.doi.org/10.1103/PhysRevD.93.084035}{{\em Phys. Rev. D}
  {\bfseries 93} no.~8, (2016) 084035},
  \href{http://arxiv.org/abs/1510.07018}{{\ttfamily arXiv:1510.07018
  [hep-th]}}\relax
\mciteBstWouldAddEndPunctfalse
\mciteSetBstMidEndSepPunct{\mcitedefaultmidpunct}
{}{\mcitedefaultseppunct}\relax
\EndOfBibitem
\bibitem{Oda:2015sma}
K.-y. Oda and M.~Yamada, ``{Non-minimal coupling in Higgs--Yukawa model with
  asymptotically safe gravity},''
  \href{http://dx.doi.org/10.1088/0264-9381/33/12/125011}{{\em Class. Quant.
  Grav.} {\bfseries 33} no.~12, (2016) 125011},
  \href{http://arxiv.org/abs/1510.03734}{{\ttfamily arXiv:1510.03734
  [hep-th]}}\relax
\mciteBstWouldAddEndPunctfalse
\mciteSetBstMidEndSepPunct{\mcitedefaultmidpunct}
{}{\mcitedefaultseppunct}\relax
\EndOfBibitem
\bibitem{Eichhorn:2016esv}
A.~Eichhorn, A.~Held, and J.~M. Pawlowski, ``{Quantum-gravity effects on a
  Higgs-Yukawa model},''
  \href{http://dx.doi.org/10.1103/PhysRevD.94.104027}{{\em Phys. Rev. D}
  {\bfseries 94} no.~10, (2016) 104027},
  \href{http://arxiv.org/abs/1604.02041}{{\ttfamily arXiv:1604.02041
  [hep-th]}}\relax
\mciteBstWouldAddEndPunctfalse
\mciteSetBstMidEndSepPunct{\mcitedefaultmidpunct}
{}{\mcitedefaultseppunct}\relax
\EndOfBibitem
\bibitem{Christiansen:2017cxa}
N.~Christiansen, D.~F. Litim, J.~M. Pawlowski, and M.~Reichert, ``{Asymptotic
  safety of gravity with matter},''
  \href{http://dx.doi.org/10.1103/PhysRevD.97.106012}{{\em Phys. Rev. D}
  {\bfseries 97} no.~10, (2018) 106012},
  \href{http://arxiv.org/abs/1710.04669}{{\ttfamily arXiv:1710.04669
  [hep-th]}}\relax
\mciteBstWouldAddEndPunctfalse
\mciteSetBstMidEndSepPunct{\mcitedefaultmidpunct}
{}{\mcitedefaultseppunct}\relax
\EndOfBibitem
\bibitem{Eichhorn:2017eht}
A.~Eichhorn and A.~Held, ``{Viability of quantum-gravity induced ultraviolet
  completions for matter},''
  \href{http://dx.doi.org/10.1103/PhysRevD.96.086025}{{\em Phys. Rev. D}
  {\bfseries 96} no.~8, (2017) 086025},
  \href{http://arxiv.org/abs/1705.02342}{{\ttfamily arXiv:1705.02342
  [gr-qc]}}\relax
\mciteBstWouldAddEndPunctfalse
\mciteSetBstMidEndSepPunct{\mcitedefaultmidpunct}
{}{\mcitedefaultseppunct}\relax
\EndOfBibitem
\bibitem{Pawlowski:2020qer}
J.~M. Pawlowski and M.~Reichert, ``{Quantum Gravity: A Fluctuating Point of
  View},'' \href{http://dx.doi.org/10.3389/fphy.2020.551848}{{\em Front. in
  Phys.} {\bfseries 8} (2021) 551848},
  \href{http://arxiv.org/abs/2007.10353}{{\ttfamily arXiv:2007.10353
  [hep-th]}}\relax
\mciteBstWouldAddEndPunctfalse
\mciteSetBstMidEndSepPunct{\mcitedefaultmidpunct}
{}{\mcitedefaultseppunct}\relax
\EndOfBibitem
\bibitem{Eichhorn:2022gku}
A.~Eichhorn and M.~Schiffer, ``{Asymptotic safety of gravity with matter},''
  \href{http://arxiv.org/abs/2212.07456}{{\ttfamily arXiv:2212.07456
  [hep-th]}}\relax
\mciteBstWouldAddEndPunctfalse
\mciteSetBstMidEndSepPunct{\mcitedefaultmidpunct}
{}{\mcitedefaultseppunct}\relax
\EndOfBibitem
\bibitem{Wetterich:1987fm}
C.~Wetterich, ``{Cosmology and the Fate of Dilatation Symmetry},''
  \href{http://dx.doi.org/10.1016/0550-3213(88)90193-9}{{\em Nucl. Phys. B}
  {\bfseries 302} (1988) 668--696},
  \href{http://arxiv.org/abs/1711.03844}{{\ttfamily arXiv:1711.03844
  [hep-th]}}\relax
\mciteBstWouldAddEndPunctfalse
\mciteSetBstMidEndSepPunct{\mcitedefaultmidpunct}
{}{\mcitedefaultseppunct}\relax
\EndOfBibitem
\bibitem{Shaposhnikov:2008xb}
M.~Shaposhnikov and D.~Zenhausern, ``{Scale invariance, unimodular gravity and
  dark energy},'' \href{http://dx.doi.org/10.1016/j.physletb.2008.11.054}{{\em
  Phys. Lett. B} {\bfseries 671} (2009) 187--192},
  \href{http://arxiv.org/abs/0809.3395}{{\ttfamily arXiv:0809.3395
  [hep-th]}}\relax
\mciteBstWouldAddEndPunctfalse
\mciteSetBstMidEndSepPunct{\mcitedefaultmidpunct}
{}{\mcitedefaultseppunct}\relax
\EndOfBibitem
\bibitem{Wetterich:2019qzx}
C.~Wetterich, ``{Quantum scale symmetry},''
  \href{http://arxiv.org/abs/1901.04741}{{\ttfamily arXiv:1901.04741
  [hep-th]}}\relax
\mciteBstWouldAddEndPunctfalse
\mciteSetBstMidEndSepPunct{\mcitedefaultmidpunct}
{}{\mcitedefaultseppunct}\relax
\EndOfBibitem
\bibitem{Wetterich:2020cxq}
C.~Wetterich, ``{Fundamental scale invariance},''
  \href{http://dx.doi.org/10.1016/j.nuclphysb.2021.115326}{{\em Nucl. Phys. B}
  {\bfseries 964} (2021) 115326},
  \href{http://arxiv.org/abs/2007.08805}{{\ttfamily arXiv:2007.08805
  [hep-th]}}\relax
\mciteBstWouldAddEndPunctfalse
\mciteSetBstMidEndSepPunct{\mcitedefaultmidpunct}
{}{\mcitedefaultseppunct}\relax
\EndOfBibitem
\bibitem{Harst:2011zx}
U.~Harst and M.~Reuter, ``{QED coupled to QEG},''
  \href{http://dx.doi.org/10.1007/JHEP05(2011)119}{{\em JHEP} {\bfseries 05}
  (2011) 119}, \href{http://arxiv.org/abs/1101.6007}{{\ttfamily arXiv:1101.6007
  [hep-th]}}\relax
\mciteBstWouldAddEndPunctfalse
\mciteSetBstMidEndSepPunct{\mcitedefaultmidpunct}
{}{\mcitedefaultseppunct}\relax
\EndOfBibitem
\bibitem{Christiansen:2017gtg}
N.~Christiansen and A.~Eichhorn, ``{An asymptotically safe solution to the U(1)
  triviality problem},''
  \href{http://dx.doi.org/10.1016/j.physletb.2017.04.047}{{\em Phys. Lett. B}
  {\bfseries 770} (2017) 154--160},
  \href{http://arxiv.org/abs/1702.07724}{{\ttfamily arXiv:1702.07724
  [hep-th]}}\relax
\mciteBstWouldAddEndPunctfalse
\mciteSetBstMidEndSepPunct{\mcitedefaultmidpunct}
{}{\mcitedefaultseppunct}\relax
\EndOfBibitem
\bibitem{Eichhorn:2017lry}
A.~Eichhorn and F.~Versteegen, ``{Upper bound on the Abelian gauge coupling
  from asymptotic safety},''
  \href{http://dx.doi.org/10.1007/JHEP01(2018)030}{{\em JHEP} {\bfseries 01}
  (2018) 030}, \href{http://arxiv.org/abs/1709.07252}{{\ttfamily
  arXiv:1709.07252 [hep-th]}}\relax
\mciteBstWouldAddEndPunctfalse
\mciteSetBstMidEndSepPunct{\mcitedefaultmidpunct}
{}{\mcitedefaultseppunct}\relax
\EndOfBibitem
\bibitem{Shaposhnikov:2009pv}
M.~Shaposhnikov and C.~Wetterich, ``{Asymptotic safety of gravity and the Higgs
  boson mass},'' \href{http://dx.doi.org/10.1016/j.physletb.2009.12.022}{{\em
  Phys. Lett. B} {\bfseries 683} (2010) 196--200},
  \href{http://arxiv.org/abs/0912.0208}{{\ttfamily arXiv:0912.0208
  [hep-th]}}\relax
\mciteBstWouldAddEndPunctfalse
\mciteSetBstMidEndSepPunct{\mcitedefaultmidpunct}
{}{\mcitedefaultseppunct}\relax
\EndOfBibitem
\bibitem{Eichhorn:2017als}
A.~Eichhorn, Y.~Hamada, J.~Lumma, and M.~Yamada, ``{Quantum gravity
  fluctuations flatten the Planck-scale Higgs potential},''
  \href{http://dx.doi.org/10.1103/PhysRevD.97.086004}{{\em Phys. Rev. D}
  {\bfseries 97} no.~8, (2018) 086004},
  \href{http://arxiv.org/abs/1712.00319}{{\ttfamily arXiv:1712.00319
  [hep-th]}}\relax
\mciteBstWouldAddEndPunctfalse
\mciteSetBstMidEndSepPunct{\mcitedefaultmidpunct}
{}{\mcitedefaultseppunct}\relax
\EndOfBibitem
\bibitem{Kwapisz:2019wrl}
J.~H. Kwapisz, ``{Asymptotic safety, the Higgs boson mass, and beyond the
  standard model physics},''
  \href{http://dx.doi.org/10.1103/PhysRevD.100.115001}{{\em Phys. Rev. D}
  {\bfseries 100} no.~11, (2019) 115001},
  \href{http://arxiv.org/abs/1907.12521}{{\ttfamily arXiv:1907.12521
  [hep-ph]}}\relax
\mciteBstWouldAddEndPunctfalse
\mciteSetBstMidEndSepPunct{\mcitedefaultmidpunct}
{}{\mcitedefaultseppunct}\relax
\EndOfBibitem
\bibitem{Eichhorn:2021tsx}
A.~Eichhorn, M.~Pauly, and S.~Ray, ``{Towards a Higgs mass determination in
  asymptotically safe gravity with a dark portal},''
  \href{http://dx.doi.org/10.1007/JHEP10(2021)100}{{\em JHEP} {\bfseries 10}
  (2021) 100}, \href{http://arxiv.org/abs/2107.07949}{{\ttfamily
  arXiv:2107.07949 [hep-ph]}}\relax
\mciteBstWouldAddEndPunctfalse
\mciteSetBstMidEndSepPunct{\mcitedefaultmidpunct}
{}{\mcitedefaultseppunct}\relax
\EndOfBibitem
\bibitem{Eichhorn:2017ylw}
A.~Eichhorn and A.~Held, ``{Top mass from asymptotic safety},''
  \href{http://dx.doi.org/10.1016/j.physletb.2017.12.040}{{\em Phys. Lett. B}
  {\bfseries 777} (2018) 217--221},
  \href{http://arxiv.org/abs/1707.01107}{{\ttfamily arXiv:1707.01107
  [hep-th]}}\relax
\mciteBstWouldAddEndPunctfalse
\mciteSetBstMidEndSepPunct{\mcitedefaultmidpunct}
{}{\mcitedefaultseppunct}\relax
\EndOfBibitem
\bibitem{Eichhorn:2018whv}
A.~Eichhorn and A.~Held, ``{Mass difference for charged quarks from
  asymptotically safe quantum gravity},''
  \href{http://dx.doi.org/10.1103/PhysRevLett.121.151302}{{\em Phys. Rev.
  Lett.} {\bfseries 121} no.~15, (2018) 151302},
  \href{http://arxiv.org/abs/1803.04027}{{\ttfamily arXiv:1803.04027
  [hep-th]}}\relax
\mciteBstWouldAddEndPunctfalse
\mciteSetBstMidEndSepPunct{\mcitedefaultmidpunct}
{}{\mcitedefaultseppunct}\relax
\EndOfBibitem
\bibitem{Alkofer:2020vtb}
R.~Alkofer, A.~Eichhorn, A.~Held, C.~M. Nieto, R.~Percacci, and M.~Schr\"ofl,
  ``{Quark masses and mixings in minimally parameterized UV completions of the
  Standard Model},'' \href{http://dx.doi.org/10.1016/j.aop.2020.168282}{{\em
  Annals Phys.} {\bfseries 421} (2020) 168282},
  \href{http://arxiv.org/abs/2003.08401}{{\ttfamily arXiv:2003.08401
  [hep-ph]}}\relax
\mciteBstWouldAddEndPunctfalse
\mciteSetBstMidEndSepPunct{\mcitedefaultmidpunct}
{}{\mcitedefaultseppunct}\relax
\EndOfBibitem
\bibitem{Kowalska:2022ypk}
K.~Kowalska, S.~Pramanick, and E.~M. Sessolo, ``{Naturally small Yukawa
  couplings from trans-Planckian asymptotic safety},''
  \href{http://dx.doi.org/10.1007/JHEP08(2022)262}{{\em JHEP} {\bfseries 08}
  (2022) 262}, \href{http://arxiv.org/abs/2204.00866}{{\ttfamily
  arXiv:2204.00866 [hep-ph]}}\relax
\mciteBstWouldAddEndPunctfalse
\mciteSetBstMidEndSepPunct{\mcitedefaultmidpunct}
{}{\mcitedefaultseppunct}\relax
\EndOfBibitem
\bibitem{DeBrito:2019rrh}
G.~P. De~Brito, Y.~Hamada, A.~D. Pereira, and M.~Yamada, ``{On the impact of
  Majorana masses in gravity-matter systems},''
  \href{http://dx.doi.org/10.1007/JHEP08(2019)142}{{\em JHEP} {\bfseries 08}
  (2019) 142}, \href{http://arxiv.org/abs/1905.11114}{{\ttfamily
  arXiv:1905.11114 [hep-th]}}\relax
\mciteBstWouldAddEndPunctfalse
\mciteSetBstMidEndSepPunct{\mcitedefaultmidpunct}
{}{\mcitedefaultseppunct}\relax
\EndOfBibitem
\bibitem{Hamada:2020vnf}
Y.~Hamada, K.~Tsumura, and M.~Yamada, ``{Scalegenesis and fermionic dark
  matters in the flatland scenario},''
  \href{http://dx.doi.org/10.1140/epjc/s10052-020-7929-3}{{\em Eur. Phys. J. C}
  {\bfseries 80} no.~5, (2020) 368},
  \href{http://arxiv.org/abs/2002.03666}{{\ttfamily arXiv:2002.03666
  [hep-ph]}}\relax
\mciteBstWouldAddEndPunctfalse
\mciteSetBstMidEndSepPunct{\mcitedefaultmidpunct}
{}{\mcitedefaultseppunct}\relax
\EndOfBibitem
\bibitem{Domenech:2020yjf}
G.~Dom\`enech, M.~Goodsell, and C.~Wetterich, ``{Neutrino masses, vacuum
  stability and quantum gravity prediction for the mass of the top quark},''
  \href{http://dx.doi.org/10.1007/JHEP01(2021)180}{{\em JHEP} {\bfseries 01}
  (2021) 180}, \href{http://arxiv.org/abs/2008.04310}{{\ttfamily
  arXiv:2008.04310 [hep-ph]}}\relax
\mciteBstWouldAddEndPunctfalse
\mciteSetBstMidEndSepPunct{\mcitedefaultmidpunct}
{}{\mcitedefaultseppunct}\relax
\EndOfBibitem
\bibitem{Grabowski:2018fjj}
F.~Grabowski, J.~H. Kwapisz, and K.~A. Meissner, ``{Asymptotic safety and
  Conformal Standard Model},''
  \href{http://dx.doi.org/10.1103/PhysRevD.99.115029}{{\em Phys. Rev. D}
  {\bfseries 99} no.~11, (2019) 115029},
  \href{http://arxiv.org/abs/1810.08461}{{\ttfamily arXiv:1810.08461
  [hep-ph]}}\relax
\mciteBstWouldAddEndPunctfalse
\mciteSetBstMidEndSepPunct{\mcitedefaultmidpunct}
{}{\mcitedefaultseppunct}\relax
\EndOfBibitem
\bibitem{deBrito:2025ges}
G.~P. de~Brito, A.~Eichhorn, A.~D. Pereira, and M.~Yamada, ``{Neutrino mass
  generation in asymptotically safe gravity},''
  \href{http://dx.doi.org/10.1103/m137-zx8f}{{\em Phys. Rev. D} {\bfseries 112}
  no.~6, (2025) 066004}, \href{http://arxiv.org/abs/2505.01422}{{\ttfamily
  arXiv:2505.01422 [hep-ph]}}\relax
\mciteBstWouldAddEndPunctfalse
\mciteSetBstMidEndSepPunct{\mcitedefaultmidpunct}
{}{\mcitedefaultseppunct}\relax
\EndOfBibitem
\bibitem{Kowalska:2020gie}
K.~Kowalska, E.~M. Sessolo, and Y.~Yamamoto, ``{Flavor anomalies from
  asymptotically safe gravity},''
  \href{http://dx.doi.org/10.1140/epjc/s10052-021-09072-1}{{\em Eur. Phys. J.
  C} {\bfseries 81} no.~4, (2021) 272},
  \href{http://arxiv.org/abs/2007.03567}{{\ttfamily arXiv:2007.03567
  [hep-ph]}}\relax
\mciteBstWouldAddEndPunctfalse
\mciteSetBstMidEndSepPunct{\mcitedefaultmidpunct}
{}{\mcitedefaultseppunct}\relax
\EndOfBibitem
\bibitem{Chikkaballi:2022urc}
A.~Chikkaballi, W.~Kotlarski, K.~Kowalska, D.~Rizzo, and E.~M. Sessolo,
  ``{Constraints on Z' solutions to the flavor anomalies with trans-Planckian
  asymptotic safety},'' \href{http://dx.doi.org/10.1007/JHEP01(2023)164}{{\em
  JHEP} {\bfseries 01} (2023) 164},
  \href{http://arxiv.org/abs/2209.07971}{{\ttfamily arXiv:2209.07971
  [hep-ph]}}\relax
\mciteBstWouldAddEndPunctfalse
\mciteSetBstMidEndSepPunct{\mcitedefaultmidpunct}
{}{\mcitedefaultseppunct}\relax
\EndOfBibitem
\bibitem{Kowalska:2020zve}
K.~Kowalska and E.~M. Sessolo, ``{Minimal models for g-2 and dark matter
  confront asymptotic safety},''
  \href{http://dx.doi.org/10.1103/PhysRevD.103.115032}{{\em Phys. Rev. D}
  {\bfseries 103} no.~11, (2021) 115032},
  \href{http://arxiv.org/abs/2012.15200}{{\ttfamily arXiv:2012.15200
  [hep-ph]}}\relax
\mciteBstWouldAddEndPunctfalse
\mciteSetBstMidEndSepPunct{\mcitedefaultmidpunct}
{}{\mcitedefaultseppunct}\relax
\EndOfBibitem
\bibitem{Boos:2022jvc}
J.~Boos, C.~D. Carone, N.~L. Donald, and M.~R. Musser, ``{Asymptotic safety and
  gauged baryon number},''
  \href{http://dx.doi.org/10.1103/PhysRevD.106.035015}{{\em Phys. Rev. D}
  {\bfseries 106} no.~3, (2022) 035015},
  \href{http://arxiv.org/abs/2206.02686}{{\ttfamily arXiv:2206.02686
  [hep-ph]}}\relax
\mciteBstWouldAddEndPunctfalse
\mciteSetBstMidEndSepPunct{\mcitedefaultmidpunct}
{}{\mcitedefaultseppunct}\relax
\EndOfBibitem
\bibitem{Boos:2022pyq}
J.~Boos, C.~D. Carone, N.~L. Donald, and M.~R. Musser, ``{Asymptotically safe
  dark matter with gauged baryon number},''
  \href{http://dx.doi.org/10.1103/PhysRevD.107.035018}{{\em Phys. Rev. D}
  {\bfseries 107} no.~3, (2023) 035018},
  \href{http://arxiv.org/abs/2209.14268}{{\ttfamily arXiv:2209.14268
  [hep-ph]}}\relax
\mciteBstWouldAddEndPunctfalse
\mciteSetBstMidEndSepPunct{\mcitedefaultmidpunct}
{}{\mcitedefaultseppunct}\relax
\EndOfBibitem
\bibitem{Pastor-Gutierrez:2022nki}
A.~Pastor-Guti\'errez, J.~M. Pawlowski, and M.~Reichert, ``{The Asymptotically
  Safe Standard Model: From quantum gravity to dynamical chiral symmetry
  breaking},'' \href{http://dx.doi.org/10.21468/SciPostPhys.15.3.105}{{\em
  SciPost Phys.} {\bfseries 15} no.~3, (2023) 105},
  \href{http://arxiv.org/abs/2207.09817}{{\ttfamily arXiv:2207.09817
  [hep-th]}}\relax
\mciteBstWouldAddEndPunctfalse
\mciteSetBstMidEndSepPunct{\mcitedefaultmidpunct}
{}{\mcitedefaultseppunct}\relax
\EndOfBibitem
\bibitem{Pastor-Gutierrez:2024sbt}
{\'A}.~Pastor-Guti{\'e}rrez, J.~M. Pawlowski, M.~Reichert, and G.~Ruisi,
  ``{e+e-{\textrightarrow}{\ensuremath{\mu}}+{\ensuremath{\mu}}- in the
  asymptotically safe standard model},''
  \href{http://dx.doi.org/10.1103/PhysRevD.111.106005}{{\em Phys. Rev. D}
  {\bfseries 111} no.~10, (2025) 106005},
  \href{http://arxiv.org/abs/2412.13800}{{\ttfamily arXiv:2412.13800
  [hep-ph]}}\relax
\mciteBstWouldAddEndPunctfalse
\mciteSetBstMidEndSepPunct{\mcitedefaultmidpunct}
{}{\mcitedefaultseppunct}\relax
\EndOfBibitem
\bibitem{Chikkaballi:2023cce}
A.~Chikkaballi, K.~Kowalska, and E.~M. Sessolo, ``{Naturally small neutrino
  mass with asymptotic safety and gravitational-wave signatures},''
  \href{http://dx.doi.org/10.1007/JHEP11(2023)224}{{\em JHEP} {\bfseries 11}
  (2023) 224}, \href{http://arxiv.org/abs/2308.06114}{{\ttfamily
  arXiv:2308.06114 [hep-ph]}}\relax
\mciteBstWouldAddEndPunctfalse
\mciteSetBstMidEndSepPunct{\mcitedefaultmidpunct}
{}{\mcitedefaultseppunct}\relax
\EndOfBibitem
\bibitem{Eichhorn:2022vgp}
A.~Eichhorn and A.~Held, ``{Dynamically vanishing Dirac neutrino mass from
  quantum scale symmetry},''
  \href{http://dx.doi.org/10.1016/j.physletb.2023.138196}{{\em Phys. Lett. B}
  {\bfseries 846} (2023) 138196},
  \href{http://arxiv.org/abs/2204.09008}{{\ttfamily arXiv:2204.09008
  [hep-ph]}}\relax
\mciteBstWouldAddEndPunctfalse
\mciteSetBstMidEndSepPunct{\mcitedefaultmidpunct}
{}{\mcitedefaultseppunct}\relax
\EndOfBibitem
\bibitem{McDonald:2001vt}
J.~McDonald, ``{Thermally generated gauge singlet scalars as selfinteracting
  dark matter},'' \href{http://dx.doi.org/10.1103/PhysRevLett.88.091304}{{\em
  Phys. Rev. Lett.} {\bfseries 88} (2002) 091304},
  \href{http://arxiv.org/abs/hep-ph/0106249}{{\ttfamily
  arXiv:hep-ph/0106249}}\relax
\mciteBstWouldAddEndPunctfalse
\mciteSetBstMidEndSepPunct{\mcitedefaultmidpunct}
{}{\mcitedefaultseppunct}\relax
\EndOfBibitem
\bibitem{Choi:2005vq}
K.-Y. Choi and L.~Roszkowski, ``{E-WIMPs},''
  \href{http://dx.doi.org/10.1063/1.2149672}{{\em AIP Conf. Proc.} {\bfseries
  805} no.~1, (2005) 30--36},
  \href{http://arxiv.org/abs/hep-ph/0511003}{{\ttfamily
  arXiv:hep-ph/0511003}}\relax
\mciteBstWouldAddEndPunctfalse
\mciteSetBstMidEndSepPunct{\mcitedefaultmidpunct}
{}{\mcitedefaultseppunct}\relax
\EndOfBibitem
\bibitem{Kusenko:2006rh}
A.~Kusenko, ``{Sterile neutrinos, dark matter, and the pulsar velocities in
  models with a Higgs singlet},''
  \href{http://dx.doi.org/10.1103/PhysRevLett.97.241301}{{\em Phys. Rev. Lett.}
  {\bfseries 97} (2006) 241301},
  \href{http://arxiv.org/abs/hep-ph/0609081}{{\ttfamily
  arXiv:hep-ph/0609081}}\relax
\mciteBstWouldAddEndPunctfalse
\mciteSetBstMidEndSepPunct{\mcitedefaultmidpunct}
{}{\mcitedefaultseppunct}\relax
\EndOfBibitem
\bibitem{Petraki:2007gq}
K.~Petraki and A.~Kusenko, ``{Dark-matter sterile neutrinos in models with a
  gauge singlet in the Higgs sector},''
  \href{http://dx.doi.org/10.1103/PhysRevD.77.065014}{{\em Phys. Rev. D}
  {\bfseries 77} (2008) 065014},
  \href{http://arxiv.org/abs/0711.4646}{{\ttfamily arXiv:0711.4646
  [hep-ph]}}\relax
\mciteBstWouldAddEndPunctfalse
\mciteSetBstMidEndSepPunct{\mcitedefaultmidpunct}
{}{\mcitedefaultseppunct}\relax
\EndOfBibitem
\bibitem{Hall:2009bx}
L.~J. Hall, K.~Jedamzik, J.~March-Russell, and S.~M. West, ``{Freeze-In
  Production of FIMP Dark Matter},''
  \href{http://dx.doi.org/10.1007/JHEP03(2010)080}{{\em JHEP} {\bfseries 03}
  (2010) 080}, \href{http://arxiv.org/abs/0911.1120}{{\ttfamily arXiv:0911.1120
  [hep-ph]}}\relax
\mciteBstWouldAddEndPunctfalse
\mciteSetBstMidEndSepPunct{\mcitedefaultmidpunct}
{}{\mcitedefaultseppunct}\relax
\EndOfBibitem
\bibitem{Ferrari:2018rey}
S.~Ferrari, T.~Hambye, J.~Heeck, and M.~H.~G. Tytgat, ``{SO(10) paths to dark
  matter},'' \href{http://dx.doi.org/10.1103/PhysRevD.99.055032}{{\em Phys.
  Rev. D} {\bfseries 99} no.~5, (2019) 055032},
  \href{http://arxiv.org/abs/1811.07910}{{\ttfamily arXiv:1811.07910
  [hep-ph]}}\relax
\mciteBstWouldAddEndPunctfalse
\mciteSetBstMidEndSepPunct{\mcitedefaultmidpunct}
{}{\mcitedefaultseppunct}\relax
\EndOfBibitem
\bibitem{Rizzo:2022lpm}
T.~G. Rizzo, ``{Toward a UV model of kinetic mixing and portal matter. II.
  Exploring unification in an SU(N) group},''
  \href{http://dx.doi.org/10.1103/PhysRevD.106.095024}{{\em Phys. Rev. D}
  {\bfseries 106} no.~9, (2022) 095024},
  \href{http://arxiv.org/abs/2209.00688}{{\ttfamily arXiv:2209.00688
  [hep-ph]}}\relax
\mciteBstWouldAddEndPunctfalse
\mciteSetBstMidEndSepPunct{\mcitedefaultmidpunct}
{}{\mcitedefaultseppunct}\relax
\EndOfBibitem
\bibitem{Cacciapaglia:2019ixa}
G.~Cacciapaglia, H.~Cai, A.~Deandrea, and A.~Kushwaha, ``{Composite Higgs and
  Dark Matter Model in SU(6)/SO(6)},''
  \href{http://dx.doi.org/10.1007/JHEP10(2019)035}{{\em JHEP} {\bfseries 10}
  (2019) 035}, \href{http://arxiv.org/abs/1904.09301}{{\ttfamily
  arXiv:1904.09301 [hep-ph]}}\relax
\mciteBstWouldAddEndPunctfalse
\mciteSetBstMidEndSepPunct{\mcitedefaultmidpunct}
{}{\mcitedefaultseppunct}\relax
\EndOfBibitem
\bibitem{Cai:2020njb}
H.~Cai and G.~Cacciapaglia, ``{Singlet dark matter in the SU(6)/SO(6) composite
  Higgs model},'' \href{http://dx.doi.org/10.1103/PhysRevD.103.055002}{{\em
  Phys. Rev. D} {\bfseries 103} no.~5, (2021) 055002},
  \href{http://arxiv.org/abs/2007.04338}{{\ttfamily arXiv:2007.04338
  [hep-ph]}}\relax
\mciteBstWouldAddEndPunctfalse
\mciteSetBstMidEndSepPunct{\mcitedefaultmidpunct}
{}{\mcitedefaultseppunct}\relax
\EndOfBibitem
\bibitem{Otsuka:2022zdy}
H.~Otsuka, T.~Shimomura, K.~Tsumura, Y.~Uchida, and N.~Yamatsu,
  ``{Pseudo-Nambu-Goldstone dark matter from non-Abelian gauge symmetry},''
  \href{http://dx.doi.org/10.1103/PhysRevD.106.115033}{{\em Phys. Rev. D}
  {\bfseries 106} no.~11, (2022) 115033},
  \href{http://arxiv.org/abs/2210.08696}{{\ttfamily arXiv:2210.08696
  [hep-ph]}}\relax
\mciteBstWouldAddEndPunctfalse
\mciteSetBstMidEndSepPunct{\mcitedefaultmidpunct}
{}{\mcitedefaultseppunct}\relax
\EndOfBibitem
\bibitem{Chiang:2023omu}
C.-W. Chiang, K.~Tsumura, Y.~Uchida, and N.~Yamatsu, ``{Pseudo-Nambu-Goldstone
  dark matter in SU(7) grand unification},''
  \href{http://dx.doi.org/10.1103/PhysRevD.109.055040}{{\em Phys. Rev. D}
  {\bfseries 109} no.~5, (2024) 055040},
  \href{http://arxiv.org/abs/2311.13753}{{\ttfamily arXiv:2311.13753
  [hep-ph]}}\relax
\mciteBstWouldAddEndPunctfalse
\mciteSetBstMidEndSepPunct{\mcitedefaultmidpunct}
{}{\mcitedefaultseppunct}\relax
\EndOfBibitem
\bibitem{Ma:2020hyy}
E.~Ma, ``{Dark matter from $SU(6) \to SU(5) \times U(1)_N$},''
  \href{http://dx.doi.org/10.1103/PhysRevD.103.L051704}{{\em Phys. Rev. D}
  {\bfseries 103} no.~5, (2021) L051704},
  \href{http://arxiv.org/abs/2011.01398}{{\ttfamily arXiv:2011.01398
  [hep-ph]}}\relax
\mciteBstWouldAddEndPunctfalse
\mciteSetBstMidEndSepPunct{\mcitedefaultmidpunct}
{}{\mcitedefaultseppunct}\relax
\EndOfBibitem
\bibitem{Codello:2007bd}
A.~Codello, R.~Percacci, and C.~Rahmede, ``{Ultraviolet properties of
  f(R)-gravity},'' \href{http://dx.doi.org/10.1142/S0217751X08038135}{{\em Int.
  J. Mod. Phys. A} {\bfseries 23} (2008) 143--150},
  \href{http://arxiv.org/abs/0705.1769}{{\ttfamily arXiv:0705.1769
  [hep-th]}}\relax
\mciteBstWouldAddEndPunctfalse
\mciteSetBstMidEndSepPunct{\mcitedefaultmidpunct}
{}{\mcitedefaultseppunct}\relax
\EndOfBibitem
\bibitem{Narain:2009qa}
G.~Narain and R.~Percacci, ``{On the scheme dependence of gravitational beta
  functions},'' {\em Acta Phys. Polon. B} {\bfseries 40} (2009) 3439--3457,
  \href{http://arxiv.org/abs/0910.5390}{{\ttfamily arXiv:0910.5390
  [hep-th]}}\relax
\mciteBstWouldAddEndPunctfalse
\mciteSetBstMidEndSepPunct{\mcitedefaultmidpunct}
{}{\mcitedefaultseppunct}\relax
\EndOfBibitem
\bibitem{Falls:2017lst}
K.~Falls, C.~R. King, D.~F. Litim, K.~Nikolakopoulos, and C.~Rahmede,
  ``{Asymptotic safety of quantum gravity beyond Ricci scalars},''
  \href{http://dx.doi.org/10.1103/PhysRevD.97.086006}{{\em Phys. Rev. D}
  {\bfseries 97} no.~8, (2018) 086006},
  \href{http://arxiv.org/abs/1801.00162}{{\ttfamily arXiv:1801.00162
  [hep-th]}}\relax
\mciteBstWouldAddEndPunctfalse
\mciteSetBstMidEndSepPunct{\mcitedefaultmidpunct}
{}{\mcitedefaultseppunct}\relax
\EndOfBibitem
\bibitem{Falls:2018ylp}
K.~G. Falls, D.~F. Litim, and J.~Schröder, ``{Aspects of asymptotic safety for
  quantum gravity},'' \href{http://dx.doi.org/10.1103/PhysRevD.99.126015}{{\em
  Phys. Rev. D} {\bfseries 99} no.~12, (2019) 126015},
  \href{http://arxiv.org/abs/1810.08550}{{\ttfamily arXiv:1810.08550
  [gr-qc]}}\relax
\mciteBstWouldAddEndPunctfalse
\mciteSetBstMidEndSepPunct{\mcitedefaultmidpunct}
{}{\mcitedefaultseppunct}\relax
\EndOfBibitem
\bibitem{deBrito:2022vbr}
G.~P. de~Brito and A.~Eichhorn, ``{Nonvanishing gravitational contribution to
  matter beta functions for vanishing dimensionful regulators},''
  \href{http://dx.doi.org/10.1140/epjc/s10052-023-11172-z}{{\em Eur. Phys. J.
  C} {\bfseries 83} no.~2, (2023) 161},
  \href{http://arxiv.org/abs/2201.11402}{{\ttfamily arXiv:2201.11402
  [hep-th]}}\relax
\mciteBstWouldAddEndPunctfalse
\mciteSetBstMidEndSepPunct{\mcitedefaultmidpunct}
{}{\mcitedefaultseppunct}\relax
\EndOfBibitem
\bibitem{Toms:2008dq}
D.~J. Toms, ``{Cosmological constant and quantum gravitational corrections to
  the running fine structure constant},''
  \href{http://dx.doi.org/10.1103/PhysRevLett.101.131301}{{\em Phys. Rev.
  Lett.} {\bfseries 101} (2008) 131301},
  \href{http://arxiv.org/abs/0809.3897}{{\ttfamily arXiv:0809.3897
  [hep-th]}}\relax
\mciteBstWouldAddEndPunctfalse
\mciteSetBstMidEndSepPunct{\mcitedefaultmidpunct}
{}{\mcitedefaultseppunct}\relax
\EndOfBibitem
\bibitem{Toms:2009vd}
D.~J. Toms, ``{Quantum gravity, gauge coupling constants, and the cosmological
  constant},'' \href{http://dx.doi.org/10.1103/PhysRevD.80.064040}{{\em Phys.
  Rev. D} {\bfseries 80} (2009) 064040},
  \href{http://arxiv.org/abs/0908.3100}{{\ttfamily arXiv:0908.3100
  [hep-th]}}\relax
\mciteBstWouldAddEndPunctfalse
\mciteSetBstMidEndSepPunct{\mcitedefaultmidpunct}
{}{\mcitedefaultseppunct}\relax
\EndOfBibitem
\bibitem{Rodigast:2009zj}
A.~Rodigast and T.~Schuster, ``{Gravitational Corrections to Yukawa and phi**4
  Interactions},'' \href{http://dx.doi.org/10.1103/PhysRevLett.104.081301}{{\em
  Phys. Rev. Lett.} {\bfseries 104} (2010) 081301},
  \href{http://arxiv.org/abs/0908.2422}{{\ttfamily arXiv:0908.2422
  [hep-th]}}\relax
\mciteBstWouldAddEndPunctfalse
\mciteSetBstMidEndSepPunct{\mcitedefaultmidpunct}
{}{\mcitedefaultseppunct}\relax
\EndOfBibitem
\bibitem{Kotlarski:2023mmr}
W.~Kotlarski, K.~Kowalska, D.~Rizzo, and E.~M. Sessolo, ``{How robust are
  particle physics predictions in asymptotic safety?},''
  \href{http://dx.doi.org/10.1140/epjc/s10052-023-11813-3}{{\em Eur. Phys. J.
  C} {\bfseries 83} no.~7, (2023) 644},
  \href{http://arxiv.org/abs/2304.08959}{{\ttfamily arXiv:2304.08959
  [hep-ph]}}\relax
\mciteBstWouldAddEndPunctfalse
\mciteSetBstMidEndSepPunct{\mcitedefaultmidpunct}
{}{\mcitedefaultseppunct}\relax
\EndOfBibitem
\bibitem{Eichhorn:2020kca}
A.~Eichhorn and M.~Pauly, ``{Safety in darkness: Higgs portal to simple Yukawa
  systems},'' \href{http://dx.doi.org/10.1016/j.physletb.2021.136455}{{\em
  Phys. Lett. B} {\bfseries 819} (2021) 136455},
  \href{http://arxiv.org/abs/2005.03661}{{\ttfamily arXiv:2005.03661
  [hep-ph]}}\relax
\mciteBstWouldAddEndPunctfalse
\mciteSetBstMidEndSepPunct{\mcitedefaultmidpunct}
{}{\mcitedefaultseppunct}\relax
\EndOfBibitem
\bibitem{Reichert:2019car}
M.~Reichert and J.~Smirnov, ``{Dark Matter meets Quantum Gravity},''
  \href{http://dx.doi.org/10.1103/PhysRevD.101.063015}{{\em Phys. Rev. D}
  {\bfseries 101} no.~6, (2020) 063015},
  \href{http://arxiv.org/abs/1911.00012}{{\ttfamily arXiv:1911.00012
  [hep-ph]}}\relax
\mciteBstWouldAddEndPunctfalse
\mciteSetBstMidEndSepPunct{\mcitedefaultmidpunct}
{}{\mcitedefaultseppunct}\relax
\EndOfBibitem
\bibitem{deBrito:2023ydd}
G.~P. de~Brito, A.~Eichhorn, M.~T. Frandsen, M.~Rosenlyst, M.~E. Thing, and
  A.~F. Vieira, ``{Ruling out models of vector dark matter in asymptotically
  safe quantum gravity},''
  \href{http://dx.doi.org/10.1103/PhysRevD.109.055022}{{\em Phys. Rev. D}
  {\bfseries 109} no.~5, (2024) 055022},
  \href{http://arxiv.org/abs/2312.02086}{{\ttfamily arXiv:2312.02086
  [hep-ph]}}\relax
\mciteBstWouldAddEndPunctfalse
\mciteSetBstMidEndSepPunct{\mcitedefaultmidpunct}
{}{\mcitedefaultseppunct}\relax
\EndOfBibitem
\bibitem{deBrito:2021akp}
G.~P. de~Brito, A.~Eichhorn, and R.~R. Lino~dos Santos, ``{Are there ALPs in
  the asymptotically safe landscape?},''
  \href{http://dx.doi.org/10.1007/JHEP06(2022)013}{{\em JHEP} {\bfseries 06}
  (2022) 013}, \href{http://arxiv.org/abs/2112.08972}{{\ttfamily
  arXiv:2112.08972 [gr-qc]}}\relax
\mciteBstWouldAddEndPunctfalse
\mciteSetBstMidEndSepPunct{\mcitedefaultmidpunct}
{}{\mcitedefaultseppunct}\relax
\EndOfBibitem
\bibitem{Eichhorn:2022ngh}
A.~Eichhorn, R.~R.~L. dos Santos, and F.~Wagner, ``{Shift-symmetric Horndeski
  gravity in the asymptotic-safety paradigm},''
  \href{http://dx.doi.org/10.1088/1475-7516/2023/02/052}{{\em JCAP} {\bfseries
  02} (2023) 052}, \href{http://arxiv.org/abs/2212.08441}{{\ttfamily
  arXiv:2212.08441 [gr-qc]}}\relax
\mciteBstWouldAddEndPunctfalse
\mciteSetBstMidEndSepPunct{\mcitedefaultmidpunct}
{}{\mcitedefaultseppunct}\relax
\EndOfBibitem
\bibitem{Eichhorn:2023gat}
A.~Eichhorn, R.~R. Lino~dos Santos, and J.~a.~L. Miqueleto, ``{From quantum
  gravity to gravitational waves through cosmic strings},''
  \href{http://dx.doi.org/10.1103/PhysRevD.109.026013}{{\em Phys. Rev. D}
  {\bfseries 109} no.~2, (2024) 026013},
  \href{http://arxiv.org/abs/2306.17718}{{\ttfamily arXiv:2306.17718
  [gr-qc]}}\relax
\mciteBstWouldAddEndPunctfalse
\mciteSetBstMidEndSepPunct{\mcitedefaultmidpunct}
{}{\mcitedefaultseppunct}\relax
\EndOfBibitem
\bibitem{Eichhorn:2019dhg}
A.~Eichhorn, A.~Held, and C.~Wetterich, ``{Predictive power of grand
  unification from quantum gravity},''
  \href{http://dx.doi.org/10.1007/JHEP08(2020)111}{{\em JHEP} {\bfseries 08}
  (2020) 111}, \href{http://arxiv.org/abs/1909.07318}{{\ttfamily
  arXiv:1909.07318 [hep-th]}}\relax
\mciteBstWouldAddEndPunctfalse
\mciteSetBstMidEndSepPunct{\mcitedefaultmidpunct}
{}{\mcitedefaultseppunct}\relax
\EndOfBibitem
\bibitem{Held:2022hnw}
A.~Held, J.~Kwapisz, and L.~Sartore, ``{Grand unification and the Planck scale:
  an SO(10) example of radiative symmetry breaking},''
  \href{http://dx.doi.org/10.1007/JHEP08(2022)122}{{\em JHEP} {\bfseries 08}
  (2022) 122}, \href{http://arxiv.org/abs/2204.03001}{{\ttfamily
  arXiv:2204.03001 [hep-ph]}}\relax
\mciteBstWouldAddEndPunctfalse
\mciteSetBstMidEndSepPunct{\mcitedefaultmidpunct}
{}{\mcitedefaultseppunct}\relax
\EndOfBibitem
\bibitem{Donoghue:2019clr}
J.~F. Donoghue, ``{A Critique of the Asymptotic Safety Program},''
  \href{http://dx.doi.org/10.3389/fphy.2020.00056}{{\em Front. in Phys.}
  {\bfseries 8} (2020) 56}, \href{http://arxiv.org/abs/1911.02967}{{\ttfamily
  arXiv:1911.02967 [hep-th]}}\relax
\mciteBstWouldAddEndPunctfalse
\mciteSetBstMidEndSepPunct{\mcitedefaultmidpunct}
{}{\mcitedefaultseppunct}\relax
\EndOfBibitem
\bibitem{Bonanno:2020bil}
A.~Bonanno, A.~Eichhorn, H.~Gies, J.~M. Pawlowski, R.~Percacci, M.~Reuter,
  F.~Saueressig, and G.~P. Vacca, ``{Critical reflections on asymptotically
  safe gravity},'' \href{http://dx.doi.org/10.3389/fphy.2020.00269}{{\em Front.
  in Phys.} {\bfseries 8} (2020) 269},
  \href{http://arxiv.org/abs/2004.06810}{{\ttfamily arXiv:2004.06810
  [gr-qc]}}\relax
\mciteBstWouldAddEndPunctfalse
\mciteSetBstMidEndSepPunct{\mcitedefaultmidpunct}
{}{\mcitedefaultseppunct}\relax
\EndOfBibitem
\bibitem{Branchina:2024xzh}
C.~Branchina, V.~Branchina, F.~Contino, and A.~Pernace, ``{Path integral
  measure and the cosmological constant},''
  \href{http://dx.doi.org/10.1103/PhysRevD.111.105018}{{\em Phys. Rev. D}
  {\bfseries 111} no.~10, (2025) 105018},
  \href{http://arxiv.org/abs/2412.10194}{{\ttfamily arXiv:2412.10194
  [hep-th]}}\relax
\mciteBstWouldAddEndPunctfalse
\mciteSetBstMidEndSepPunct{\mcitedefaultmidpunct}
{}{\mcitedefaultseppunct}\relax
\EndOfBibitem
\bibitem{Branchina:2024lai}
C.~Branchina, V.~Branchina, F.~Contino, and A.~Pernace, ``{Path integral
  measure and RG equations for gravity},''
  \href{http://dx.doi.org/10.1103/wqv2-j5dt}{{\em Phys. Rev. D} {\bfseries 111}
  no.~12, (2025) 125021}, \href{http://arxiv.org/abs/2412.14108}{{\ttfamily
  arXiv:2412.14108 [hep-th]}}\relax
\mciteBstWouldAddEndPunctfalse
\mciteSetBstMidEndSepPunct{\mcitedefaultmidpunct}
{}{\mcitedefaultseppunct}\relax
\EndOfBibitem
\bibitem{Bonanno:2025xdg}
A.~Bonanno, K.~Falls, and R.~Ferrero, ``{Path integral measures and
  diffeomorphism invariance},''
  \href{http://dx.doi.org/10.1007/JHEP05(2025)164}{{\em JHEP} {\bfseries 05}
  (2025) 164}, \href{http://arxiv.org/abs/2503.02941}{{\ttfamily
  arXiv:2503.02941 [hep-th]}}\relax
\mciteBstWouldAddEndPunctfalse
\mciteSetBstMidEndSepPunct{\mcitedefaultmidpunct}
{}{\mcitedefaultseppunct}\relax
\EndOfBibitem
\bibitem{Held:2025vkd}
A.~Held, B.~Knorr, J.~M. Pawlowski, A.~Platania, M.~Reichert, F.~Saueressig,
  and M.~Schiffer, ``{Comment on Path integral measure and RG equations for
  gravity},'' \href{http://arxiv.org/abs/2504.12006}{{\ttfamily
  arXiv:2504.12006 [hep-th]}}\relax
\mciteBstWouldAddEndPunctfalse
\mciteSetBstMidEndSepPunct{\mcitedefaultmidpunct}
{}{\mcitedefaultseppunct}\relax
\EndOfBibitem
\bibitem{Fehre:2021eob}
J.~Fehre, D.~F. Litim, J.~M. Pawlowski, and M.~Reichert, ``{Lorentzian Quantum
  Gravity and the Graviton Spectral Function},''
  \href{http://dx.doi.org/10.1103/PhysRevLett.130.081501}{{\em Phys. Rev.
  Lett.} {\bfseries 130} no.~8, (2023) 081501},
  \href{http://arxiv.org/abs/2111.13232}{{\ttfamily arXiv:2111.13232
  [hep-th]}}\relax
\mciteBstWouldAddEndPunctfalse
\mciteSetBstMidEndSepPunct{\mcitedefaultmidpunct}
{}{\mcitedefaultseppunct}\relax
\EndOfBibitem
\bibitem{Pawlowski:2023gym}
J.~M. Pawlowski and M.~Reichert, {\em {Quantum Gravity from Dynamical Metric
  Fluctuations}}.
\newblock 2024.
\newblock \href{http://arxiv.org/abs/2309.10785}{{\ttfamily arXiv:2309.10785
  [hep-th]}}\relax
\mciteBstWouldAddEndPunctfalse
\mciteSetBstMidEndSepPunct{\mcitedefaultmidpunct}
{}{\mcitedefaultseppunct}\relax
\EndOfBibitem
\bibitem{DAngelo:2023wje}
E.~D'Angelo, ``{Asymptotic safety in Lorentzian quantum gravity},''
  \href{http://dx.doi.org/10.1103/PhysRevD.109.066012}{{\em Phys. Rev. D}
  {\bfseries 109} no.~6, (2024) 066012},
  \href{http://arxiv.org/abs/2310.20603}{{\ttfamily arXiv:2310.20603
  [hep-th]}}\relax
\mciteBstWouldAddEndPunctfalse
\mciteSetBstMidEndSepPunct{\mcitedefaultmidpunct}
{}{\mcitedefaultseppunct}\relax
\EndOfBibitem
\bibitem{Poole:2019kcm}
C.~Poole and A.~E. Thomsen, ``{Constraints on 3- and 4-loop $\beta$-functions
  in a general four-dimensional Quantum Field Theory},''
  \href{http://dx.doi.org/10.1007/JHEP09(2019)055}{{\em JHEP} {\bfseries 09}
  (2019) 055}, \href{http://arxiv.org/abs/1906.04625}{{\ttfamily
  arXiv:1906.04625 [hep-th]}}\relax
\mciteBstWouldAddEndPunctfalse
\mciteSetBstMidEndSepPunct{\mcitedefaultmidpunct}
{}{\mcitedefaultseppunct}\relax
\EndOfBibitem
\bibitem{Sartore:2020gou}
L.~Sartore and I.~Schienbein, ``{PyR@TE 3},''
  \href{http://dx.doi.org/10.1016/j.cpc.2020.107819}{{\em Comput. Phys.
  Commun.} {\bfseries 261} (2021) 107819},
  \href{http://arxiv.org/abs/2007.12700}{{\ttfamily arXiv:2007.12700
  [hep-ph]}}\relax
\mciteBstWouldAddEndPunctfalse
\mciteSetBstMidEndSepPunct{\mcitedefaultmidpunct}
{}{\mcitedefaultseppunct}\relax
\EndOfBibitem
\bibitem{Thomsen:2021ncy}
A.~E. Thomsen, ``{Introducing RGBeta: a Mathematica package for the evaluation
  of renormalization group $ \beta $-functions},''
  \href{http://dx.doi.org/10.1140/epjc/s10052-021-09142-4}{{\em Eur. Phys. J.
  C} {\bfseries 81} no.~5, (2021) 408},
  \href{http://arxiv.org/abs/2101.08265}{{\ttfamily arXiv:2101.08265
  [hep-ph]}}\relax
\mciteBstWouldAddEndPunctfalse
\mciteSetBstMidEndSepPunct{\mcitedefaultmidpunct}
{}{\mcitedefaultseppunct}\relax
\EndOfBibitem
\bibitem{Esteban:2024eli}
I.~Esteban, M.~C. Gonzalez-Garcia, M.~Maltoni, I.~Martinez-Soler, J.~a.~P.
  Pinheiro, and T.~Schwetz, ``{NuFit-6.0: updated global analysis of
  three-flavor neutrino oscillations},''
  \href{http://dx.doi.org/10.1007/JHEP12(2024)216}{{\em JHEP} {\bfseries 12}
  (2024) 216}, \href{http://arxiv.org/abs/2410.05380}{{\ttfamily
  arXiv:2410.05380 [hep-ph]}}\relax
\mciteBstWouldAddEndPunctfalse
\mciteSetBstMidEndSepPunct{\mcitedefaultmidpunct}
{}{\mcitedefaultseppunct}\relax
\EndOfBibitem
\bibitem{Lindner:2014oea}
M.~Lindner, S.~Schmidt, and J.~Smirnov, ``{Neutrino Masses and Conformal
  Electro-Weak Symmetry Breaking},''
  \href{http://dx.doi.org/10.1007/JHEP10(2014)177}{{\em JHEP} {\bfseries 10}
  (2014) 177}, \href{http://arxiv.org/abs/1405.6204}{{\ttfamily arXiv:1405.6204
  [hep-ph]}}\relax
\mciteBstWouldAddEndPunctfalse
\mciteSetBstMidEndSepPunct{\mcitedefaultmidpunct}
{}{\mcitedefaultseppunct}\relax
\EndOfBibitem
\bibitem{Wetterich:2016uxm}
C.~Wetterich and M.~Yamada, ``{Gauge hierarchy problem in asymptotically safe
  gravity--the resurgence mechanism},''
  \href{http://dx.doi.org/10.1016/j.physletb.2017.04.049}{{\em Phys. Lett. B}
  {\bfseries 770} (2017) 268--271},
  \href{http://arxiv.org/abs/1612.03069}{{\ttfamily arXiv:1612.03069
  [hep-th]}}\relax
\mciteBstWouldAddEndPunctfalse
\mciteSetBstMidEndSepPunct{\mcitedefaultmidpunct}
{}{\mcitedefaultseppunct}\relax
\EndOfBibitem
\bibitem{Pawlowski:2018ixd}
J.~M. Pawlowski, M.~Reichert, C.~Wetterich, and M.~Yamada, ``{Higgs scalar
  potential in asymptotically safe quantum gravity},''
  \href{http://dx.doi.org/10.1103/PhysRevD.99.086010}{{\em Phys. Rev. D}
  {\bfseries 99} no.~8, (2019) 086010},
  \href{http://arxiv.org/abs/1811.11706}{{\ttfamily arXiv:1811.11706
  [hep-th]}}\relax
\mciteBstWouldAddEndPunctfalse
\mciteSetBstMidEndSepPunct{\mcitedefaultmidpunct}
{}{\mcitedefaultseppunct}\relax
\EndOfBibitem
\bibitem{Wetterich:2019zdo}
C.~Wetterich and M.~Yamada, ``{Variable Planck mass from the gauge invariant
  flow equation},'' \href{http://dx.doi.org/10.1103/PhysRevD.100.066017}{{\em
  Phys. Rev. D} {\bfseries 100} no.~6, (2019) 066017},
  \href{http://arxiv.org/abs/1906.01721}{{\ttfamily arXiv:1906.01721
  [hep-th]}}\relax
\mciteBstWouldAddEndPunctfalse
\mciteSetBstMidEndSepPunct{\mcitedefaultmidpunct}
{}{\mcitedefaultseppunct}\relax
\EndOfBibitem
\bibitem{Basile:2021krr}
I.~Basile and A.~Platania, ``{Asymptotic Safety: Swampland or Wonderland?},''
  \href{http://dx.doi.org/10.3390/universe7100389}{{\em Universe} {\bfseries 7}
  no.~10, (2021) 389}, \href{http://arxiv.org/abs/2107.06897}{{\ttfamily
  arXiv:2107.06897 [hep-th]}}\relax
\mciteBstWouldAddEndPunctfalse
\mciteSetBstMidEndSepPunct{\mcitedefaultmidpunct}
{}{\mcitedefaultseppunct}\relax
\EndOfBibitem
\bibitem{Knorr:2024yiu}
B.~Knorr and A.~Platania, ``{Unearthing the intersections: positivity bounds,
  weak gravity conjecture, and asymptotic safety landscapes from
  photon-graviton flows},''
  \href{http://dx.doi.org/10.1007/JHEP03(2025)003}{{\em JHEP} {\bfseries 03}
  (2025) 003}, \href{http://arxiv.org/abs/2405.08860}{{\ttfamily
  arXiv:2405.08860 [hep-th]}}\relax
\mciteBstWouldAddEndPunctfalse
\mciteSetBstMidEndSepPunct{\mcitedefaultmidpunct}
{}{\mcitedefaultseppunct}\relax
\EndOfBibitem
\bibitem{Eichhorn:2024rkc}
A.~Eichhorn, A.~Hebecker, J.~M. Pawlowski, and J.~Walcher, ``{The absolute
  swampland},'' \href{http://dx.doi.org/10.1209/0295-5075/ada1f3}{{\em EPL}
  {\bfseries 149} no.~3, (2025) 39001},
  \href{http://arxiv.org/abs/2405.20386}{{\ttfamily arXiv:2405.20386
  [hep-th]}}\relax
\mciteBstWouldAddEndPunctfalse
\mciteSetBstMidEndSepPunct{\mcitedefaultmidpunct}
{}{\mcitedefaultseppunct}\relax
\EndOfBibitem
\bibitem{Basile:2025zjc}
I.~Basile, B.~Knorr, A.~Platania, and M.~Schiffer, ``{Asymptotic safety,
  quantum gravity, and the swampland: a conceptual assessment},''
  \href{http://arxiv.org/abs/2502.12290}{{\ttfamily arXiv:2502.12290
  [hep-th]}}\relax
\mciteBstWouldAddEndPunctfalse
\mciteSetBstMidEndSepPunct{\mcitedefaultmidpunct}
{}{\mcitedefaultseppunct}\relax
\EndOfBibitem
\bibitem{Drees:1992am}
M.~Drees and M.~M. Nojiri, ``{The Neutralino relic density in minimal $N=1$
  supergravity},'' \href{http://dx.doi.org/10.1103/PhysRevD.47.376}{{\em Phys.
  Rev. D} {\bfseries 47} (1993) 376--408},
  \href{http://arxiv.org/abs/hep-ph/9207234}{{\ttfamily
  arXiv:hep-ph/9207234}}\relax
\mciteBstWouldAddEndPunctfalse
\mciteSetBstMidEndSepPunct{\mcitedefaultmidpunct}
{}{\mcitedefaultseppunct}\relax
\EndOfBibitem
\bibitem{Baer:1995nc}
H.~Baer and M.~Brhlik, ``{Cosmological relic density from minimal supergravity
  with implications for collider physics},''
  \href{http://dx.doi.org/10.1103/PhysRevD.53.597}{{\em Phys. Rev. D}
  {\bfseries 53} (1996) 597--605},
  \href{http://arxiv.org/abs/hep-ph/9508321}{{\ttfamily
  arXiv:hep-ph/9508321}}\relax
\mciteBstWouldAddEndPunctfalse
\mciteSetBstMidEndSepPunct{\mcitedefaultmidpunct}
{}{\mcitedefaultseppunct}\relax
\EndOfBibitem
\bibitem{Bai:2014osa}
Y.~Bai and J.~Berger, ``{Lepton Portal Dark Matter},''
  \href{http://dx.doi.org/10.1007/JHEP08(2014)153}{{\em JHEP} {\bfseries 08}
  (2014) 153}, \href{http://arxiv.org/abs/1402.6696}{{\ttfamily arXiv:1402.6696
  [hep-ph]}}\relax
\mciteBstWouldAddEndPunctfalse
\mciteSetBstMidEndSepPunct{\mcitedefaultmidpunct}
{}{\mcitedefaultseppunct}\relax
\EndOfBibitem
\bibitem{Cirelli:2024ssz}
M.~Cirelli, A.~Strumia, and J.~Zupan, ``{Dark Matter},''
  \href{http://arxiv.org/abs/2406.01705}{{\ttfamily arXiv:2406.01705
  [hep-ph]}}\relax
\mciteBstWouldAddEndPunctfalse
\mciteSetBstMidEndSepPunct{\mcitedefaultmidpunct}
{}{\mcitedefaultseppunct}\relax
\EndOfBibitem
\bibitem{Okada:2018tgy}
N.~Okada, S.~Okada, and D.~Raut, ``{Natural Z' -portal Majorana dark matter in
  alternative U(1) extended standard model},''
  \href{http://dx.doi.org/10.1103/PhysRevD.100.035022}{{\em Phys. Rev. D}
  {\bfseries 100} no.~3, (2019) 035022},
  \href{http://arxiv.org/abs/1811.11927}{{\ttfamily arXiv:1811.11927
  [hep-ph]}}\relax
\mciteBstWouldAddEndPunctfalse
\mciteSetBstMidEndSepPunct{\mcitedefaultmidpunct}
{}{\mcitedefaultseppunct}\relax
\EndOfBibitem
\bibitem{Okada:2020cue}
N.~Okada, S.~Okada, and Q.~Shafi, ``{Light $Z?$ and dark matter from U(1)$_X$
  gauge symmetry},''
  \href{http://dx.doi.org/10.1016/j.physletb.2020.135845}{{\em Phys. Lett. B}
  {\bfseries 810} (2020) 135845},
  \href{http://arxiv.org/abs/2003.02667}{{\ttfamily arXiv:2003.02667
  [hep-ph]}}\relax
\mciteBstWouldAddEndPunctfalse
\mciteSetBstMidEndSepPunct{\mcitedefaultmidpunct}
{}{\mcitedefaultseppunct}\relax
\EndOfBibitem
\bibitem{CMS:2021ctt}
{\bfseries CMS} Collaboration, A.~M. Sirunyan {\em et~al.}, ``{Search for
  resonant and nonresonant new phenomena in high-mass dilepton final states at
  $ \sqrt{s} $ = 13 TeV},''
  \href{http://dx.doi.org/10.1007/JHEP07(2021)208}{{\em JHEP} {\bfseries 07}
  (2021) 208}, \href{http://arxiv.org/abs/2103.02708}{{\ttfamily
  arXiv:2103.02708 [hep-ex]}}\relax
\mciteBstWouldAddEndPunctfalse
\mciteSetBstMidEndSepPunct{\mcitedefaultmidpunct}
{}{\mcitedefaultseppunct}\relax
\EndOfBibitem
\bibitem{Roszkowski:2017nbc}
L.~Roszkowski, E.~M. Sessolo, and S.~Trojanowski, ``{WIMP dark matter
  candidates and searches\textemdash{}current status and future prospects},''
  \href{http://dx.doi.org/10.1088/1361-6633/aab913}{{\em Rept. Prog. Phys.}
  {\bfseries 81} no.~6, (2018) 066201},
  \href{http://arxiv.org/abs/1707.06277}{{\ttfamily arXiv:1707.06277
  [hep-ph]}}\relax
\mciteBstWouldAddEndPunctfalse
\mciteSetBstMidEndSepPunct{\mcitedefaultmidpunct}
{}{\mcitedefaultseppunct}\relax
\EndOfBibitem
\bibitem{Kowalska:2018toh}
K.~Kowalska and E.~M. Sessolo, ``{The discreet charm of higgsino dark matter -
  a pocket review},'' \href{http://dx.doi.org/10.1155/2018/6828560}{{\em Adv.
  High Energy Phys.} {\bfseries 2018} (2018) 6828560},
  \href{http://arxiv.org/abs/1802.04097}{{\ttfamily arXiv:1802.04097
  [hep-ph]}}\relax
\mciteBstWouldAddEndPunctfalse
\mciteSetBstMidEndSepPunct{\mcitedefaultmidpunct}
{}{\mcitedefaultseppunct}\relax
\EndOfBibitem
\bibitem{Nagata:2014wma}
N.~Nagata and S.~Shirai, ``{Higgsino Dark Matter in High-Scale
  Supersymmetry},'' \href{http://dx.doi.org/10.1007/JHEP01(2015)029}{{\em JHEP}
  {\bfseries 01} (2015) 029}, \href{http://arxiv.org/abs/1410.4549}{{\ttfamily
  arXiv:1410.4549 [hep-ph]}}\relax
\mciteBstWouldAddEndPunctfalse
\mciteSetBstMidEndSepPunct{\mcitedefaultmidpunct}
{}{\mcitedefaultseppunct}\relax
\EndOfBibitem
\bibitem{Gondolo:1990dk}
P.~Gondolo and G.~Gelmini, ``{Cosmic abundances of stable particles: Improved
  analysis},'' \href{http://dx.doi.org/10.1016/0550-3213(91)90438-4}{{\em Nucl.
  Phys. B} {\bfseries 360} (1991) 145--179}\relax
\mciteBstWouldAddEndPunctfalse
\mciteSetBstMidEndSepPunct{\mcitedefaultmidpunct}
{}{\mcitedefaultseppunct}\relax
\EndOfBibitem
\bibitem{ArkaniHamed:2006mb}
N.~Arkani-Hamed, A.~Delgado, and G.~Giudice, ``{The Well-tempered
  neutralino},'' \href{http://dx.doi.org/10.1016/j.nuclphysb.2006.02.010}{{\em
  Nucl. Phys. B} {\bfseries 741} (2006) 108--130},
  \href{http://arxiv.org/abs/hep-ph/0601041}{{\ttfamily
  arXiv:hep-ph/0601041}}\relax
\mciteBstWouldAddEndPunctfalse
\mciteSetBstMidEndSepPunct{\mcitedefaultmidpunct}
{}{\mcitedefaultseppunct}\relax
\EndOfBibitem
\end{mcitethebibliography}

\end{document}